\pdfoutput=1
\documentclass[11pt,reqno,twoside]{article}

\usepackage{cmap}            
\usepackage[T1]{fontenc}     
\usepackage[utf8]{inputenc}  
\usepackage{lmodern}         
\usepackage[scaled=0.88]{helvet} 
\usepackage{times}           
\usepackage{enumitem}

\usepackage{setspace}
\usepackage[centering,top=1in,bottom=1in,left=1in,right=1in]{geometry}
\usepackage{titlesec}
\usepackage{titling}
\usepackage{amsmath,amssymb,amsthm,amsfonts}
\usepackage{bm}              
\usepackage{mathtools}       
\mathtoolsset{centercolon}   
\usepackage{mathrsfs}        

\usepackage{algorithm,algcompatible}
\usepackage{graphicx,subcaption}
\usepackage{placeins}
\usepackage{array}
\usepackage{microtype}
\usepackage{booktabs}
\usepackage[font=small,margin=12pt,labelfont={sf,bf},labelsep=space]{caption}

\usepackage[dvipsnames]{xcolor}
\definecolor{dkblue}{rgb}{0,0,.5}
\definecolor{rust}{rgb}{0.5,0.1,0.1}

\usepackage{hyperref}
\hypersetup{
    pdftitle={Offline Learning of Reserve Prices from Censored Bids},
    pdfauthor={Lavanya Marla, Yunzong Xu, Boxuan Zhou},
    colorlinks=true,
    linkcolor=dkblue,
    citecolor=rust,
    urlcolor=rust
}
\usepackage{url}

\usepackage{fancyhdr}
\newtheoremstyle{myThm}
    {\topsep}{\topsep}{\itshape}{}{\sffamily\bfseries}{.}{.5em}{}

\newtheoremstyle{myRem}
    {\topsep}{\topsep}{}{}{\sffamily}{.}{.5em}{}

\newtheoremstyle{myDef}
    {\topsep}{\topsep}{}{}{\sffamily\bfseries}{.}{.5em}{}

\theoremstyle{myThm}
\newtheorem{theorem}{Theorem}[section]
\newtheorem{lemma}[theorem]{Lemma}
\newtheorem{proposition}[theorem]{Proposition}
\newtheorem{corollary}[theorem]{Corollary}
\newtheorem{assumption}[theorem]{Assumption}

\theoremstyle{myRem}
\newtheorem{remark}[theorem]{Remark}

\theoremstyle{myDef}
\newtheorem{definition}[theorem]{Definition}
\newtheorem{example}[theorem]{Example}

\usepackage[numbers,square]{natbib}
\usepackage{chngcntr}
\counterwithin{table}{section}
\counterwithin{figure}{section}
\counterwithin{algorithm}{section}
\numberwithin{equation}{section}

\renewcommand{\hat}{\widehat}

\DeclareMathOperator*{\argmax}{arg\,max}

\providecommand{\keywords}[1]{{\small\textbf{\textit{Keywords:}} #1\par}}

\title{Offline Learning of Reserve Prices from Censored Bids}
\author{Lavanya Marla, Yunzong Xu, Boxuan Zhou}
\date{}

\begin{document}
\thispagestyle{plain}
{\centering\Large\bfseries\thetitle\par}
\vspace{0.9em}
{\centering\large Lavanya Marla\qquad Yunzong Xu\qquad Boxuan Zhou{\renewcommand{\thefootnote}{\fnsymbol{footnote}}\footnote{Authors are listed in alphabetical order. Boxuan Zhou is the lead author and main contributor to this work. Correspondence: \href{mailto:bzhou23@illinois.edu}{bzhou23@illinois.edu}.}}\par}
\setcounter{footnote}{0}
\vspace{0.3em}
{\centering\small Department of Industrial and Enterprise Systems Engineering,
 University of Illinois Urbana-Champaign\par}
\vspace{0.9em}
\begin{abstract}
We study how a seller can learn a reserve price for a second-price auction
from censored historical bid data. Because values below past reserves are
unobserved, even unlimited data may not reveal which reserve earns the
most revenue. At the population level, we characterize exactly when an
optimal reserve can be identified under a monotone hazard rate (MHR)
assumption, even though the value distribution may remain unidentified.
{Identification holds in two situations. In the first, revenue has not
yet begun to fall at the lowest historical reserve, so the best reserve
lies among the observed bids. In the second, the censored fraction is as
large as MHR allows given the observed bids, which pins down the entire
value distribution.}
When identification fails,
we derive the reserve that minimizes worst-case revenue regret over
distributions consistent with the data. {We prove that both the unique
minimax reserve and its regret depend on just two observable quantities:
the censored fraction and the hazard just above the lowest historical
reserve.}

With finite data, we estimate these two quantities from $N$ historical bidders
who faced the lowest reserve. {In the identifiable regime the optimal
reserve itself is learned, with expected worst-case regret $O(N^{-1/2})$
from the observed bids when it lies among them, and $O(N^{-1})$ from the censored count
alone when the censored fraction is as large as MHR allows and the
optimum lies at or below the lowest reserve.}
In the nonidentifiable regime, we show that
{the rule that plugs these estimates into the minimax reserve has}
expected worst-case revenue regret within $O(N^{-2/5})$
of the population minimum under local density smoothness. We prove a matching minimax lower bound on a smooth class in this regime.
Beyond MHR, we solve the minimax problem for regular and unrestricted
distributions, and prove an $O(N^{-2/5})$ learning upper bound under smoothness and known bounds on the boundary parameters in the regular nonidentifiable regime.
Numerical experiments {on synthetic and real-world eBay data} show gains from
exploiting MHR and the benefits of our robust pricing rule when parametric
models are misspecified.
\end{abstract}


\section{Introduction}\label{sec:intro}

Prices shape both trade and what a seller can learn about demand.
A buyer who turns down an offer may be willing to pay just a little less,
or much less. The refusal alone does not tell us which. Repeating the offer
reveals how often buyers reject it, but leaves open which lower prices
they would accept.

Reserve prices make this tension concrete in auctions. A reserve is the
lowest price at which the seller will sell. In a second-price auction,
a winner pays the reserve if no other bid meets it, and otherwise pays
the second-highest bid. A bidder's own bid determines whether it wins, but not
the price it pays if it wins. Bidding its true value is therefore optimal
regardless of the other bids~\citep{vickrey1961counterspeculation}.
Raising the reserve can increase the payment or
prevent a sale. Classical auction theory characterizes the best reserve
when the value distribution is known~\citep{myerson1981optimal,bulow1989simple}.
Historical bids offer a way to learn this choice: in a field experiment
on search-advertising auctions, reserves learned from past bids increased
revenue~\citep{ostrovsky2023reserve}.

A past reserve, however, can hide precisely the values needed to evaluate
a lower one. We study historical bid data in which a bidder reports a value when it
meets the reserve and submits no bid otherwise. The seller knows how many
bidders faced each reserve, including those who did not bid. The task is
offline: choose a new reserve from these data, without collecting
new observations at lower prices.

\subsection{Learning despite censoring}

\begin{figure}[htbp]
 \centering
 \includegraphics[width=0.45\textwidth]{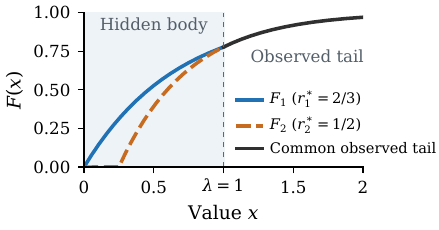}
 \caption{A nonidentifiable instance. Two value distributions agree at
 and above the historical reserve $\lambda=1$, so they induce the same
 distribution of censored records.
 Their unique optimal reserves are $2/3$ and $1/2$.
 The seller observes the total probability below the boundary, but not
 its distribution within the shaded region.
 Both distributions have a monotone hazard rate
 (see Section~\ref{sec:identifiability}).}\label{fig:running-example}
\end{figure}

Let $\lambda$ be the lowest historical reserve{, which we call the
\emph{censoring boundary}}. Even with unlimited data,
the seller learns only the distribution at and above $\lambda$---the
\emph{observed tail}---and the probability of a value below it.
The distribution of these lower values---the hidden
\emph{body}---remains unidentified without further assumptions.
Figure~\ref{fig:running-example} shows two
value distributions that produce indistinguishable censored records but different
optimal reserves. No number of additional bids collected at the same
historical reserves can distinguish them.
Does this make it impossible to learn an effective reserve from censored bids?

Fortunately, choosing a reserve can require less information than
recovering the whole distribution: different bodies can share an optimal
reserve. When their optima disagree, we adopt the minimax viewpoint developed for
offline pricing and censored inventory
problems~\citep{bu2023offline,hssaine2024data}. We compare the revenue of a chosen reserve with each
compatible distribution's own optimal revenue.
We call this difference \emph{revenue regret}, whether the reserve is
chosen using exact information or learned from samples.
When the censored fraction (the share of bidders who submit no bid) and
the observed tail are known exactly, the
smallest achievable worst-case regret is the \emph{population minimax regret}.
It is the benchmark for learning from samples: more records at the same
historical reserves cannot reduce worst-case regret below it.
For censored auctions, this leads to our research questions:
\begin{quote}
\itshape
When do censored bids identify an optimal reserve? When they do not,
what revenue regret is unavoidable, and how can the seller learn the
best reserve despite the hidden values?
\end{quote}

\subsection{Main results}

Our main results exploit monotone hazard rate (MHR), a standard shape
constraint on demand in auction and pricing
theory~\citep{dhangwatnotai2015revenue,cole2014sample,allouah2022pricing}.
The hazard rate measures the relative rate at which demand falls as the
price rises and governs the tradeoff between a higher payment and a lost sale.
MHR requires this rate to be nondecreasing and covers a broad class of
distributions, including all distributions with
log-concave densities~\citep{bagnoli2005logconcave}.
MHR excludes some hidden bodies that the censored data alone cannot
rule out. To quantify the benefit of this constraint, we also solve the
problem for regular and unrestricted distributions
(Section~\ref{sec:extensions} and Table~\ref{tab:exp-shape}).
Under MHR, we first solve the problem with the tail known exactly, then
learn the required quantities from samples.

\paragraph{Identification}
We distinguish two regimes according to whether censored bids identify
an optimal reserve shared by all compatible distributions
(Theorem~\ref{thm:frontier}). In the \emph{identifiable regime}, this can
happen in two ways. If revenue has not yet begun to fall at $\lambda$,
the hidden body cannot change the best choice in the observed tail.
We call this the \emph{tail-optimal case}.
Alternatively, if the censored fraction is as large as MHR allows for
the hazard just above $\lambda$, MHR forces the body to have constant hazard.
This \emph{fully identified case} determines the entire distribution
and is a special case of the identifiable regime. The two cases can overlap.
In the remaining \emph{nonidentifiable regime}, compatible distributions
prefer different reserves, leaving a positive minimax regret $\Delta$
that more bids at the same prices cannot remove.

\paragraph{Minimax pricing}
When compatible distributions prefer different reserves, pricing too high
loses sales while pricing too low gives up revenue. We prove that a unique
minimax reserve balances the largest possible losses from these mistakes
(Theorem~\ref{thm:mhr-equalizer}). {For fixed bidder count $n$ and historical
reserve $\lambda$, this reserve and $\Delta$ depend on only two observable quantities:
the censored fraction and the hazard just above $\lambda$, the right-hand
hazard of the observed tail, which we call the boundary hazard
(Proposition~\ref{thm:two-scalar}).} {The minimax reserve lies strictly
between the reciprocal of the boundary hazard, the optimum if the hazard
stayed at its boundary value, and the historical reserve. Reusing
$\lambda$ is therefore strictly worse in every nonidentifiable instance
(Theorem~\ref{thm:mhr-equalizer} and Proposition~\ref{prop:rdagger}).}

{In the nonidentifiable regime every compatible optimum lies at or below
$\lambda$, so only reserves up to $\lambda$ need to be compared.} {The
revenue difference between two such reserves depends on the values above
$\lambda$ only through how often they occur, not through their size. An
auction with at least two values above $\lambda$ pays the same
second-highest value under either reserve.} {The censored fraction gives
this frequency, and the boundary hazard bounds the hidden bodies through
MHR, which is why two observable quantities suffice.}
For each fixed reserve, the worst case is attained by a body
with at most two constant hazard levels (Theorem~\ref{thm:two-level}), reducing the
infinite-dimensional search over compatible distributions to a
two-dimensional optimization over these levels.

\paragraph{Learning from finite data}
The learning rate depends on what censoring leaves us to estimate.
In the tail-optimal case,
an empirical tail rule has $O(N^{-1/2})$ expected regret
(Proposition~\ref{prop:tail-rate}). Here $N$ counts historical bidders
facing $\lambda$. In the nonidentifiable regime, we instead estimate the
censored fraction and the hazard just above $\lambda$. Under local density
smoothness, the rule that plugs these estimates into the minimax reserve,
the \emph{plug-in rule}, has expected worst-case regret
$\Delta+O(N^{-2/5})$ (Theorem~\ref{thm:finite-sample}). A matching lower bound on a smooth class of nonidentifiable MHR instances shows that the exponent $2/5$ cannot be improved uniformly
(Proposition~\ref{prop:interior-lower-bound}).

In the special fully
identified case with an optimum at or below $\lambda$, {the \emph{linear-completion rule},
which uses only the censored count, the number of bidders who submit no
bid,} gives
$O(N^{-1})$ expected total regret without estimating a density
(Proposition~\ref{prop:edge}). {Small departures from full identification
add only a regret term that is quadratic in the distance to full
identification. That distance is the gap between the largest cumulative
hazard that MHR allows below $\lambda$ and the one the censored fraction
implies, $\lambda h_0-A$ in the notation of Section~\ref{sec:identifiability}.}

To choose a reserve without knowing the regime, the \emph{auto-selector}
(Algorithm~\ref{alg:pretest}) first tests for a tail optimum using the
estimated boundary hazard, then selects the tail rule or the plug-in
rule. Under common smoothness and parameter bounds, it has
$O(N^{-2/5})$ excess worst-case regret with high probability
(Proposition~\ref{prop:pretest-uniform}).

\paragraph{Extensions beyond MHR}
Regular distributions, a broader class, rule out a revenue decline
followed by an increase as the reserve rises. Their nonidentifiable regime retains the reduction to two observable quantities and an
$O(N^{-2/5})$ upper bound on expected excess regret under the smoothness
and parameter conditions in Section~\ref{sec:extensions}. They do not
retain MHR's fully identified case or its fast $O(N^{-1})$ guarantee.
We also solve the minimax problem without shape restrictions.

\paragraph{Numerical experiments}
We compare the reserves chosen under MHR, regular distributions, and no
shape restrictions for the same censored records. Exploiting MHR
reduces actual regret relative to the unrestricted reserve in all three
synthetic examples of Section~\ref{sec:exp-shape}, alongside much smaller minimax regret (Table~\ref{tab:exp-shape}).
We then compare the robust rule with maximum-likelihood estimators (MLEs),
which fit the parameters of an assumed distribution family to the
censored records. A correct family can help recover the optimal reserve.
But a misspecified family can match the observed records while assigning
the wrong distribution to the hidden values, causing regret that persists
as the sample grows.
Our experiments quantify the robust rule's protection against this error
and its cost under individual distributions (Section~\ref{sec:experiments}).
{On eBay bids, the rule's held-out regret is below every censored
parametric fit's at two of the three Palm Pilot reference boundaries,
and below the Weibull fit's at seven of the eight most frequent Palm Pilot opening bids (four to seven as the imputed bidder count varies). Reusing the opening bid loses up to $54\%$ of the optimal revenue
(Section~\ref{sec:exp-real}).}

\subsection{Related work}\label{sec:related}

Our identification and learning questions build on the theory of offline
decisions from censored data. Bu et al.~\citep{bu2023offline} study pricing
from historical prices, inventories, and censored sales; they introduced
the notion of problem identifiability, linked it to vanishing minimax
regret, and developed a policy approaching minimax regret. The closest problem is the censored newsvendor of Hssaine and
Sinclair~\citep{hssaine2024data}. They characterize minimax regret,
solve the population minimax problem, and give finite-sample guarantees with matching lower bounds up to logarithmic factors. As in
both papers, we evaluate worst-case regret at the decision produced by the
sample. We also build on Hssaine and Sinclair's use of
samples at the censoring boundary and a regime test before selecting a rule.
{Kumar and Mouchtaki~\citep{kumar2026value} compute the exact worst-case
regret of standard censored newsvendor policies at every sample size.}

In the newsvendor model of~\citep{hssaine2024data}, costs specify the
optimal demand quantile, so the censored fraction determines whether
it is observed. Optimal auction reserves instead depend on the hazard rate:
the density relative to the probability of exceeding the reserve.
The censored fraction alone cannot determine this tradeoff between
higher payments and lost sales. {Accordingly, the population minimax
problem is no longer one-dimensional in the censored fraction: under MHR
it becomes a two-dimensional optimization over hazard levels. We solve
it with certified error bounds and match the learning rate of the resulting plug-in rule with a
lower bound.}
We also seek to exploit economically meaningful shape constraints on
demand. Under MHR or regular distributions, a worst-case body must join
the observed tail without violating the shape restriction. Constructing
such bodies is an additional analytical challenge. Exploiting these
constraints can lower regret relative to
each compatible distribution's own optimal revenue. We quantify the
benefit by also solving the unrestricted auction
problem (Section~\ref{sec:unrestricted} and Table~\ref{tab:exp-shape}).

The dependence on the hazard also affects learning. In the nonidentifiable MHR
regime, we must estimate a density at the censoring
boundary. {We show how its estimation error affects revenue regret. The
$N^{-2/5}$ upper bound is uniform over classes with common local density
smoothness bounds and is matched by a minimax lower bound on a smooth interior class; the newsvendor
rate in~\citep{hssaine2024data} is $N^{-1/2}$, up to logarithmic factors.}

Allouah et al.~\citep{allouah2023single} maximize the guaranteed fraction
of optimal posted-price revenue when the sale probability at one historical
price is known. Two constant hazard levels also appear in their MHR
analysis. For posted pricing, sale probabilities at two prices fix their
revenue difference. In a second-price auction, this difference also depends
on the distribution of values between the reserves. Our two-level
characterization follows from a separate comparison of auction revenues
while preserving the observed tail (Theorem~\ref{thm:two-level}).
In the nonidentifiable MHR regime,
this reduction determines the smallest worst-case revenue regret and
the reserve that achieves it. {Anunrojwong~\citep{anunrojwong2026scalefree}
also builds worst cases for MHR bidders from constant hazards, for a
revenue ratio and without data; in our problem the levels must also
join the observed tail and account for the censored fraction.}

Online reserve learning from censored auction feedback is well
established~\citep{cesabianchi2015regret}, including with
strategic bidders or context~\citep{kanoria2021incentive,golrezaei2021dynamic}.
Online learning has a fundamental advantage: the seller can deliberately
lower reserves to explore values hidden by earlier reserves.
{Zhang et al.~\citep{zhang2024thresholds} learn a threshold from censored
feedback by choosing the thresholds they query, and Cherapanamjeri et
al.~\citep{cherapanamjeri2022estimation} estimate auction models by setting
reserves; both can sample below any earlier threshold.} In offline
learning, values below the lowest historical reserve remain hidden even
with unlimited data. We address this obstruction directly.

With uncensored values, auction-learning bounds quantify the samples
needed for near-optimal revenue~\citep{cole2014sample,guo2019settling,
morgenstern2015pseudo,mohri2016learning,dhangwatnotai2015revenue}.
Jin et al.~\citep{jin2022learning} establish tight sample-complexity bounds
for learning reserves in second-price auctions. {Methods for personalized
reserves in advertising also optimize over fully logged
bids~\citep{paesleme2016field,derakhshan2022linear}.} In our setting, by
contrast, censored records can leave positive regret even with unlimited
data. Bid-landscape forecasting also
learns from censored auction logs~\citep{ren2019landscape,ghosh2019landscape}.
Those works estimate a bidder's distribution of market prices from
winning and losing auctions; our decision is the seller's reserve.
{Chahuara et al.~\citep{chahuara2017realtime} model bids censored by a
publisher's reserves with a survival model inside an online engine,
without identification or rate guarantees.}

Auction econometrics studies what incomplete bids reveal about values
and counterfactual revenues~\citep{haile2003inference,tang2011bounds,
aradillas2013identification,guerre2000optimal,athey2002identification}.
{With a binding public reserve,
Paarsch~\citep{paarsch1997deriving} estimates the optimal reserve from a
parametric family; Freyberger and Larsen~\citep{freyberger2022identification}
and Hickman et al.~\citep{hickman2017identification} give point
identification when the number of bidders is unknown or only bounded
below; and Qi~\citep{qi2025truncated} shows that truncated second-price
transaction prices identify the censored fraction and the value distribution above
the reserve, taking the posted reserve to be optimal. These are the
observables of our model; we ask what they imply for a new reserve when
the values below the old one stay hidden.}

Reserve selection under partial
identification has also been studied~\citep{aryal2013point,jun2024information}.
{Kim et al.~\citep{kim2026partial} compute
maximin-revenue reserves from valuation bounds. Ponomarev and
Wen~\citep{ponomarev2026identification} use minimax regret, our criterion,
but their ambiguity concerns the highest value under non-binding reserves,
restricted by likelihood ratios rather than hazards, and their reserve
depends on a whole estimated distribution. Tang and
Wang~\citep{tang2026anonymous} design maximin mechanisms from the
distribution of one order statistic.}
Coey et al.~\citep{coey2021scalable} show that the top two bids can
identify an optimal reserve without the whole value distribution, using
zero or secret reserves that leave those bids observable. A positive
public reserve can instead hide both highest values and leave revenue
at a lower reserve unidentified even with unlimited data. We determine
when this hidden information still permits an optimal decision;
otherwise, we characterize the reserve with the smallest regret relative
to each compatible distribution's own optimum, and the rate at which it
can be learned.

Robust pricing and auction design also use finite samples, shape
information, moments, or support
restrictions~\citep{allouah2022pricing,tang2026hidden,allouah2020prior,
suzdaltsev2022distributionally,che2022robustly,kocyigit2020distributionally}.
Minimax regret also appears in posted pricing, screening, and
auctions~\citep{bergemann2008pricing,wang2024minimax,anunrojwong2024robustness}.
{Anunrojwong et al.~\citep{anunrojwong2026support} find the minimax-regret
mechanism when only the support of values is known, measuring regret
against the revenue obtainable with known values. We measure it against
each compatible distribution's own optimal revenue, as
in~\citep{bu2023offline,hssaine2024data}.}
Censored inventory research also studies exploration, adaptive ordering,
and confidence intervals for learned
policies~\citep{besbes2013implications,huh2009nonparametric,ban2020confidence}.

\subsection{Organization and notation}\label{sec:organization-notation}
Section~\ref{sec:setting} introduces the auction, historical data, and regret criterion.
Sections~\ref{sec:identifiability}--\ref{sec:finite-data} determine when
an optimal reserve can be identified, solve the minimax problem, and
develop learning guarantees. Section~\ref{sec:extensions} extends the
analysis to regular and unrestricted distributions.
Section~\ref{sec:experiments} reports experiments, and Section~\ref{sec:conclusion} concludes.
{The appendices contain proofs, additional theoretical results, and further experimental details; Appendix~\ref{app:notation} lists the notation.}

Throughout, $n$ counts bidders in the target auction, and $N$ counts
historical bidders who faced the lowest reserve $\lambda$.
We write $G$ for the true cumulative distribution function and $F$ for
a candidate, with densities $g$ and $f$. Also,
$(x)_+=\max\{x,0\}$, and $\mathbf1\{E\}$ is the indicator of an event $E$.

\section{Model and Objective}\label{sec:setting}

\subsection{Auction model and revenue}\label{subsec:auction}

The seller offers one item in a second-price auction with reserve
$r\ge0$ to $n\ge2$ risk-neutral bidders. Their private values are independent
draws from an unknown absolutely continuous distribution $G$ on
$[0,\infty)$ with finite mean. A bidder reports its value if it meets the reserve and submits
no bid otherwise; this is optimal regardless of the other
bids~\citep{vickrey1961counterspeculation}.
With exactly one qualifying bid, the winner pays $r$; with at least two,
it pays the second-highest value. With none, revenue is zero.

The admissible value distributions form the class
\begin{equation}\label{eq:G-class}
 \mathcal G:=\{F:\ \mathrm{supp}(F)\subseteq[0,\infty),\quad
 F\text{ absolutely continuous},\quad \mathbb E_F[V]<\infty\}.
\end{equation}
Let $f$ be the density of a candidate $F$, $\mathbf V=(V_1,\ldots,V_n)$
the values, and $V_{(1)}\ge V_{(2)}$ for the two highest. The auction's payment
can be written as
\begin{equation}\label{eq:realized-revenue}
 \mathrm{rev}(r,\mathbf V)=r\mathbf1\{V_{(1)}\ge r\}+(V_{(2)}-r)_+.
\end{equation}
Let $R_F(r)$ be its expectation under $F$. Splitting by one or at least two
qualifying bids gives
\begin{equation}\label{eq:revenue}
 R_F(r)=nrF(r)^{n-1}(1-F(r))
 +\int_r^\infty u\,n(n-1)F(u)^{n-2}(1-F(u))f(u)\,du.
\end{equation}
For a known distribution $F$, define its optimal revenue and reserve by
\begin{equation}\label{eq:R-star-def}
 R_F^*:=\max_{r\ge0}R_F(r),\qquad
 r_F^*:=\min\argmax_{r\ge0}R_F(r).
\end{equation}
Finite mean ensures that the maximum exists (Lemma~\ref{lem:r^*_exists});
the smallest maximizer resolves ties. The identification notion of
Definition~\ref{def:id} below asks only for a shared maximizer and does
not depend on this tie rule.

\subsection{Censored bid data}\label{subsec:censored-data}

The historical data may contain several reserves. Auction
$k\in\{1,\ldots,K\}$ has a known reserve $r_k^{\mathrm{off}}$ and
bidder count $N_k$, which need not equal the target count $n$.
The reserves and counts are fixed before values are drawn. The values
$V_{ki}^{\mathrm{off}}$, $i=1,\ldots,N_k$, are i.i.d.\ from $G$ across
all $k,i$, independently of this design. A randomized design can be
treated conditionally on its realization. The seller observes
\[
 Y_{ki}^{\mathrm{off}}=\begin{cases}
 V_{ki}^{\mathrm{off}},&V_{ki}^{\mathrm{off}}\ge r_k^{\mathrm{off}},\\
 \varnothing,&V_{ki}^{\mathrm{off}}<r_k^{\mathrm{off}}.
 \end{cases}
\]
An observed bid reveals a value; $\varnothing$ reveals only that it lies
below the reserve. The observed bids and the number of bidders who submit no bid determine $N_k$; their identities are unnecessary. Such counts are
available, for example, when participation is registered before bidding.

Write $Y^{\mathrm{off}}=(Y_{ki}^{\mathrm{off}})_{k,i}$ and
$r^{\mathrm{off}}=(r_k^{\mathrm{off}})_{k\in\{1,\ldots,K\}}$.
\begin{definition}[Censoring boundary and sample size]\label{def:observable-boundary}
The lowest historical reserve and the number of bidders who faced it are
\begin{equation}\label{eq:lambda-def}
 \lambda:=\min_k r_k^{\mathrm{off}},\qquad
 N:=\sum_{k:r_k^{\mathrm{off}}=\lambda}N_k.
\end{equation}
We assume $\lambda>0$ and call it the \emph{censoring boundary}.
\end{definition}
The lowest reserve determines what remains hidden even with unlimited
data. In the population limit, records at $\lambda$ reveal $G(x)$ for every $x\ge\lambda$.
Higher-reserve records can improve finite-sample tail estimates but add
no population information. Our rules use the $N$ lowest-reserve records,
and their learning bounds are expressed in terms of $N$.
Reserve choices that depend on the values they censor are
outside this sampling model (Remark~\ref{rem:endogenous}).

For these $N$ records, we suppress the auction index, write $V_i$, $i=1,\ldots,N$, for their values, and set $Y_i=V_i$ if $V_i\ge\lambda$, and $Y_i=\varnothing$ otherwise.
The number of no-bid records divided by $N$ estimates the censored fraction
\begin{equation}\label{eq:alpha-def}
 \alpha:=G(\lambda)\in(0,1).
\end{equation}
The observed bids alone would not determine this fraction.
Censoring does not reveal how this probability is distributed below
$\lambda$. We call this hidden part the \emph{hidden body}, or simply the body, and a distribution
that agrees with the observed tail a \emph{compatible completion}, or a
compatible distribution.

\subsection{Minimax regret}\label{sec:criterion}

Following the offline censored-data framework of Bu et al.~\citep{bu2023offline}
and Hssaine and Sinclair~\citep{hssaine2024data}, we group distributions
that the historical data cannot distinguish and evaluate reserves by
worst-case revenue regret.
At reserves at least $\lambda$, all compatible distributions have the same
revenue. Below $\lambda$, their revenues and optimal reserves can differ.
The distributions indistinguishable from $G$ in the population limit form
the following set.
\begin{definition}[Ambiguity set]\label{df:ambiguity_set}
\begin{equation}\label{eq:ambiguity_set}
 \mathcal F(\lambda;G):=\{F\in\mathcal G:F(x)=G(x)\text{ for all }x\ge\lambda\}.
\end{equation}
\end{definition}
A shape assumption narrows this set. For any resulting
$\mathcal A\subseteq\mathcal F(\lambda;G)$ containing $G$, we evaluate a
reserve by its revenue regret relative to each completion's own optimum.
\begin{definition}[Minimax regret]\label{def:minimax_risk}
\begin{equation}\label{eq:minimax-auction}
 \Delta(\mathcal A):=\inf_{r\ge0}\sup_{F\in\mathcal A}\{R_F^*-R_F(r)\}.
\end{equation}
\end{definition}
The seller chooses a reserve to minimize the largest revenue regret over
compatible distributions. The value $\Delta(\mathcal A)$ measures the
irreducible regret caused by censoring when the tail is known exactly,
and does not depend on $N$. Since $G\in\mathcal A$, regret under the true
distribution can be smaller than the worst case at the same reserve:
\begin{equation}\label{eq:true-regret-bound}
 R_G^*-R_G(r)\le\sup_{F\in\mathcal A}\{R_F^*-R_F(r)\}\qquad(r\ge0).
\end{equation}
In particular, a reserve's regret under $G$ can fall below $\Delta(\mathcal A)$.

\subsection{Identification and learning}\label{subsec:decision}

An optimal reserve can be identifiable even when the value distribution
is not fully identified: it suffices that all compatible distributions share one.
\begin{definition}[Identifiability]\label{def:id}
The distribution of censored records \emph{identifies an optimal reserve} over
$\mathcal A$ if its compatible distributions share an optimum:
\begin{equation}\label{eq:id-def}
 \text{there exists }r\ge0\text{ such that }R_F(r)=R_F^*
 \quad\text{for every }F\in\mathcal A.
\end{equation}
\end{definition}
With finite data, a reserve rule is a measurable map
\begin{equation}\label{eq:policy-def}
 \pi:(Y^{\mathrm{off}},r^{\mathrm{off}})\longmapsto r^\pi\in[0,\infty).
\end{equation}
We write $\Pi$ for these rules. A rule's regret under the true distribution is
\begin{equation}\label{eq:reg1}
 \mathrm{Regret}(r^\pi):=R_G^*-R_G(r^\pi).
\end{equation}
Following Bu et al.~\citep{bu2023offline} and Hssaine and
Sinclair~\citep{hssaine2024data}, we evaluate each sample's reserve by its
worst-case revenue regret over compatible distributions. The expected
excess regret is
\begin{equation}\label{eq:learning-criterion}
 \mathbb E_G\!\left[\sup_{F\in\mathcal A}
       \{R_F^*-R_F(r^\pi)\}\right]-\Delta(\mathcal A).
\end{equation}
\begingroup The sample produces a reserve, which is then evaluated
against every compatible distribution. Averaging this worst-case regret
over samples and subtracting $\Delta(\mathcal A)$ isolates the additional
cost of learning from finite data. Taking the worst case before averaging
lets the worst compatible distribution vary with the realized reserve;
averaging first would instead evaluate the rule against a single
completion fixed across samples. \endgroup

The connection between identifiability and zero minimax regret underlies
the offline pricing analysis of Bu et al.~\citep{bu2023offline} and is
stated explicitly for the censored newsvendor problem by Hssaine and
Sinclair~\citep{hssaine2024data}. The following proposition establishes the
corresponding connection for auctions.
\begin{proposition}[Identification and vanishing regret]\label{prop:identifiable-iff}
Let $G\in\mathcal A\subseteq\mathcal F(\lambda;G)$.
The minimax infimum is attained, and identifiability is equivalent to
$\Delta(\mathcal A)=0$. For the unrestricted, MHR, and regular classes studied here
(Definition~\ref{def:regular-mhr}), each zero-regret case admits one
sample-based rule whose worst-case regret vanishes in probability for every
$G$ in that case. For regular distributions, the right density at $\lambda$ is
assumed to exist, as in Theorem~\ref{thm:frontier-reg}.
\end{proposition}
Thus a common optimum can be learned even when the body remains unidentified.
The population claim is proved in Appendix~\ref{app:proofs-ident}, the
learning claims in Appendices~\ref{app:ident-policies}
and~\ref{app:ident-body-policies}.
When the compatible optima disagree,
the goal is to approach the positive population value $\Delta(\mathcal A)$.

\section{Identifiability under MHR}\label{sec:identifiability}\label{sec:shape-restricted}

Our main analysis exploits MHR to link the hidden body to the observed tail, excluding distributions
that the censored data alone cannot rule out. This restriction can identify
an optimal reserve even when the body remains unknown
(Theorem~\ref{thm:frontier}), and can lower minimax regret when no common
optimum is identifiable (Table~\ref{tab:exp-shape}). We first define the
shape restrictions, then determine when MHR identifies an optimal reserve.
Section~\ref{sec:extensions} shows what survives for regular and unrestricted distributions.

\subsection{Revenue and the hazard rate}\label{sec:shared-lemmas}

Consider a small increase from $r$ to $r+\varepsilon$. To first order,
the revenue change comes from auctions with exactly one bid above $r$.
The winner pays an additional $\varepsilon$ if its value clears the new
reserve; otherwise the sale is lost, reducing revenue by about $r$.
Conditional on clearing $r$, the probability that this value lies in
$[r,r+\varepsilon)$ is approximately $f(r)\varepsilon/(1-F(r))$.
The coefficient $f(r)/(1-F(r))$ is the \emph{hazard rate}, which measures
the rate at which qualifying bids cease to qualify as the reserve rises.
\begin{definition}[Hazard and cumulative hazard]\label{def:hazard}
For a distribution $F$ with density $f$, where $F(x)<1$, define
\[
 h_F(x):=\frac{f(x)}{1-F(x)},\qquad
 H_F(x):=\int_0^x h_F(u)\,du=-\log(1-F(x)).
\]
Thus $1-F(x)=e^{-H_F(x)}$.
\end{definition}
The payment gain and the risk of losing the sale give, almost everywhere
on the interior of the effective support,
\[
 R_F'(r)=nF(r)^{n-1}(1-F(r))\bigl(1-rh_F(r)\bigr).
\]
The positive factor is the probability of exactly one qualifying bid.
Thus revenue rises when $rh_F(r)<1$ and falls when $rh_F(r)>1$.
For a fixed completion, the bidder count changes the magnitude of this
effect but not its direction.

Where $0<F(r)<1$ and $f(r)>0$, define the \emph{virtual value}
$\phi_F(r):=r-(1-F(r))/f(r)=r-1/h_F(r)$.
The revenue derivative can then be written as
$R_F'(r)=-nF(r)^{n-1}f(r)\phi_F(r)$.
We consider two restrictions on the distribution's shape.
\begin{definition}[MHR and regular distributions]\label{def:regular-mhr}
A distribution has \emph{monotone hazard rate} (MHR)
if $h_F$ is nondecreasing, or equivalently if $H_F$ is convex.
It is \emph{regular} if its virtual value $\phi_F$ is nondecreasing.
\end{definition}
For both classes, we assume $f>0$ on the interior of the effective support
$\{x:0<F(x)<1\}$, so virtual value is defined there and prices can be
indexed by quantiles.

Every MHR distribution is regular because both $r$ and $-1/h_F(r)$ are nondecreasing,
but the converse does not hold. Section~\ref{sec:regular-summary} compares
the identification and regret guarantees under the two assumptions.
\begin{definition}[Shape-restricted ambiguity sets]\label{def:shape-restricted}
\begin{equation}\label{eq:mhr_ambiguity_set}
 \mathcal F_{\mathrm{MHR}}(\lambda;G)
 :=\{F\in\mathcal F(\lambda;G):F\text{ is MHR}\}.
\end{equation}
Define $\mathcal F_{\mathrm{reg}}(\lambda;G)$ similarly over regular distributions.
Then $\mathcal F_{\mathrm{MHR}}\subseteq\mathcal F_{\mathrm{reg}}$.
\end{definition}
\begin{assumption}[MHR]\label{ass:mhr}
The true value distribution $G$ is MHR.
\end{assumption}
Sections~\ref{sec:identifiability}--\ref{sec:finite-data} use this assumption,
so $G$ belongs to the MHR ambiguity set. Results for regular and unrestricted
distributions in Section~\ref{sec:extensions} and the appendices state their
own distributional conditions.

\subsection{Zero-regret frontier}\label{sec:auction-mhr-ident}\label{sec:shared-boundary}

For now, take the censored fraction and observed tail as known exactly.
They fix the cumulative hazard's height and its right slope at the censoring boundary:
\begin{equation}\label{eq:Ah0-def}
 A:=H_G(\lambda)=-\log(1-\alpha)>0,
\end{equation}
\begin{equation}\label{eq:h0-def}
 h_0:=H_G'(\lambda^+)\in(0,\infty).
\end{equation}
We assume that this right hazard is finite; when the right density exists,
$h_0=g(\lambda^+)/(1-\alpha)$, with $g$ the density of $G$. At a jump of the hazard we use its
right-continuous version. For an MHR distribution this is the right
derivative of the convex function $H_G$, as in \eqref{eq:h0-def}. For a
regular distribution it is the right density divided by the survival
probability, as Appendix~\ref{app:regular} assumes. For a compatible body, $A/\lambda$ is the
average hazard on $[0,\lambda]$, and MHR bounds its hazard above by $h_0$.
The average cannot exceed this upper bound, so feasibility requires
\begin{equation}\label{eq:left-feasible}
 A\le\lambda h_0.
\end{equation}
This condition is also sufficient: since $G$ is MHR, its tail cumulative
hazard is convex with hazard at least $h_0$ above $\lambda$, so the body
with constant hazard $A/\lambda\le h_0$ (the chord of
Figure~\ref{fig:ambiguity_sets}) pastes to it and the result is MHR.

The zero-regret frontier has two sources. In the \emph{tail-optimal case},
$\lambda h_0\le1$, so $rh_F(r)\le1$ throughout every compatible body. Raising the reserve
toward $\lambda$ cannot lower revenue, so the best reserve in the observed
tail is optimal regardless of the hidden body.

If $A=\lambda h_0$, the average body hazard equals its upper bound,
forcing $h_F(x)=h_0$ almost everywhere below $\lambda$. Every compatible
body therefore has $F(x)=1-e^{-h_0x}$ for $0\le x\le\lambda$.
Together with the observed tail, this determines the entire distribution;
we call it the \emph{fully identified case}. The optimum is $1/h_0$ when
$1/h_0\le\lambda$; otherwise it lies in the observed tail above $\lambda$.
The two cases overlap when $A=\lambda h_0\le1$. When neither identification
condition holds, compatible bodies can place
the crossing of $rh_F(r)=1$ at different reserves.

\begin{example}[Nonidentifiable MHR completions]\label{ex:two-completions}
Figure~\ref{fig:running-example} illustrates the nonidentifiable regime.
Set $\lambda=1$, $A=3/2$, and $h_0=2$, so $A<\lambda h_0$ and
$\lambda h_0>1$. Two compatible body cumulative hazards are
\[
 H_1(x)=\tfrac32x,\qquad
 H_2(x)=\max\{0,2x-\tfrac12\},\qquad 0\le x\le1,
\]
continued by $H_i(x)=\tfrac32+2(x-1)$ above $1$.
Both distributions $F_i=1-e^{-H_i}$ are MHR and censor the same fraction
$1-e^{-3/2}$. In the first body the hazard is $3/2$ throughout; in the
second it is zero up to $1/4$ and then $2$. The product $rh_{F_i}(r)$
crosses one at $2/3$ and $1/2$, respectively; since the sign of the
revenue derivative does not depend on the bidder count
(Section~\ref{sec:shared-lemmas}), these are the unique optimal reserves
of $F_1$ and $F_2$ for every $n\ge2$. The historical record nevertheless
has the same law under both distributions.
Section~\ref{sec:dro} returns to this example to
choose a reserve that minimizes regret over all compatible bodies.
\end{example}

For the formal characterization, write
\begin{equation}\label{eq:qlam}
 q_\lambda:=1-\alpha=e^{-A},
\end{equation}
\begin{equation}\label{eq:h0-auction}
 \phi_0:=\phi_G(\lambda)=\lambda-1/h_0.
\end{equation}
Here $q_\lambda$ is the probability of an observed bid, and $\phi_0$ is the
right boundary virtual value. The condition $\phi_0\le0$ is exactly
$\lambda h_0\le1$.
\begin{theorem}[Zero-regret frontier]\label{thm:frontier}
Under Assumption~\ref{ass:mhr}, suppose $0<h_0<\infty$. Then
\[
 \Delta(\mathcal F_{\mathrm{MHR}}(\lambda;G))=0
 \quad\Longleftrightarrow\quad
 \phi_0\le0\ \text{or}\ A=\lambda h_0.
\]
If $\phi_0\le0$, every maximizer of $R_G$ over $[\lambda,\infty)$ is
optimal for every compatible MHR completion.
\end{theorem}
In the remaining \emph{nonidentifiable regime}, $h_0>1/\lambda$ and
$A<\lambda h_0$. {Theorem~\ref{thm:frontier} thus sorts instances into two
regimes and three cases: the identifiable regime, which holds the
tail-optimal and fully identified cases (they can overlap), and the
nonidentifiable regime, the third case.}
The proof constructs two completions with different
unique optima and establishes positive minimax regret
(Appendix~\ref{app:proofs-ident}). Since the censored record has the same
law under these completions, additional observations at the same reserves
cannot distinguish their optimal decisions.

\begin{corollary}[Censored fraction and tail optimality]\label{rem:heavy-forces-hard}
Under Assumption~\ref{ass:mhr}, if $\alpha>1-e^{-1}\approx0.632$,
the instance cannot be in the tail-optimal case.
It is nonidentifiable unless $A=\lambda h_0$.
\end{corollary}
Indeed, $A>1$ and \eqref{eq:left-feasible} force $\lambda h_0>1$.
The censored count can rule out the tail-optimal case, but cannot distinguish
the nonidentifiable regime from the fully identified case.

\subsection{Feasible cumulative hazards}

\begin{figure}[htbp]
 \centering
 \includegraphics[width=0.29\textwidth]{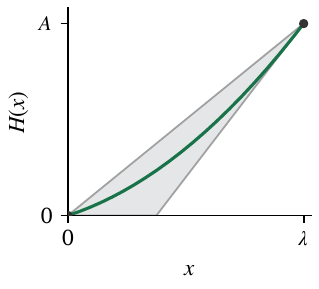}
 \caption{The observed tail fixes the endpoint $(\lambda,A)$ and caps the
 body's slope at $h_0$. Every MHR body is a convex nondecreasing curve
 between the chord and the backward tangent. The bounds coincide in the
 fully identified case, $A=\lambda h_0$.}
 \label{fig:ambiguity_sets}\label{fig:mhr_hazard_space}
\end{figure}

The feasible bodies admit a geometric description that will be useful
for the minimax problem. A compatible $H$ is a convex nondecreasing curve
joining $(0,0)$ to $(\lambda,A)$, with terminal slope at most $h_0$.
Convexity bounds it above by the chord, while the terminal slope bound
places it above the line drawn backward from $(\lambda,A)$
(Figure~\ref{fig:ambiguity_sets}). Hence
\begin{equation}\label{eq:H-interval}
 (A-h_0(\lambda-r))_+\le H(r)\le rA/\lambda,
 \qquad 0\le r\le\lambda.
\end{equation}
In survival probabilities, this is
\begin{equation}\label{eq:S-interval}
 e^{-rA/\lambda}\le1-F(r)\le e^{-(A-h_0(\lambda-r))_+}.
\end{equation}
Both bounds are attained (Lemma~\ref{lem:left-pointwise-H}).
The two bodies in Example~\ref{ex:two-completions} are precisely the chord
and backward tangent: they share the observed endpoint but differ below
it. The next section uses this geometry to minimize revenue regret over
the full set of compatible bodies.

\section{Minimax Reserve under MHR}\label{sec:dro}

In the nonidentifiable regime, no reserve is optimal for every compatible
distribution. We choose one that minimizes the largest revenue regret.
Throughout this section, $0<A<\lambda h_0$ and $h_0>1/\lambda$.
The key is to compare two reserves directly: their common tail payments
cancel, and the worst comparisons come from bodies with at most two
constant hazard levels.

\subsection{Reduction to two boundary parameters}\label{sec:oracle-minimax}\label{sec:two-scalar-red}

Above $\lambda$, MHR gives $h_F(r)\ge h_0>1/\lambda$, so revenue
cannot increase there. The reserve $\lambda$ is therefore at least as good
as every higher reserve. We need only compare reserves $r,s\in[0,\lambda]$:
$r$ is the seller's choice, and $s$ is an alternative that may earn more
under a compatible distribution.

Consider the payments in a single auction. If at least two values exceed
$\lambda$, both reserves collect the same second-highest value. If exactly
one does, the winner pays the larger of the reserve and the second-highest
value, which is below $\lambda$. If none does, every value is in the hidden
body. Thus the magnitudes of values above $\lambda$ do not affect the
payment difference; only their frequency matters.

The revenue formula makes this cancellation explicit. Write
$\psi(p)=1-np^{n-1}+(n-1)p^n$ for the probability that at least two bidders
exceed $x$ when $p=F(x)$. The payment identity
\eqref{eq:realized-revenue} gives
$R_F(r)=r[1-F(r)^n]+\int_r^\infty\psi(F(x))\,dx$, and hence
\begin{equation}\label{eq:mhr-cdf-gap}
 R_F(s)-R_F(r)=s[1-F(s)^n]-r[1-F(r)^n]
              +\int_s^r\psi(F(x))\,dx.
\end{equation}
The comparison uses $F$ only at the two reserves and between them.

The tail still limits which hidden bodies are possible. The censored
fraction fixes $H(\lambda)=A$, while MHR bounds the body's hazard by $h_0$.
Hence the feasible cumulative hazards form $\mathcal H(A,h_0)$: the convex
nondecreasing curves joining $(0,0)$ to $(\lambda,A)$ with final slope at
most $h_0$. Both the allowed bodies and the revenue comparisons depend
on the tail only through these two quantities.

For a posted reserve $r$, let $W_G(r)$ be its largest revenue regret over
compatible distributions, and let $\Delta$ be the smallest such regret:
\begin{equation}\label{eq:W-def}
 W_G(r):=\sup_{F\in\mathcal F_{\mathrm{MHR}}(\lambda;G)}
 \{R_F^*-R_F(r)\},\qquad \Delta:=\min_{r\in[0,\lambda]}W_G(r).
\end{equation}
Restricting the minimum to $[0,\lambda]$ loses nothing, because no reserve
above $\lambda$ improves on $\lambda$ under any compatible completion
(Lemma~\ref{lem:auction-tail-domination}). Throughout, $\Delta$
abbreviates $\Delta(\mathcal F_{\mathrm{MHR}}(\lambda;G))$ of
Definition~\ref{def:minimax_risk}; we write $\Delta(A,h_0)$ for the same
quantity when its dependence on the boundary pair matters, and
$\Delta(G)$ (Section~\ref{sec:finite-data}) when its dependence on the
population $G$ matters.
Computing $W_G(r)$ means choosing both a compatible distribution and the
alternative reserve that makes $r$ look worst:
\begin{equation}\label{eq:first-stage-DRO2}
 W_G(r)=\sup_{F\in\mathcal F_{\mathrm{MHR}}(\lambda;G)}
 \sup_{s\in[0,\lambda]}\{R_F(s)-R_F(r)\}.
\end{equation}
By \eqref{eq:mhr-cdf-gap}, the inner difference uses $F$ only on
$[0,\lambda]$, where $F=1-e^{-H}$ for a feasible body $H$. We therefore
define the body objective
\begin{equation}\label{eq:W-body-def}
 \mathsf W(r;A,h_0):=\sup_{H\in\mathcal H(A,h_0)}\ \sup_{s\in[0,\lambda]}
 \{R_F(s)-R_F(r)\},\qquad F=1-e^{-H}\ \text{on }[0,\lambda],
\end{equation}
the largest revenue advantage of an alternative reserve in $[0,\lambda]$
over $r$ across feasible bodies. It is defined for every
$0\le A\le\lambda h_0$, including the identifiable cases, where
Section~\ref{sec:finite-data} uses it, and it equals $W_G(r)$ whenever
$h_0\ge1/\lambda$, since then no alternative reserve above $\lambda$ is
needed.
\begin{proposition}[Two-parameter reduction]\label{thm:two-scalar}
Let $\mathcal D=\{(A,h_0):0<A<\lambda h_0,\ h_0>1/\lambda\}$.
Under Assumption~\ref{ass:mhr}, for $(A,h_0)\in\mathcal D$ and $r\in[0,\lambda]$,
$W_G(r)=\mathsf W(r;A,h_0)$ of \eqref{eq:W-body-def}, which depends only
on $A,h_0$ and the known $n,\lambda$. Its minimum $\Delta(A,h_0)$ and its unique
minimizing reserve $r^\dagger(A,h_0)$ also depend only on these quantities.
Uniqueness is established in Theorem~\ref{thm:mhr-equalizer} below.
\end{proposition}

For example, keep $\lambda=1$, $A=3/2$, and $h_0=2$ from
Example~\ref{ex:two-completions}, but replace the common tail by
$H(1+u)=3/2+2u+u^2/2$ for $u\ge0$. This changes the observed bid values
and the absolute revenues. It changes neither the allowed bodies nor the
revenue differences between reserves below $\lambda$, so the minimax
reserve and minimax regret remain the same.

\subsection{Two-level worst-case bodies}

The remaining search is over an entire cumulative-hazard curve.
Convexity lets us replace it by two straight pieces without decreasing
the revenue advantage of an alternative reserve $s$ over the posted
reserve $r$.

If $s>r$, join $(0,0)$ to $(s,H(s))$ and then to $(\lambda,A)$.
These chords lie above $H$ between the reserves. They preserve $F(s)$,
increase $F(r)$, and decrease $\psi(F(x))$ on $[r,s]$. In
\eqref{eq:mhr-cdf-gap}, the endpoint term at $r$ and the negatively signed
integral can therefore only increase the revenue difference.

\begin{figure}[htbp]
 \centering
 \begin{subfigure}[b]{0.45\textwidth}
  \includegraphics[width=\textwidth]{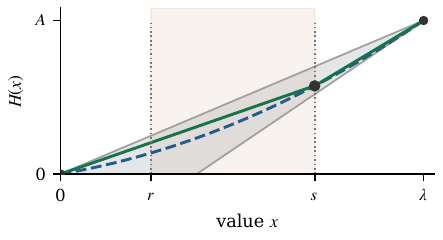}
  \caption{A higher alternative reserve: $s>r$.}\label{fig:two-level-left}
 \end{subfigure}\hspace{0.04\textwidth}
 \begin{subfigure}[b]{0.45\textwidth}
  \includegraphics[width=\textwidth]{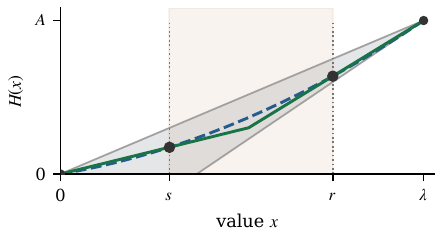}
  \caption{A lower alternative reserve: $s<r$.}\label{fig:two-level-right}
 \end{subfigure}
 \caption{Straight-line replacements of a compatible cumulative hazard.
 When the alternative reserve is higher, two chords raise the curve
 between the reserves. When it is lower, two extended lines lower it.
 Each replacement can only increase the alternative reserve's revenue
 advantage and gives a body hazard with at most one upward jump. Dashed: the original curve; gray: the band of Figure~\ref{fig:ambiguity_sets}; the shaded strip lies between the two reserves.}\label{fig:two-level}
\end{figure}

If $s<r$, extend the chord from $(0,0)$ to $(s,H(s))$ to the right and
the chord from $(r,H(r))$ to $(\lambda,A)$ to the left, then take the
larger of the two lines. The resulting curve lies below $H$ between the
reserves and agrees with it at both endpoints. The endpoint terms stay
fixed, while the positively signed integral increases. Both constructions
preserve convexity and the slope bound; Figure~\ref{fig:two-level} shows
the geometry. Appendix~\ref{app:proofs-dro} treats the endpoint cases
and proves attainment.

Write $a$ for the first body hazard level and $b$ for the second. Their
cumulative hazard is
\[
 H_{a,b}(x)=\max\{ax,A-b(\lambda-x)\},\qquad
 (a,b)\in[0,A/\lambda]\times[A/\lambda,h_0].
\]
The average hazard $A/\lambda$ lies between the two levels. Once $a<b$
are chosen, the total cumulative hazard fixes the jump location:
$a\tau+b(\lambda-\tau)=A$, so $\tau=(b\lambda-A)/(b-a)$, which lies
in $[0,\lambda]$ because $a\le A/\lambda\le b$.
If $a=b=A/\lambda$, the hazard is constant. Let $\mathcal T(A,h_0)$
denote these bodies joined to the observed tail. Thus $A,h_0$ are the
observed inputs, while $a,b$ are the variables in the worst-case search.

\begin{theorem}[Two-level worst-case completions]\label{thm:two-level}
Under Assumption~\ref{ass:mhr}, let $(A,h_0)\in\mathcal D$.
For every posted reserve $r\in[0,\lambda]$,
\[
 W_G(r)=\max_{F\in\mathcal T(A,h_0)}\{R_F^*-R_F(r)\}.
\]
For a fixed alternative reserve $s>r$, a maximizing body has its break at $s$;
for $s<r$, a maximizing body has its break in $[s,r]$.
\end{theorem}
This reduces the infinite-dimensional search over hidden bodies to an
exact two-dimensional search over $a,b$. The straight-line replacement
works whenever $0\le A\le\lambda h_0$; the nonidentifiable-regime condition
is used only to ensure that an optimal alternative reserve lies in
$[0,\lambda]$.

\subsection{Balancing the two sources of regret}\label{sec:rdagger}

A posted reserve can lose revenue in two ways: it may be too high for one
compatible distribution and too low for another. We separate these risks
by comparing $r$ with lower and higher alternatives:
\begin{equation}\label{eq:directional-losses}
\begin{aligned}
 L(r)&:=\sup_F\max_{0\le s\le r}\{R_F(s)-R_F(r)\},\\
 U(r)&:=\sup_F\max_{r\le s\le\lambda}\{R_F(s)-R_F(r)\},
\end{aligned}
\end{equation}
where $F\in\mathcal F_{\mathrm{MHR}}(\lambda;G)$. Here $L(r)$ is the
worst-case loss from posting too high, and $U(r)$ is the worst-case loss
from posting too low. Their maximum is $\mathsf W(r;A,h_0)$.

Each completion's revenue is nondecreasing up to an optimum and
nonincreasing after it. As $r$ rises, regret against lower reserves can
therefore only increase, while regret against higher reserves can only
decrease. Taking the worst case preserves these directions. The best
compromise is the reserve where the two losses are equal. We call it
the \emph{equalizer}.
\begin{theorem}[Uniqueness of the minimax reserve]\label{thm:mhr-equalizer}
Under Assumption~\ref{ass:mhr}, for $(A,h_0)\in\mathcal D$,
the minimax reserve $r^\dagger$ is unique and
\[
 L(r^\dagger)=U(r^\dagger)=\Delta>0.
\]
Both directional regrets are continuous; $L$ is nondecreasing and $U$ nonincreasing,
strictly so wherever they are positive. Consequently $\mathsf W$ decreases
strictly before $r^\dagger$ and increases strictly after it.
\end{theorem}

\begin{figure}[htbp]
 \centering
 \includegraphics[width=0.50\textwidth]{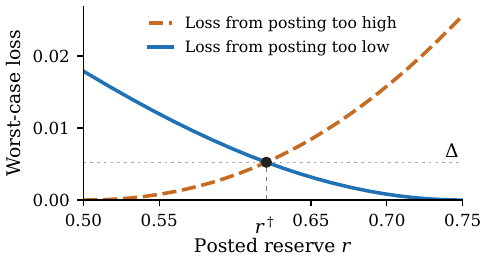}
 \caption{The two worst-case losses cross at $r^\dagger\approx0.621$.
 Raising the reserve beyond this point increases the loss from posting
 too high; lowering it increases the loss from posting too low. Their
 common value is $\Delta\approx0.00524$ for $\lambda=1,A=1.5,h_0=2,n=3$.}\label{fig:equalizer-example}
\end{figure}

Using $n=3$ and the boundary parameters of Example~\ref{ex:two-completions},
numerical maximization over the two-level bodies gives
$r^\dagger\approx0.621$ and $\Delta\approx0.00524$
(Figure~\ref{fig:equalizer-example}). At this reserve, one compatible
distribution would prefer $0.5$, while another would prefer approximately
$0.720$. Their revenue regrets are equal. The first of these is the
body $H_2$ of Example~\ref{ex:two-completions}. The minimax choice thus balances losses under different
distributions, rather than finding the optimum of one fitted distribution.

This balance matters for learning. At $r^\dagger$, one adverse distribution
has a lower optimum and another has a higher optimum. Moving the reserve
in either direction increases one of their revenue regrets to first order.
The worst-case regret can therefore have a corner at its minimum, even
if the individual revenue curves are smooth. Formally, for an
instance-dependent $c>0$,
\[
 \mathsf W(r;A,h_0)-\Delta(A,h_0)\ge c|r-r^\dagger|
 \qquad(0\le r\le\lambda)
\]
(Lemma~\ref{lem:kink}). Excess regret at most $\varepsilon$ thus requires
reserve error at most $\varepsilon/c$. The next section studies how
accurately samples allow us to locate this reserve.

\begin{proposition}[Location and monotonicity]\label{prop:rdagger}
Under Assumption~\ref{ass:mhr}, let $(A,h_0)\in\mathcal D$.
The minimax reserve lies strictly inside
\begin{equation}\label{eq:rdagger-interval}
 \frac1{h_0}<r^\dagger<
 \min\{\lambda,\lambda-(A-1)/h_0\}.
\end{equation}
This interval has length at most $\lambda-A/h_0$.
At fixed $A$, $\Delta(A,h_0)$ is nondecreasing in $h_0$.
\end{proposition}
The interval shrinks to a point as $A$ reaches $\lambda h_0$, matching
the fully identified case of Section~\ref{sec:auction-mhr-ident}.
At fixed $A$, increasing $h_0$ allows more hidden bodies and cannot
decrease minimax regret. In every nonidentifiable instance, reusing the
historical reserve $\lambda$ gives strictly greater worst-case regret than
$r^\dagger$. Appendix~\ref{app:proofs-dro} proves these results.

\paragraph{Numerical computation}\label{sec:mhr-computation}
For each pair $a,b$, we compute an optimal reserve where $rh_F(r)$
reaches or jumps past one, then compare its revenue with that of the
posted reserve. These comparisons have closed forms. We maximize them
over a grid of hazard pairs and minimize the resulting worst-case regret
over a grid of posted reserves. Corollary~\ref{cor:two-level-oracle} bounds
the errors from both grids; Appendix~\ref{app:core-computation} gives the
formulas. With finite data, we use the same calculation with estimates
of $A,h_0$.

\section{Learning from Censored Bids}\label{sec:finite-data}

{We now build on the population analysis to learn a reserve from
$N$ censored records. The two regimes of Section~\ref{sec:identifiability}
leave three learning tasks.} In the tail-optimal case,
we learn a common optimum from the empirical CDF. If the optimum
is not identifiable, we estimate the boundary hazard to find the minimax
reserve. If MHR determines the hidden body and its optimum lies at or below
the boundary, the censored count alone suffices. Table~\ref{tab:learning-rates}
summarizes the corresponding rates. Each task requires a different kind
of estimate from the same censored records.

We measure expected excess regret by
$\mathbb E[W_G(r^\pi)-\Delta(G)]$ for a rule $\pi$, where
$\Delta(G):=\Delta(\mathcal F_{\mathrm{MHR}}(\lambda;G))$ is the minimax
regret of Definition~\ref{def:minimax_risk} for the population $G$,
keeping the worst case inside the sample expectation as in
\eqref{eq:learning-criterion}. In the nonidentifiable regime it equals
$\Delta$ of \eqref{eq:W-def}; in the identifiable regime
$\Delta(G)=0$ (Proposition~\ref{prop:identifiable-iff}), so this is total regret.
Throughout, let $m$ be the number of uncensored bids and set
$\hat A_N=-\log(m/N)$, with $\hat A_N=+\infty$ when $m=0$.

\begin{table}[htbp]
 \centering\small
 \caption{{Two regimes, three cases:} expected excess worst-case regret
 of the corresponding rules under MHR, under the conditions stated below.
 At $A=\lambda h_0=1$, the linear-completion guarantee also applies.}
 \label{tab:learning-rates}
 \begin{tabular}{llllc}
  \toprule
  {Regime} & Case & Boundary condition & Rule & Regret rate \\
  \midrule
  {Identifiable} & Tail-optimal & $\lambda h_0\le1$ & Empirical tail & $O(N^{-1/2})$ \\
  & Fully identified & $A=\lambda h_0\ge1$ & Linear completion & $O(N^{-1})$ \\
  {Nonidentifiable} & --- & $A<\lambda h_0$, $\lambda h_0>1$ & Plug-in & $O(N^{-2/5})$ \\
  \bottomrule
 \end{tabular}
\end{table}

\subsection{Learning in the tail-optimal case}\label{sec:tail-learning}

In the tail-optimal case, $h_0\le1/\lambda$, and every compatible distribution shares an optimal
reserve in the observed tail. We can compare reserves there without
reconstructing the body. Let $\widehat G_N^-(x)$ be the fraction of
historical bidders whose value is below $x\ge\lambda$, counting censored
bidders as below. Fix a known lower bound $h_{\min}>0$ on $h_0$ and set
$\bar\lambda=\max\{\lambda,1/h_{\min}\}$. MHR makes revenue nonincreasing
beyond $\bar\lambda$, so this is a bounded search interval.

The empirical tail rule $\hat r_N^{\rm tail}$ maximizes the estimated
revenue gain over posting $\lambda$, using $\widehat G_N^-$ on
$[\lambda,\bar\lambda]$ (equation \eqref{eq:tail-empirical-gap} in
Appendix~\ref{app:ident-policies}). Subtracting
revenue at $\lambda$ cancels payments from values beyond this interval.
Only the empirical CDF is needed, and its $N^{-1/2}$ estimation error
transfers to revenue regret.

\begin{proposition}[Learning in the observed tail]\label{prop:tail-rate}
Under Assumption~\ref{ass:mhr}, suppose $h_0\ge h_{\min}>0$.
For every $\delta\in(0,1)$, with probability at least $1-\delta$,
\[
 \sup_{r\ge\lambda}R_G(r)-R_G(\hat r_N^{\rm tail})
 \le4n\bar\lambda\sqrt{\frac{\log(2/\delta)}{2N}}.
\]
When $h_0\le1/\lambda$, this is total worst-case regret and $\Delta(G)=0$.
In this case, $\mathbb E[W_G(\hat r_N^{\rm tail})]=O(N^{-1/2})$.
No density smoothness is needed. Appendix~\ref{app:ident-policies} gives the proof.
\end{proposition}

\subsection{Learning in the nonidentifiable regime}\label{sec:nonid-learning}

\subsubsection{Estimating the boundary parameters}\label{sec:est-boundary}

\begingroup In outline: the censored fraction gives
$A$ directly from the count of no-bids. The boundary hazard $h_0$ is a
density at the edge of the observed tail, so it must be estimated from
the bids just above $\lambda$, and that estimate is what sets the
$N^{-2/5}$ rate. Sampling error can produce a pair that no MHR body is
compatible with, so we project the estimates onto the feasible set.
Everything else rests on one fact: the worst-case regret is Lipschitz in
the pair $(A,h_0)$, so an error in the estimates moves the worst case by
at most a constant times that error, and the rule that posts the minimax
reserve for the estimated pair inherits their rate. \endgroup

Conditional on observing a bid, its density is $\tilde g=g/q_\lambda$,
so $\tilde g(\lambda^+)=h_0$.
\begin{assumption}[Boundary density smoothness]\label{ass:boundary-smoothness}
The conditional bid density $\tilde g$ is twice continuously differentiable
on a fixed interval $[\lambda,\lambda+\bar b]$, with $\bar b>0$.
\end{assumption}
We use this condition only for results that estimate the boundary density.
No smoothness is imposed on the hidden body.

We fit a line to histogram density estimates over a window of width $b_N$,
using distance from $\lambda$ as the predictor. Its slope accounts for
the density's local linear change, and its intercept $\hat h_0$ estimates $h_0$.
Its remaining bias is $O(b_N^2)$, while its sampling error is
$O((Nb_N)^{-1/2})$, since the expected number of bids in the window is
of order $Nb_N$. Balancing
these terms gives $b_N=c_bN^{-1/5}$ and error $N^{-2/5}$
\citep{fan1996local,cheng1997boundary}. We use a fixed number of
histogram cells. The weights, concentration bounds, and fixed
tuning choices are specified in Appendix~\ref{app:learning-details}; the
values used in the experiments are in Appendix~\ref{app:exp-methods}.

As noted above, sampling error can produce estimates that are
incompatible with MHR, such as $\hat h_0<0$ or $\hat A_N>\lambda\hat h_0$,
and we project the estimated pair onto the feasible parameter set. This changes the
set of compatible bodies, so the regret bound must control the effect
of changing that set. Recall the
body objective $\mathsf W(r;A,h_0)$ of \eqref{eq:W-body-def}, defined for
every $0\le A\le\lambda h_0$. It equals $W_G(r)$ there when $h_0\ge1/\lambda$;
when $h_0<1/\lambda$, the full objective also compares with tail reserves.

\begin{lemma}[Lipschitz stability of the body objective]\label{lem:stability}
Let $\lambda>0$ and $n\ge2$. For pairs
$(A,h_0),(A',h_0')\in[0,\infty)^2$ satisfying
$A\le\lambda h_0$ and $A'\le\lambda h_0'$, including equality,
\begin{equation}\label{eq:global-stability}
 \sup_{r\in[0,\lambda]}|\mathsf W(r;A,h_0)-\mathsf W(r;A',h_0')|
 \le3\lambda\bigl(|A-A'|+\lambda|h_0-h_0'|\bigr).
\end{equation}
The bound is independent of the target bidder count $n$. Moreover,
$r\mapsto\mathsf W(r;A,h_0)$ is Lipschitz on $[0,\lambda]$ with constant
$\mathrm{Lip}_r=\max\{1,\lambda h_0\}$.
\end{lemma}

The proof constructs corresponding bodies by adjusting their total
hazard and hazard cap. It also applies in the fully identified case
(Appendix~\ref{app:proofs-dro}). For $h_0\ge1/\lambda$, we
project the scaled estimates $(\hat A_N,\lambda\hat h_0)$ onto
\begin{equation}\label{eq:closed-wedge}
 \mathcal P_\lambda=\{(x,y):0\le x\le y,\ y\ge1\}.
\end{equation}
In the scaled coordinates $(x,y)=(A,\lambda h_0)$, the bound
\eqref{eq:global-stability} reads $3\lambda(|x-x'|+|y-y'|)$, so projection
error and stability are measured in the same units.
When $\lambda h_0\ge1$, the closed convex set $\mathcal P_\lambda$ contains the true pair, so projection cannot move the estimate away from it. If too few bids are observed, we instead use the reserve
\begin{equation}\label{eq:chord-reserve}
 r_c(A)=\min\{\lambda,\lambda/A\}\quad(0\le A<\infty),
 \qquad r_c(+\infty)=0,
\end{equation}
with $\lambda/0=+\infty$.
This is the \emph{linear-completion reserve}: the optimal reserve when the
hidden body is completed with the constant hazard $A/\lambda$, so that its
cumulative hazard is the straight line (the chord) from $(0,0)$ to $(\lambda,A)$ in Figure~\ref{fig:ambiguity_sets}.
It uses the censored fraction alone and never estimates $h_0$.

\begin{definition}[Projected plug-in rule]\label{def:plug-in}
Fix $m_{\min}\ge1$, $c_b,c_{\mathrm g}>0$, and $M\ge2$ independently of the data.
Here $M$ is the number of histogram cells. The projected plug-in rule,
or \emph{plug-in} for short, is $\hat r_N^\dagger$ of Algorithm~\ref{alg:plug-in}.
\end{definition}
\begingroup The rule uses the target bidder count $n$, the boundary $\lambda$, the
record count $N$, the tuning constants $m_{\min}$, $M$, $c_{\mathrm g}$, and the
bandwidth constant $c_b$, or the fixed grid from which it is selected
(Section~\ref{sec:fs-guarantees}). \endgroup

\begin{algorithm}[htbp]
\caption{Projected plug-in reserve}\label{alg:plug-in}\label{alg:plugin-sketch}
\small
\begin{algorithmic}[1]
\STATE Count uncensored bids $m$ and set $\hat A_N=-\log(m/N)$, with $-\log0=+\infty$.
\STATE \textbf{if} $m<m_{\min}$ \textbf{then return} $r_c(\hat A_N)$.
\STATE Estimate $\hat h_0$ by the local-linear histogram fit on $[\lambda,\lambda+c_bN^{-1/5}]$.
\STATE $(\bar A_N,\lambda\bar h_N)\leftarrow\operatorname{proj}_{\mathcal P_\lambda}(\hat A_N,\lambda\hat h_0)$.
\STATE \textbf{if} $\bar A_N=0$ or $\lambda\bar h_N=1$ \textbf{then return} $\lambda$.
\STATE \textbf{if} $\bar A_N=\lambda\bar h_N$ \textbf{then return} $r_c(\bar A_N)$.
\STATE Grid $[0,\lambda]$, $[0,\bar A_N/\lambda]$, and $[\bar A_N/\lambda,\bar h_N]$ with endpoints and mesh at most $\varepsilon_N=c_{\mathrm g}N^{-1/2}$.
\STATE \textbf{return} the smallest reserve minimizing the finite two-level worst case of Corollary~\ref{cor:two-level-oracle}.
\end{algorithmic}
\end{algorithm}

\subsubsection{Regret of the plug-in rule}\label{sec:fs-guarantees}

\begin{theorem}[Regret of the plug-in rule]\label{thm:finite-sample}
Under Assumptions~\ref{ass:mhr} and~\ref{ass:boundary-smoothness},
suppose $h_0\ge1/\lambda$. For every $\delta\in(0,1)$ and
$N\ge \underline N(\delta)$, with probability at least $1-\delta$,
\begin{equation}\label{eq:finite-sample-hp}
 W_G(\hat r_N^\dagger)
 \le\Delta(G)+C_\star\sqrt{\log(6/\delta)}\,N^{-2/5}
       +c_{\mathrm g}C_{\rm num}N^{-1/2}.
\end{equation}
Here $C_\star$ and $C_{\rm num}$ bound statistical and discretization
errors, respectively. Appendix~\ref{app:learning-details} gives their
formulas and their dependence on the true distribution and tuning choices.
The required sample-size threshold $\underline N(\delta)$ is specified in
\eqref{eq:N0-explicit}.
Moreover, $\mathbb E[W_G(\hat r_N^\dagger)]\le\Delta(G)+O(N^{-2/5})$.
Both bounds are uniform on classes with $q_\lambda$ bounded away from
zero and common bounds on $\tilde g$ and $|\tilde g''|$ over the same
interval. They require no positive distance from either zero-regret frontier.
\end{theorem}

\begingroup Theorem~\ref{thm:finite-sample} fixes the bandwidth constant $c_b$ in
advance. In practice we select it from the fixed grid
$\mathcal C_b=\{\lambda/4,\lambda/2,\lambda,2\lambda\}$ by a data-driven
rule, and the theorem extends to that selection with the same rate, a larger constant, and $\log(24/\delta)$ in place of $\log(6/\delta)$
(Corollary~\ref{cor:selected-bandwidth} in
Appendix~\ref{app:learning-details}). We note this because the
experiments of Section~\ref{sec:experiments} use the selected bandwidth. \endgroup

\begin{proposition}[Matching interior lower bound]\label{prop:interior-lower-bound}
For every fixed $n\ge2$ and $\lambda>0$, there exist a class
$\mathcal G_{\rm int}$ whose members satisfy Assumptions~\ref{ass:mhr}
and~\ref{ass:boundary-smoothness}, and a constant $c>0$ such that, for all sufficiently
large $N$,
\[
 \inf_\pi\sup_{G\in\mathcal G_{\rm int}}
 \mathbb E_G[W_G(r^\pi)-\Delta(G)]\ge cN^{-2/5}.
\]
The class consists of one base distribution, with an exponential body and
a smooth tail, together with a sequence of local perturbations of its
boundary hazard near a single interior boundary pair. Its members have
common boundary smoothness bounds and boundary pairs in
a compact subset of $0<A<\lambda h_0$, $h_0>1/\lambda$.
The infimum includes randomized rules based only on $N$ independent
records censored at $\lambda$.
\end{proposition}

\begingroup The upper bound follows by applying the stability bound
\eqref{eq:global-stability} to the two estimation errors. Projection keeps the combined error at order
$N^{-2/5}$, and discretizing
the reserve and hazard levels adds the smaller $N^{-1/2}$ term. The
lower bound uses a smooth hazard perturbation that changes $h_0$ and
the minimax reserve by $N^{-2/5}$ while producing censored-record laws
that are difficult to distinguish. The linear cost of reserve error
then gives revenue regret of the same order
(Appendix~\ref{app:lb-statistical-reconstruction}). Matching refers to
the rate on this family: the upper bound is uniform over classes with
common bounds, and no rule improves on the exponent $2/5$ over this
family, although individual distributions may admit faster rates. \endgroup

\begingroup A natural alternative to estimating $(A,h_0)$ is to fit a parametric
family to the censored records and post the fitted optimum. Such a rule
converges to a fixed reserve, and the next proposition shows that no such
rule can attain the minimax regret unless that reserve happens to be
$r^\dagger$.

\begin{proposition}[Parametric reserve rules]\label{rem:parametric-limit}
Under the conditions of Theorem~\ref{thm:finite-sample}, let
$(A,h_0)\in\mathcal D$, and let $\hat r_N^{\rm par}$ be any reserve rule
that converges in probability to a constant $r^{\rm par}\ge0$. Then
\[
 \lim_{N\to\infty}\mathbb E\bigl[W_G(\hat r_N^{\rm par})\bigr]=W_G(r^{\rm par})
 >\Delta(G)=\lim_{N\to\infty}\mathbb E\bigl[W_G(\hat r_N^\dagger)\bigr]
 \qquad\text{unless }r^{\rm par}=r^\dagger .
\]
Moreover, $\mathbb E[R_G^*-R_G(\hat r_N^\dagger)]\le\Delta(G)+O(N^{-2/5})$,
whereas the limit $R_G^*-R_G(r^{\rm par})$ is not bounded by $\Delta(G)$
in general.
\end{proposition} \endgroup

\begingroup A censored maximum-likelihood fit within a parametric family, followed by
the fitted distribution's optimal reserve, is an example. Under the
conditions of White~\citep{white1982maximum}, the estimate converges to
the family member whose censored-record law is closest to that of $G$ in
Kullback--Leibler divergence, and $r^{\rm par}$ is that member's optimal
reserve when it is unique. This holds whether or not the family contains
$G$. If it does, $r^{\rm par}$ is the optimal reserve $r_G^*$ of $G$
itself, which is in general not the minimax reserve $r^\dagger$, so even a
correctly specified family has limit worst-case regret above $\Delta(G)$.
Section~\ref{sec:exp-truths} quantifies this limit for censored
parametric fits. The proof is in Appendix~\ref{app:learning-details}. \endgroup

\subsection{Learning in the fully identified case}\label{sec:edge-learning}

In the fully identified case, $A=\lambda h_0$ and every completion has the same body,
$H(x)=Ax/\lambda$. When $h_0\ge1/\lambda$, its optimal reserve is
$r_c(A)\le\lambda$, so the linear-completion rule $\hat r_c=r_c(\hat A_N)$
requires no density estimate. To study departures from full identification,
write $d=\lambda h_0-A$.

\begin{proposition}[Regret near full identification]\label{prop:edge}
Under Assumption~\ref{ass:mhr}, let $h_0\ge1/\lambda$,
$0\le A\le\lambda h_0$, $d=\lambda h_0-A$, and
$k_1=\max_{0\le z\le1}nz^{n-1}(1-z)=((n-1)/n)^{n-1}$. Then
\begin{equation}\label{eq:chord-expected}
 \mathbb E[W_G(\hat r_c)]
 \le\frac{k_1d^2}{h_0}
 +\frac{4k_1\lambda^2h_0\alpha}{q_\lambda N}
 +\lambda e^{-Nq_\lambda/8}.
\end{equation}
At the population level,
\begin{equation}\label{eq:quadratic-edge}
 \Delta(G)\le W_G\!\left(\min\{\lambda,(1+d/2)/h_0\}\right)
 \le\frac{k_1d^2}{8h_0},\qquad
 W_G(r_c(A))\le\frac{k_1d^2}{2h_0}.
\end{equation}
No density smoothness is needed. On classes with bounded $h_0$ and
$q_\lambda$ bounded away from zero, at fixed $\lambda$, total expected
worst-case revenue regret is
$O(d^2+N^{-1})$, giving the fast $O(N^{-1})$ rate in the fully identified case.
\end{proposition}

\begingroup Every completion has an optimal reserve between $1/h_0$ and
$\min\{\lambda,(1+d)/h_0\}$, an interval of length at most $d/h_0$;
integrating the revenue derivative over it gives the quadratic bounds on
the population minimax regret and on the regret of $r_c(A)$. At $d=0$
the only error is the $N^{-1/2}$ error in estimating $A$ from the
censored fraction, which produces $O(N^{-1})$ revenue regret
(Appendix~\ref{app:edge-pretest}), and Proposition~\ref{prop:lower-bound}
shows that this scale cannot be improved for instances within
$O(N^{-1/2})$ of the fully identified case. For $d>0$ the same statistical
term is added to the population term of order $d^2$ in
\eqref{eq:chord-expected}, so the total rate is still $N^{-1}$ when
$d=O(N^{-1/2})$. \endgroup

\subsection{Selecting a rule from the data}\label{sec:pretest}

Algorithm~\ref{alg:plug-in} assumes $h_0\ge1/\lambda$ and posts a reserve
in $[0,\lambda]$. In the tail-optimal case with $h_0<1/\lambda$, all compatible distributions share an
optimal reserve above $\lambda$, in the observed tail, and a reserve in
$[0,\lambda]$ forfeits it. The seller usually does not know the regime,
so we first check the estimated boundary hazard against $1/\lambda$ to
select the reserve rule. We call the rule that performs this check and
then posts the selected reserve the \emph{auto-selector}.
\begin{algorithm}[htbp]
\caption{Auto-selector reserve}\label{alg:pretest}
\small
\begin{algorithmic}[1]
\STATE Count uncensored bids $m$ and set $\hat A_N=-\log(m/N)$, as in Algorithm~\ref{alg:plug-in}.
\STATE \textbf{if} $m<m_{\min}$ \textbf{then return} $r_c(\hat A_N)$.
\STATE Estimate $\hat h_0$ as in Algorithm~\ref{alg:plug-in}, before projection.
\STATE \textbf{if} $\hat h_0\le1/\lambda$ \textbf{then return} the tail rule $\hat r_N^{\rm tail}$.
\STATE \textbf{else return} the plug-in reserve $\hat r_N^\dagger$ of Algorithm~\ref{alg:plug-in}.
\end{algorithmic}
\end{algorithm}

The three rates above are guarantees for separate rules. The auto-selector
selects between the tail rule and the plug-in rule.
On the estimation event of Lemma~\ref{lem:est-rates}, which has
probability at least $1-\delta$, a wrong branch can be selected only
near $h_0=1/\lambda$.
At this threshold, $\lambda$ is a common optimum. Nearby, stability
controls body regret, while the gain from a missed tail optimum is bounded
by a quadratic term in $\lambda h_0-1$. When survival, hazard, and local smoothness have common
bounds, the auto-selector has uniform $O_P(N^{-2/5})$ excess regret across both
zero-regret frontiers (Proposition~\ref{prop:pretest-uniform}), without
prior knowledge of the regime. Here $O_P$ denotes a rate in probability.
The fast $O(N^{-1})$ guarantee in the
fully identified case is for the separate linear-completion rule of
Section~\ref{sec:edge-learning}.
This construction follows the regime-testing approach of
Hssaine and Sinclair~\citep{hssaine2024data}; their test estimates a
probability, whereas ours estimates a density.

\begingroup \paragraph{What each assumption carries}
The bidder count, including no-bids, is the only source of the censored
fraction $\alpha$, and hence of $A$. Every rule uses it, and without it
the fully identified case cannot be recognized
(Table~\ref{tab:exp-real-natural-count} varies the imputed count). MHR
ties the hidden body to the observed tail; regular distributions keep the plug-in and, given a known parameter
rectangle, its $O(N^{-2/5})$ rate but lose full identification, and
unrestricted distributions lose the link altogether
(Section~\ref{sec:extensions}). The bound $h_{\min}$ sets the tail rule's
search interval and, with an upper bound $h_{\max}$ on $h_0$, the constants of the
auto-selector's guarantee. Smoothness serves only the $O(N^{-2/5})$ rate,
and common bounds make the guarantees uniform over a class; the tail
rule needs no smoothness, and the identification results of
Section~\ref{sec:identifiability} need only MHR. The $O(N^{-2/5})$ rate is
minimax optimal on a smooth interior class, and the $O(N^{-1})$ scale is
unavoidable within $O(N^{-1/2})$ of full identification
(Propositions~\ref{prop:interior-lower-bound} and~\ref{prop:lower-bound}). \endgroup

\section{Beyond MHR}\label{sec:extensions}

\begingroup MHR does two jobs: it ties the hidden body to the observed tail, and it keeps revenue single-peaked in the reserve. Regular distributions keep the
second job and only a weaker link to the tail; unrestricted distributions
keep neither. The regular class keeps the two-quantity reduction and the
learning rate but loses full identification; the unrestricted class
reduces the worst case to three simple scenarios, in which the observed tail starts to matter.
Section~\ref{sec:exp-shape} shows what each relaxation costs. \endgroup

\paragraph{Regular distributions}\label{sec:regular-summary}
\begingroup A regular distribution lets the hazard fall, but never so far that
revenue, once falling, rises again, so the balancing argument of
Section~\ref{sec:rdagger} still applies. \endgroup Recall $q_\lambda=e^{-A}$ and the boundary
virtual value $\phi_0=\lambda-1/h_0$.
\begin{proposition}[Reserve characterization for regular distributions]\label{prop:regular-summary}
Suppose $G$ is regular, $A>0$, its right density at $\lambda$ exists,
and $0<h_0<\infty$. Then
$\Delta(\mathcal F_{\mathrm{reg}}(\lambda;G))=0$ exactly when $\phi_0\le0$.
If $\phi_0>0$, worst-case regret for each $r\in[0,\lambda]$, its minimum,
and the minimizing reserve depend only on $A,h_0$ and the known $n,\lambda$.
In this case, the minimax reserve is the unique equalizer of the worst losses from
posting too high and too low, and lies in $(\lambda q_\lambda,\lambda)$.
\end{proposition}
\begingroup Compared with MHR, two things change. The reserve is now located by the censored fraction
alone, between $\lambda$ times the probability of an observed bid and
$\lambda$; the MHR interval of Proposition~\ref{prop:rdagger} also uses the boundary hazard and collapses to a point as $A$ approaches $\lambda h_0$. That collapse is full identification. Regular
distributions leave many bodies compatible even then, so the fully
identified case and its fast $O(N^{-1})$ rate (Proposition~\ref{prop:edge})
do not extend. The worst case still has a finite description, so the
minimax reserve is a scalar computation with error bounds
(Appendix~\ref{app:regular-computation}), and the plug-in carries over
with $O(N^{-2/5})$ expected excess worst-case regret under
Assumption~\ref{ass:boundary-smoothness} and a known bounded parameter
rectangle, away from $A=0$ and $h_0=1/\lambda$, containing the true pair
in its interior (Proposition~\ref{prop:reg-finite};
Appendix~\ref{app:regular} gives the rule and proofs). \endgroup

\paragraph{Unrestricted distributions}\label{sec:unrestricted}
\begingroup Without a shape restriction, the censored fraction $\alpha$ can sit
anywhere below $\lambda$: just under the posted reserve $r$, just under
$\lambda$, or so that a reserve above $\lambda$ is best. Each scenario
gives one term of the worst-case regret. \endgroup
Let $r_{\mathrm T}$ maximize revenue on $[\lambda,\infty)$, and let
$B_\lambda=R_G(r_{\mathrm T})-R_G(\lambda)$ be its gain over $\lambda$.
Write $k_\alpha=\max_{0\le z\le\alpha}nz^{n-1}(1-z)$.
\begin{proposition}[Unrestricted solution]\label{prop:unrestricted-summary}
Let $G$ satisfy the model of Section~\ref{sec:setting}, without a shape
restriction. For $0\le r\le\lambda$, unrestricted worst-case regret is
\[
 W_{\mathrm{all}}(r)=\max\{r\alpha^n,(\lambda-r)k_\alpha,
 B_\lambda+(\lambda-r)n\alpha^{n-1}(1-\alpha)\}.
\]
The best tail reserve has regret
$W_{\mathrm{all}}(r_{\mathrm T})=(\lambda\alpha^n-B_\lambda)_+$.
Thus $\Delta(\mathcal F(\lambda;G))=0$ exactly when
$B_\lambda\ge\lambda\alpha^n$.
\end{proposition}
\begingroup The first term is the loss when every hidden value sits just below $r$:
with probability $\alpha^n$ nobody clears $r$, and a slightly lower
reserve would have sold. The second is the loss against a reserve between
$r$ and $\lambda$ that a lone bidder would have paid. The third is the
loss against the best tail reserve: $B_\lambda$, plus the gap
$\lambda-r$ whenever exactly one bidder clears $\lambda$. The maximum is
piecewise linear, so the minimizer comes from a few breakpoints compared
with $r_{\mathrm T}$. Zero regret occurs exactly when $B_\lambda$ is at
least $\lambda\alpha^n$, the most any reserve below $\lambda$ can gain
when every bidder is hidden; the boundary hazard plays no role.
Theorem~\ref{thm:unrestricted} and
Corollary~\ref{cor:unrestricted-computation} give the proof and the
minimizer. \endgroup

\paragraph{Randomization and competition}\label{sec:randomization-competition}
\begingroup Two facts hold for MHR and regular distributions. Randomizing the reserve cannot remove the cost of nonidentification:
a random reserve mixes fixed ones, so it gains nothing under our
criterion, and under the weaker criterion that averages before taking the
worst case it lowers minimax regret by at most half
(Proposition~\ref{prop:randomization-bound}); the $N^{-2/5}$
lower bound already allows randomized rules
(Proposition~\ref{prop:interior-lower-bound}). Competition caps the cost
of censoring: lowering the reserve below $\lambda$ can improve on posting
$\lambda$ only when every value is hidden, an event of probability
$\alpha^n$, so minimax regret is at most $\lambda\alpha^n$
(Proposition~\ref{prop:competition}). The body
calculations also extend to an observed tail without shape restrictions,
with slope caps imposed separately and the tail's best gain over
$\lambda$ as an extra input (Proposition~\ref{prop:local-tail-extension}). \endgroup

\section{Computational Experiments}\label{sec:experiments}

{We ask three questions. What does the shape constraint buy
(Section~\ref{sec:exp-shape})? Does the learned rule approach the minimax
regret at the predicted rate (Section~\ref{sec:exp-synthetic})? Does it
beat parametric fits to the censored records
(Section~\ref{sec:exp-truths})? We answer them on synthetic instances,
where the true distribution is known, and then ask the same questions of
recorded eBay bids (Section~\ref{sec:exp-real}).}
All comparisons use $n=3$ bidders. Within each replication,
learned rules receive the same $N$ independent censored records.
The plug-in estimates $(A,h_0)$ and posts the minimax reserve for the
estimates; the auto-selector
can instead learn a reserve from the observed tail
(Algorithms~\ref{alg:plug-in} and~\ref{alg:pretest}).
The plug-in selects its bandwidth constant $c_b$ from a fixed grid by the
data-driven rule of Corollary~\ref{cor:selected-bandwidth}.

We compare these rules with \emph{maximum-likelihood estimators} (MLEs),
which fit Weibull, gamma, or lognormal parameters to make the censored
records most likely, then post the fitted distribution's optimal reserve.
An \emph{oracle MLE} uses the correct family and estimates its parameters
from the same sample.
We distinguish two measures. \emph{Actual regret} under a distribution
$F$ is $R_F^*-R_F(r)$: the revenue lost by posting $r$ instead of that
distribution's optimal reserve. \emph{Worst-case regret} $W_G(r)$ is
the largest such loss over compatible MHR distributions. Section~\ref{sec:exp-shape}
also compares the regular and unrestricted classes.
For learned reserves, we average each measure over sample replications.
We report excess worst-case regret $W_G-\Delta$ when studying convergence
to the minimax reserve; actual regret can be smaller than $\Delta$ \eqref{eq:true-regret-bound}.
Percentages use the reference revenue specified in each table or figure.
{Appendix~\ref{app:experiments} gives the experimental settings, numerical checks,
the studies that extend each design, and the data statement.}

\subsection{The value of shape constraints}\label{sec:exp-shape}

With the censored fraction and observed tail known exactly, we choose
the minimax reserve under MHR, regular distributions, and no shape
restrictions. Table~\ref{tab:exp-shape} reports its worst-case regret
over the chosen class, followed by its actual regret under each listed
distribution. The MHR reserve has lower actual regret than the
unrestricted reserve in all three examples. For Weibull$(2.5,1.5)$,
regret falls from $0.58\%$ to $0.19\%$ of optimal revenue.
The regular reserve does still better on the two Weibull distributions:
minimizing worst-case regret need not minimize regret at any one distribution.

\begin{table}[htbp]
  \centering\small
  \caption{Regret of the minimax reserves under three distribution
  classes ($n=3$), with exact censored fraction and observed tail.
  Worst-case regret ranges over compatible distributions in the indicated
  class; actual regret uses the listed distribution.
  All percentages use the listed distribution's $R_G^*$.
  Appendix~\ref{app:exp-regular-more} gives the reserves and computation details.}
  \label{tab:exp-shape}
  \begin{tabular}{lc*{3}{rr}}
    \toprule
    & & \multicolumn{2}{c}{MHR} & \multicolumn{2}{c}{Regular} & \multicolumn{2}{c}{Unrestricted} \\
    \cmidrule(lr){3-4}\cmidrule(lr){5-6}\cmidrule(lr){7-8}
Instance & $\lambda$ & Worst-case & Actual & Worst-case & Actual & Worst-case & Actual \\
    \midrule
Rayleigh$(1)$ & $1.5$ & $1.65\%$ & $0.20\%$ & $4.66\%$ & $0.49\%$ & $22.07\%$ & $0.24\%$ \\
Weibull$(1.5,1)$ & $1$ & $0.80\%$ & $0.18\%$ & $3.98\%$ & $0.05\%$ & $18.41\%$ & $0.56\%$ \\
Weibull$(2.5,1.5)$ & $2$ & $3.37\%$ & $0.19\%$ & $9.74\%$ & $<0.01\%$ & $40.09\%$ & $0.58\%$ \\
    \bottomrule
  \end{tabular}
\end{table}

MHR also provides much stronger worst-case protection. In the Rayleigh
example, minimax regret is $1.65\%$ under MHR, $4.66\%$ under regular
distributions, and $22.07\%$ without shape restrictions. The censored records are the same; the difference comes from excluding hidden
bodies that violate the chosen constraint.

\subsection{Learning the minimax reserve}\label{sec:exp-synthetic}

\begin{figure}[htbp]
  \centering
  \begin{subfigure}[b]{0.66\textwidth}
    \centering\includegraphics[width=\textwidth]{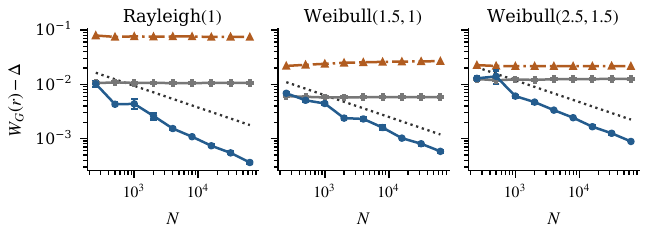}
    \caption{Excess worst-case regret against sample size, one panel per instance.}\label{fig:exp-syn-rate}
  \end{subfigure}\hfill
  \begin{subfigure}[b]{0.32\textwidth}
    \centering\includegraphics[width=\textwidth]{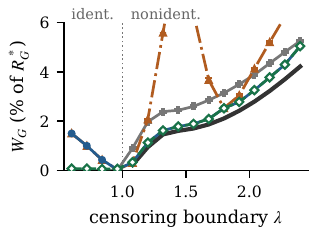}
    \caption{Regret across the frontier.}\label{fig:exp-syn-lambda}
  \end{subfigure}
  \par\smallskip
  \includegraphics[width=0.92\textwidth]{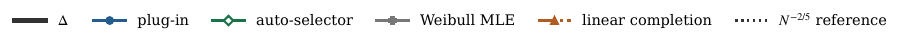}
  \caption{\textbf{Learning the robust reserve ($n=3$).}
  (a) Mean excess worst-case regret $W_G-\Delta$ in revenue units, over $200$ replications.
  (b) Mean worst-case regret in percent of $R_G^*$ for Rayleigh$(1)$,
  with $N=2000$ and $80$ replications per $\lambda$, on a linear axis
  capped at $6\%$; the linear-completion rule exceeds it at $\lambda=1.44$ (its peak, $9.1\%$), $1.56$, $2.28$ and $2.40$.
  Bars show one Monte-Carlo standard error; some are smaller than the
  markers, and the linear-completion rule in (a) is drawn without them.}
  \label{fig:exp-syn}
\end{figure}

Figure~\ref{fig:exp-syn-rate} uses the three instances of
Table~\ref{tab:exp-shape}. The plug-in's excess worst-case regret falls
toward zero as $N$ grows. {Its fitted slopes over $N=250$ to $64{,}000$, $-0.58$, $-0.45$ and $-0.51$ in panel order (standard errors at most $0.03$), are steeper than the reference $-2/5$: the
$N^{-2/5}$ rate is a worst case over the smoothness class, and a fixed
distribution can be learned faster.} The Weibull MLE is an oracle here, yet it
learns each distribution's own optimal reserve, whose worst-case regret
can exceed $\Delta$. Another compatible body may favor a different
reserve. Thus fitting the correct family and minimizing worst-case
regret remain different objectives. The linear-completion rule, which posts the
reserve $r_c(\hat A_N)$ of Section~\ref{sec:nonid-learning} from the
censored fraction alone, stays flat at a higher level because it ignores
the boundary hazard.

\begingroup Figure~\ref{fig:exp-syn-lambda} varies the historical reserve for
Rayleigh$(1)$. The auto-selector performs well on both sides of the
zero-regret frontier $\lambda=1$. In the tail-optimal case it learns the
optimum from the observed tail, where the plug-in's search below
$\lambda$ cannot reach it. In the nonidentifiable regime it coincides
with the plug-in and stays close to $\Delta$, below the correctly
specified Weibull MLE and below the linear-completion rule, except just
above the frontier and near $\lambda=1.8$, where the linear-completion
reserve is itself nearly minimax. This sweep suggests that, under MHR, a seller who does not know the
regime should
post the auto-selector's reserve: it loses little on either side of the
frontier, whereas the plug-in and the linear-completion rule each fail
on one side, and even the correctly specified Weibull MLE stays about a
point above $\Delta$ beyond the frontier. \endgroup

\subsection{Robustness to model misspecification}\label{sec:exp-truths}\label{sec:exp-parametric}

We censor Weibull$(2.5,1)$, denoted by $G$, at its median
$\lambda=(\log2)^{2/5}\approx0.864$, then replace its hidden body in three
ways. The \emph{exponential body} has constant hazard $A/\lambda$;
the \emph{equalizer body} makes $r^\dagger$ optimal; the \emph{tangent
body} has hazard zero followed by $h_0$.
These names describe the hidden bodies. The four completed distributions
are MHR and share the same censored fraction and Weibull tail, but favor
different reserves (Appendix~\ref{app:exp-truths}).

The Weibull family correctly describes the distribution of observed
records in all four cases. Weibull MLE is an oracle only under $G$, and
\emph{misspecified} under the other three because it imposes the wrong
hidden body.
Table~\ref{tab:exp-truths} reports both measures. Weibull MLE has almost
zero actual regret under $G$, but $2.49\%$ under the exponential completion,
compared with the plug-in's $1.36\%$. These columns evaluate each rule
under a fixed distribution. The worst-case column instead maximizes
regret over all compatible MHR distributions for each learned reserve,
then averages over samples. The plug-in's numerical worst-case
regret is $1.46\%$, close to minimax regret $1.35\%$.
Plug-in and auto-selector agree at the displayed precision and share a row.

\begin{table}[t]
  \centering\small
  \caption{\textbf{Four distributions with indistinguishable censored records.}
  Mean worst-case and actual regret in percent of the original Weibull
  distribution's $R_G^*$ ($N=3000$, $200$ paired replications).
  Actual regret compares each distribution with its own optimal revenue.
  Numerical minimax regret is $\Delta=1.35\%$; Monte-Carlo standard errors
  are at most $0.02$ percentage points.}
  \label{tab:exp-truths}
  \begin{tabular}{lccccc}
    \toprule
    & Worst-case regret & \multicolumn{4}{c}{Actual regret under each distribution} \\
    \cmidrule(lr){3-6}
    Rule & $W_G$ & Weibull $G$ & Equalizer & Exponential & Tangent \\
    \midrule
    Plug-in, Auto-selector & $1.46$ & $0.32$ & $0.00$ & $1.36$ & $1.35$ \\
    Plug-in (Weibull hazard) & $1.43$ & $0.31$ & $0.00$ & $1.36$ & $1.34$ \\
    \addlinespace
    Weibull MLE & $2.49$ & $0.00$ & $0.12$ & $2.49$ & $0.43$ \\
    Gamma MLE & $2.97$ & $0.04$ & $0.22$ & $2.97$ & $0.22$ \\
    Lognormal MLE & $3.17$ & $0.07$ & $0.26$ & $3.17$ & $0.16$ \\
    \bottomrule
  \end{tabular}
\end{table}

Using the Weibull fit only to estimate the boundary hazard gives a
similar worst-case regret ($1.43\%$ versus $1.46\%$).
This suggests that the reserve objective matters more than the hazard
estimator on this instance. The gaps between parametric fits and the
plug-in persist as samples grow (Appendix~\ref{app:exp-truths}):
additional censored records cannot check a fitted model's assumptions
about hidden values.

\begingroup \subsection{Experiments on eBay bids}\label{sec:exp-real}

We now ask the same questions of recorded bids from eBay auctions
\citep{jank2010modeling}.
The data record every bid in $628$ auctions of three standardized items:
Palm Pilot M515 handhelds, Xbox game consoles, and Cartier wristwatches.
eBay's proxy bidding is a second-price mechanism, so a bidder's highest
proxy bid in an auction serves as a value proxy. The seller's opening bid
acts as a reserve: bidders with lower values never appear. We therefore
form each item's reference sample from the auctions whose opening bid is
at most one dollar: $1{,}652$ bidder records in $132$ Palm Pilot auctions,
$460$ in $42$ Xbox auctions, and $392$ in $40$ Cartier auctions. eBay
reveals the winner's bid only up to the closing price, so one record per
auction is right-censored; Appendix~\ref{app:exp-real} describes its
imputation. These records serve as the values in what follows.

A shape-constrained refit $G$ of each reference sample, a kernel density
with its hazard projected onto the nondecreasing cone, supplies the
minimax regret $\Delta$ and the regime at each boundary. The boundaries
$155$, $199$, and $219$ of Table~\ref{tab:exp-ebay-holdout} are the
whole-dollar quantiles $0.5$, $0.7$, and $0.8$ of the Palm Pilot refit
(all nine cells: Appendix~\ref{app:exp-real}).

A kernel-hazard diagnostic with bidder-bootstrap bands
(Appendix~\ref{app:exp-real}) finds no significant hazard decrease in the
Palm Pilot records at any bandwidth. For Xbox it finds a significant
decrease at every bandwidth, and for Cartier a simultaneous-band decrease at the narrowest
bandwidth and no MHR-feasible kernel fit at any boundary. Palm Pilot
therefore carries
the tables below; Appendix~\ref{app:exp-real} reports Xbox and Cartier.
One lesson from Xbox is worth stating here: at its highest boundary the
jointly projected pair lands on the face $\bar A_N=\lambda\bar h_N$ in every split,
and the plug-in, which then posts the linear-completion reserve
$r_c(\bar A_N)$ at the projected estimate, loses heavily. The
linear-completion rule $r_c(\hat A_N)$, which keeps the original
censored-count estimate rather than adjusting it through the joint
projection, loses far less (Appendix~\ref{app:exp-real}). Here the projection reaches the face because the shape constraint fails, and this rule is the better fallback; reaching the face
does not by itself show that the true distribution is near full
identification.
Appendix~\ref{app:exp-real-mirror} repeats the three synthetic experiments
with the refit as the true distribution, with the same conclusions. The
two studies below also evaluate reserves on the recorded bids. \endgroup

\begingroup \paragraph{Learning from censored records, scored on the recorded bids}
Each of $1000$ random splits assigns $70\%$ of the item's bidders to
training and the rest to testing. The rules see the training records
censored at $\lambda$, so the censored count is known exactly; $N$ is the
number of training records, since every bidder in a one-dollar auction
faced no effective reserve. We evaluate a
posted reserve by its exact mean payment on $20{,}000$ simulated
three-bidder auctions among distinct test bidders and compare it with the
test optimum. This evaluates reserves on the observed and imputed bid
records rather than on a refit. The
full-information reserve (Full-info), fitted to the uncensored training
records, is the oracle row of Table~\ref{tab:exp-ebay-holdout};
Table~\ref{tab:exp-policies} defines every rule.

On the Palm Pilot cells the plug-in and auto-selector have lower held-out
regret than every MLE at $\lambda=155$ and $219$ ($1.4\%$ and $0.6\%$
against $2.3$--$2.8\%$ and $1.1$--$1.9\%$); at $219$ the boundary rule (post $\lambda$) and
Drop (maximize empirical revenue over the observed bids only) lose $28\%$. At $\lambda=199$ the MLEs are ahead ($1.25$--$1.52\%$
against $1.68\%$): the data-driven bandwidth rule selects a twelve-dollar
window there, and the plug-in with the fixed window $c_b=\lambda/2$
loses $1.0\%$ ($1.1\%$ and $0.6\%$ at $155$ and $219$). The gap comes
from the narrow window, not from whole-dollar price points
(Appendix~\ref{app:exp-real}). Full-info loses $0.08\%$.

\begin{table}[htbp]
  \centering\footnotesize
\setlength{\tabcolsep}{3pt}
\caption{\textbf{Held-out regret on the recorded Palm Pilot bids}, in
  percent of each split's test optimum ($n=3$, $1000$ bidder splits,
  $20{,}000$ test auctions per split; Monte-Carlo standard errors at most
  $0.13$ points). $N$ is the mean number of training records. Full-info
  fits a reserve to the uncensored training records.}
  \label{tab:exp-ebay-holdout}
\begin{tabular}{lrrccccccccc}
  \toprule
  & & & & & & \multicolumn{3}{c}{Censored MLE} & & & \\
  \cmidrule(lr){7-9}
  Item & $\lambda$ & $N$ & Full-info & Plug-in & Auto-selector & Weibull & Gamma & Lognormal & Linear compl. & Boundary & Drop \\
  \midrule
    Palm Pilot & $155$ & $1156$ & $0.08$ & $1.39$ & $1.44$ & $2.26$ & $2.61$ & $2.77$ & $1.64$ & $1.64$ & $2.34$ \\
    Palm Pilot & $199$ & $1156$ & $0.08$ & $1.68$ & $1.68$ & $1.52$ & $1.30$ & $1.25$ & $3.64$ & $10.84$ & $11.14$ \\
    Palm Pilot & $219$ & $1157$ & $0.08$ & $0.63$ & $0.63$ & $1.05$ & $1.41$ & $1.90$ & $1.53$ & $27.73$ & $27.98$ \\
  \bottomrule
\end{tabular}
\end{table} \endgroup

\begingroup \paragraph{Sellers' own reserves}
The remaining $211$ Palm Pilot auctions have opening bids above one
dollar: reserves chosen by sellers. eBay records only the bidders who
appeared, so the censored fraction must come from participation
(Section~\ref{sec:setting}). The reference auctions have $12.5$ records
each, so $K$ auctions with opening bid $\lambda$ are taken to have faced
$\hat N=12.5K$ bidders, of whom $m$ appeared; the rules receive the $m$
records and $\hat N-m$ no-bids. This assumes that the number of potential
bidders does not depend on the opening bid; the comparison with
$G(\lambda)$ below shows where it fails. Table~\ref{tab:exp-real-natural} uses every opening bid shared by at least eight auctions (opening bids
within one cent), eight in all. Every rule's search
bound is the larger of $4\lambda$ and twice the largest observed
record, since the optimum lies far
above a ten-dollar opening bid.

At opening bids of $150$ and above, the participation-based censored
fraction $\hat\alpha$ is within $0.03$ of $G(\lambda)$ except at $175$,
where it is $0.70$ against $0.58$. At $10$, $50$, and $100$ it overstates
censoring ($0.13$ against $0.02$, $0.35$ against $0.14$, $0.45$ against $0.31$): auctions with a
positive opening bid draw fewer bidders for reasons beyond censoring.
Varying the rate from $8$ to $20$ records per auction leaves the detected regime unchanged in every cell, keeps the auto-selector below the Weibull fit in four to seven of the eight cells, and moves the reserve by
two to four dollars per record in the nonidentifiable cells
(Appendix~\ref{app:exp-real}).

The auto-selector's branch agrees with the refit's regime in seven of the
eight cells. At $50$ the data-driven bandwidth rule selects its narrowest window
($c_b=\lambda/4$), whose boundary-hazard estimate exceeds $1/\lambda$, so the
auto-selector takes the plug-in branch, and the overstated count leads it
to post about $49$ dollars, near the opening bid; it loses $6.9\%$ on the
reference records against $3.4\%$ for the Weibull MLE. With the fixed
window $c_b=\lambda/2$ it takes the tail branch and loses $0.04\%$. In the
five nonidentifiable cells, with $100$ to $488$ inferred bidders, its
worst-case regret is $0.3$ to $2.8$ points above $\Delta$ and below the
Weibull MLE's in every cell. Scored on the reference records it loses
$0.4$--$2.5\%$, against $1.6$--$3.0\%$ for the Weibull MLE; the boundary
loses from $0.04\%$ at $150$, where the opening bid sits at the optimum,
to $54\%$ at $240$. The refit's optimal reserve itself scores $0.76\%$
on these test auctions, so the auto-selector's $0.4$--$0.9\%$ at $200$ to
$240$ is the attainable level. Its $2.5\%$ at $175$ comes from the
overstated count, which pulls the reserve down to $126$ dollars.

\begin{table}[htbp]
  \centering\scriptsize
\setlength{\tabcolsep}{2.5pt}
  \caption{\textbf{Sellers' own opening bids on Palm Pilot} ($n=3$). $K$
  auctions with opening bid within one cent of $\lambda$, $m$ records, $\hat N=12.5K$ inferred bidders, and $\hat\alpha=1-m/\hat N$; $G(\lambda)$ and the
  regime are those of the refit, and Detected is the auto-selector's branch.
  Weibull is the censored Weibull MLE. Worst-case regret $W_G$ in percent of $R_G^*$ and held-out regret on
  $20{,}000$ auctions among reference bidders in percent of the test
  optimum, means over $50$ draws of the winner imputation (standard errors at most $0.26$ points, conditional on the recorded auctions); $\Delta$ in percent of $R_G^*$.}
  \label{tab:exp-real-natural}
\begin{tabular}{lrrrrcclcc*{2}{c}*{3}{c}}
  \toprule
  & & & & & & & & & & \multicolumn{2}{c}{Worst-case} & \multicolumn{3}{c}{Held-out} \\
  \cmidrule(lr){11-12}\cmidrule(lr){13-15}
  Item & $\lambda$ & $K$ & $m$ & $\hat N$ & $\hat\alpha$ & $G(\lambda)$ & Regime & Detected & $\Delta$ & Auto-selector & Weibull & Auto-selector & Weibull & Boundary \\
  \midrule
    Palm Pilot & $10$ & $32$ & $347$ & $400$ & $0.13$ & $0.02$ & ident. & ident. & $0$ & $0.03$ & $1.07$ & $0.15$ & $2.52$ & $7.56$ \\
    Palm Pilot & $50$ & $9$ & $73$ & $113$ & $0.35$ & $0.14$ & ident. & nonident. & $0$ & $4.82$ & $1.57$ & $6.88$ & $3.39$ & $6.84$ \\
    Palm Pilot & $100$ & $9$ & $62$ & $113$ & $0.45$ & $0.31$ & ident. & ident. & $0$ & $0.06$ & $2.19$ & $0.04$ & $4.18$ & $3.57$ \\
    Palm Pilot & $150$ & $14$ & $90$ & $175$ & $0.49$ & $0.48$ & nonident. & nonident. & $0.14$ & $0.94$ & $2.38$ & $1.17$ & $2.32$ & $0.04$ \\
    Palm Pilot & $175$ & $39$ & $145$ & $488$ & $0.70$ & $0.58$ & nonident. & nonident. & $1.65$ & $3.00$ & $3.89$ & $2.52$ & $2.95$ & $2.41$ \\
    Palm Pilot & $200$ & $8$ & $27$ & $100$ & $0.73$ & $0.70$ & nonident. & nonident. & $2.18$ & $2.46$ & $3.31$ & $0.94$ & $1.96$ & $11.06$ \\
    Palm Pilot & $215$ & $11$ & $28$ & $138$ & $0.80$ & $0.78$ & nonident. & nonident. & $3.05$ & $3.48$ & $3.74$ & $0.83$ & $1.63$ & $25.41$ \\
    Palm Pilot & $240$ & $14$ & $22$ & $175$ & $0.87$ & $0.90$ & nonident. & nonident. & $5.34$ & $8.12$ & $12.09$ & $0.40$ & $2.10$ & $53.52$ \\
  \bottomrule
\end{tabular}
\end{table} \endgroup

\section{Conclusion}\label{sec:conclusion}

\begingroup More bids at the same reserves can reduce estimation error, but cannot
directly reveal how values are distributed below the lowest historical
reserve. Our main results show how to exploit MHR to choose a reserve
even when these hidden values prevent identification of the optimum. \endgroup

\begingroup In the nonidentifiable regime, two observable quantities determine the
reserve with the smallest worst-case revenue regret, and our learning
bounds quantify the extra regret from estimating them. The analysis
extends to regular and unrestricted distributions: regular distributions
retain the two-quantity reduction, while MHR additionally permits full
identification and can reduce minimax regret. \endgroup

\begingroup Experiments on synthetic instances show the learned rule approaching
the minimax regret at moderate sample sizes, while the parametric fits'
worst-case regret stays above it. On recorded eBay bids its held-out
regret is below every parametric fit's at two of the three Palm Pilot
reference boundaries and below the Weibull fit's at seven of the eight most frequent Palm Pilot opening bids (four to seven under other imputed bidder counts), and reusing the historical reserve can lose more than half the optimal revenue. Two practical lessons follow: MHR can be checked on the
observed tail, and when the projected pair lands on the face $\bar A_N=\lambda\bar h_N$, as on one Xbox cell, the linear-completion rule at the original censored-count estimate can be the better fallback. \endgroup

Our analysis focuses on second-price auctions and assumes independent
values from a common distribution. It also requires participation
counts to estimate the censored fraction. The $N^{-2/5}$ learning guarantees
require a smooth density just above the censoring boundary. We leave
these limitations to future work.

\section*{Acknowledgments}
\paragraph{Use of AI tools}
The authors proposed all research questions, designed all experiments, and
derived the original proofs of the theoretical results, except for a few
steps in the lower-bound proof of Proposition~\ref{prop:interior-lower-bound}.
We used ChatGPT and Claude to complete those steps
(Appendix~\ref{app:lb-statistical-reconstruction}), polish the writing,
implement experimental code, and assist with proof checks and simplification.
The authors verified the results and take full responsibility for the manuscript.

\bibliography{references}

\clearpage
\appendix
\section{Proofs for the Population Problem}\label{app:proofs}

The first results concern what can be learned from the population record.
We then prove the MHR reduction, its equalizer, and the numerical error bounds.
Throughout, $n\ge2$ is an integer, $\lambda>0$, and $G\in\mathcal G$.
Each result states the shape restrictions it uses.

\begingroup \subsection{Notation}\label{app:notation}
Table~\ref{tab:notation} collects the recurring notation of the paper.

\begin{table}[htbp]
  \centering\scriptsize
  \caption{Recurring notation. Boundary quantities belong to the observed tail and are common to all compatible completions.}
  \label{tab:notation}
  \begin{tabular}{@{}p{0.24\textwidth}>{\raggedright\arraybackslash}p{0.71\textwidth}@{}}
    \toprule
    Symbol & Meaning\\
    \midrule
    $n$, $N$ & Number of bidders in the target auction; number of historical records subject to censoring at the lowest reserve $\lambda$, one per bidder who faced it.\\
    $G$, $F$, $\lambda$ & True distribution, a compatible completion, and the censoring boundary (the lowest historical reserve).\\
    $\mathcal F(\lambda;G)$ & Ambiguity set: all completions of the observed tail; subscripts restrict the shape.\\
    $h_F$, $H_F$ & Hazard and cumulative hazard; MHR means $H_F$ convex.\\
    $\phi_F$, $\Phi_F$ & Virtual value and single-bidder revenue curve in quantile coordinates; regularity means $\Phi_F$ concave.\\
    $R_F(r)$, $R_F^*$ & Second-price auction revenue at reserve $r$, and its optimal value.\\
    $\Delta(\mathcal A)$, $\Delta_{\mathrm{rand}}(\mathcal A)$ & Minimax regret over the completion class $\mathcal A$; its value when the seller may randomize the reserve. $\Delta$, $\Delta(A,h_0)$ and $\Delta(G)$ all denote the MHR value $\Delta(\mathcal F_{\mathrm{MHR}}(\lambda;G))$ (Section~\ref{sec:dro}).\\
    $r^\dagger$, $W_G(r)$ & The minimax reserve and the worst-case regret of reserve $r$ over compatible MHR completions.\\
    $\mathsf W(r;A,h_0)$ & Body objective, Equation~\eqref{eq:W-body-def}; equals $W_G(r)$ when $h_0\ge1/\lambda$.\\
    $\mathsf W_{\rm reg}$ & The regular worst case, Equation~\eqref{eq:W-reg}.\\
    $A$, $h_0$ & Boundary cumulative hazard $H_G(\lambda)=-\log(1-\alpha)$ and boundary hazard, the right derivative $H_G'(\lambda^+)$.\\
    $\alpha$, $q_\lambda$ & Censored fraction $G(\lambda)=1-e^{-A}$ and probability of an observed bid $1-\alpha=e^{-A}$.\\
    $B_\lambda$ & Tail gain: revenue of the best reserve in $[\lambda,\infty)$ minus revenue at $\lambda$.\\
    $c_\alpha$, $p_\alpha$, $k_\alpha$ & Probability $\alpha^n$ that all $n$ values lie below $\lambda$, probability $n\alpha^{n-1}(1-\alpha)$ that exactly one lies above, and $\max_{0\le z\le\alpha}nz^{n-1}(1-z)$.\\
    $\phi_0$, $c_0$ & Boundary virtual value $\lambda-1/h_0$ and revenue-curve height $\lambda q_\lambda$ at the boundary.\\
    $\mathcal H(A,h_0)$ & Feasible body cumulative hazards: convex nondecreasing curves from $(0,0)$ to $(\lambda,A)$ with final slope at most $h_0$.\\
    $\mathcal T(A,h_0)$ & Two-level completions: bodies in $\mathcal H(A,h_0)$ with two hazard levels, joined to the observed tail; they attain the MHR worst case of every reserve. As a superscript ($\mathsf W^{\mathcal T}$, $\Delta^{\mathcal T}$), restriction to a finite grid of levels.\\
    $\mathcal C(A,h_0)$ & Feasible regular body revenue curves; $S_s$ and $E_t$ are the straight and capped kinked families.\\
    $L(r)$, $U(r)$ & Worst-case losses against lower and higher alternative reserves: from posting too high and from posting too low.\\
    $\mathsf L_{\mathrm{reg}}$, $\mathsf U_{\mathrm{reg}}$ & The same directional losses over regular completions.\\
    $d$, $k_1$ & Distance to full identification $\lambda h_0-A$ and the constant $k_1=((n-1)/n)^{n-1}$, the value of $k_\alpha$ at $\alpha=1$.\\
    $m$, $\mathcal U_N$ & Number and multiset of uncensored bids among the $N$ records.\\
    $\hat A_N$, $\hat h_0$ & Raw boundary estimates, $-\log(m/N)$ and the local-linear histogram estimate of $h_0$; bars denote the projected estimates.\\
    $\mathcal P_\lambda$ & Projection set for the scaled estimates $(\hat A_N,\lambda\hat h_0)$, Equation~\eqref{eq:closed-wedge}.\\
    $\mathcal D$ & Parameter domain $\{(A,h_0):0<A<\lambda h_0,\ h_0>1/\lambda\}$ of the nonidentifiable regime.\\
    $\theta$, $\hat\theta$, $\bar\theta$ & Scaled boundary pair $(A,\lambda h_0)$, its raw estimate and its projection onto $\mathcal P_\lambda$; Appendix~\ref{app:regular} uses the unscaled pair $(A,h_0)$.\\
    $\psi(p)$ & $1-np^{n-1}+(n-1)p^n$, the probability that at least two of $n$ values exceed the $p$-quantile.\\
    $\tilde g$, $\mathcal B_N$ & Conditional density $g/q_\lambda$ of an observed bid; the event $\{m\ge Nq_\lambda/2\}$.\\
    $T_\lambda$, $r_{\mathrm T}$ & Second-highest-value contribution above $\lambda$, common to all completions; the reserve maximizing $R_G$ on $[\lambda,\infty)$.\\
    $\mathcal C_b$, $\hat c$ & Grid of bandwidth constants of the data-driven rule and the selected constant.\\
    $\mathcal Q$ & Quantile grid of the regular scalar search, Corollary~\ref{cor:regular-scalar-grid}.\\
    $\mathcal K_{\rm reg}$ & Parameter rectangle of the regular plug-in.\\
    $r_c(A)$ & Linear-completion reserve $\min\{\lambda,\lambda/A\}$, optimal under constant body hazard $A/\lambda$; $r_c(0)=\lambda$, $r_c(+\infty)=0$.\\
    $h_{\min}$, $\bar\lambda$ & Known lower bound on $h_0$ and the end of the tail search interval, $\max\{\lambda,1/h_{\min}\}$.\\
    $h_{\max}$, $q_{\min}$ & Upper bound on $h_0$ and lower bound on $q_\lambda$ in the auto-selector's guarantee, Proposition~\ref{prop:pretest-uniform}.\\
    $m_{\min}$, $M$, $b_N$, $c_{\mathrm g}$ & Fewest uncensored bids before the plug-in estimates $h_0$ (below it, the MHR plug-in posts the linear-completion reserve and the regular rule of Definition~\ref{def:plug-in-reg} posts $\lambda$); number of histogram cells; bandwidth $c_bN^{-1/5}$; mesh constant of the reserve and hazard-level grids, $\varepsilon_N=c_{\mathrm g}N^{-1/2}$.\\
    $\kappa_A$, $\kappa_h$ & Constants in the boundary concentration bounds, Lemma~\ref{lem:est-rates}.\\
    $\mathrm{Lip}_\Theta$, $\mathrm{Lip}_r$ & MHR stability constants $3\lambda\max\{1,\lambda\}$ and $\max\{1,\lambda h_0\}$.\\
    $C_{\mathrm{reg}}$, $\mathrm{Lip}_r^{\mathrm{reg}}$ & Parameter-stability and reserve-Lipschitz constants of the regular worst case, Proposition~\ref{prop:reg-finite}.\\
    $C_\star$, $C_{\rm num}$ & Statistical and discretization constants in Theorem~\ref{thm:finite-sample}.\\
    \bottomrule
  \end{tabular}
\end{table} \endgroup

\subsection{Revenue and the unrestricted benchmark}\label{app:proofs-setting}

\begin{lemma}[Existence of an optimal reserve]\label{lem:r^*_exists}
Let $F\in\mathcal G$ and $n\ge2$. Then the nonnegative function $R_F$ in
\eqref{eq:revenue} attains its supremum on $[0,\infty)$.
\end{lemma}

\begin{proof}
Finite mean bounds the tail integrand in \eqref{eq:revenue} by an
integrable multiple of $uf(u)$, so $R_F$ is continuous. Moreover,
$r(1-F(r))\le\mathbb E_F[V\mathbf1\{V>r\}]\to0$, and the integral
also tends to zero. Since $R_F(0)=\mathbb E_F[V_{(2)}]>0$, the maximum
is attained on a compact interval.
\end{proof}

Recall $\alpha=G(\lambda)\in(0,1)$ and put
\[
 c_\alpha=\alpha^n,\qquad p_\alpha=n\alpha^{n-1}(1-\alpha),\qquad
 k_\alpha=\max_{0\le z\le\alpha}nz^{n-1}(1-z).
\]
The last maximum occurs at $z=\min\{\alpha,(n-1)/n\}$.
Let $r_{\rm T}$ maximize the common tail revenue, and write
$B_\lambda=R_G(r_{\rm T})-R_G(\lambda)$.
Here $c_\alpha$ is the all-censored probability, $p_\alpha$ the
exactly-one-uncensored probability, and $B_\lambda$ the observable tail gain.

\begin{theorem}[Unrestricted minimax regret]\label{thm:unrestricted}
Let $G\in\mathcal G$, $n\ge2$, $\lambda>0$, and
$\alpha=G(\lambda)\in(0,1)$. Use the quantities
$c_\alpha,p_\alpha,k_\alpha,r_{\mathrm T},B_\lambda$ defined above, and set
$W_{\mathrm{all}}(r):=\sup_{F\in\mathcal F(\lambda;G)}
\{R_F^*-R_F(r)\}$. For $0\le r\le\lambda$,
\begin{equation}\label{eq:unrestricted-risk}
  W_{\mathrm{all}}(r)
  =\max\{r c_\alpha,\ (\lambda-r)k_\alpha,
                    \ B_\lambda+(\lambda-r)p_\alpha\},
\end{equation}
and $W_{\mathrm{all}}(r_{\mathrm T})=(\lambda c_\alpha-B_\lambda)_+$.
For $r\ge\lambda$, every completion agrees with $G$, so
$W_{\mathrm{all}}(r)=\max\{B_\lambda,\lambda c_\alpha\}+R_G(\lambda)-R_G(r)$,
which is minimized over the tail at $r_{\mathrm T}$.
Consequently,
\begin{equation}\label{eq:unrestricted-frontier}
  \Delta\bigl(\mathcal F(\lambda;G)\bigr)=0
  \quad\Longleftrightarrow\quad
  B_\lambda\ge\lambda\alpha^n.
\end{equation}
\end{theorem}

\begin{figure}[htbp]
     \centering
     \includegraphics[width=0.78\textwidth]{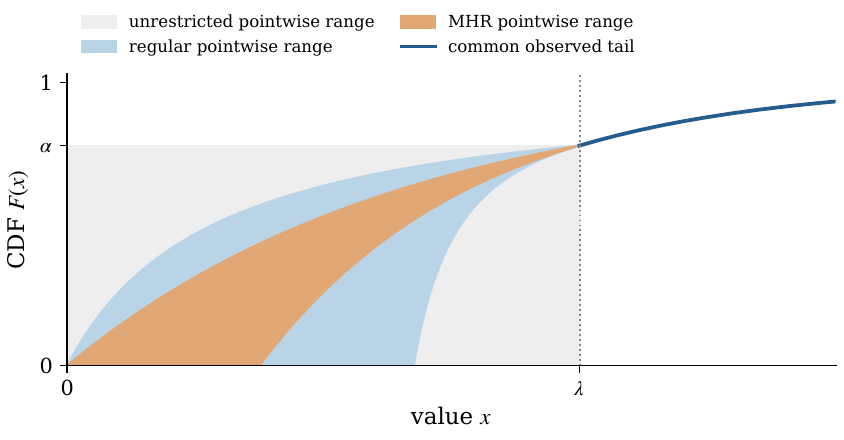}
     \caption{Pointwise CDF ranges for unrestricted, regular, and MHR
completions of the same observed tail. A curve inside a shaded band must
still satisfy its class's shape constraint; unattained unrestricted
boundaries are limits of absolutely continuous completions.}
     \label{fig:general_set}
\end{figure}

The three branches compare with reserves below $r$, elsewhere in the
body, and in the tail. The optimal body reserve equalizes the increasing
branch with the larger decreasing branch. Figure~\ref{fig:general_set}
shows the pointwise CDF ranges of the three classes for one observed tail.

\begin{corollary}[Computing the unrestricted reserve]\label{cor:unrestricted-computation}
Under the hypotheses and notation of Theorem~\ref{thm:unrestricted},
if $B_\lambda<\lambda c_\alpha$, set
\[
  r_{\mathrm b}:=\max\left\{
       \frac{\lambda k_\alpha}{c_\alpha+k_\alpha},\,
       \frac{B_\lambda+\lambda p_\alpha}{c_\alpha+p_\alpha}
                         \right\}.
\]
Then $r_{\mathrm b}\in[0,\lambda]$ minimizes the body objective, and
\[
  \Delta\bigl(\mathcal F(\lambda;G)\bigr)
  =\min\{c_\alpha r_{\mathrm b},\ \lambda c_\alpha-B_\lambda\}.
\]
In this case, an optimal reserve is the better of $r_{\mathrm b}$ and
$r_{\mathrm T}$. If instead $B_\lambda\ge\lambda c_\alpha$, posting
$r_{\mathrm T}$ gives zero regret.
\end{corollary}

\begin{proof}[Proof of Theorem~\ref{thm:unrestricted} and Corollary~\ref{cor:unrestricted-computation}]
Let $T_\lambda=\mathbb E_F[V_{(2)}\mathbf1\{V_{(2)}\ge\lambda\}]$,
which is common to all completions. Then
$R_G(\lambda)=T_\lambda+\lambda p_\alpha$. Fix $r\in[0,\lambda]$.

For a lower competitor $s\le r$, reserve $r$ weakly dominates $s$
whenever the largest value clears $r$. Otherwise the payment at $s$
is at most $r$. Thus the gain is at most $rF(r)^n\le rc_\alpha$.
For $r>0$, put all body mass in $(r-\varepsilon,r)$ and post
$s=r-\varepsilon$. The gain tends to $rc_\alpha$; at $r=0$ it is zero.

For $r\le s\le\lambda$, the revenue derivative gives
\[
 R_F(s)-R_F(r)=\int_r^s nF(u)^{n-1}(1-F(u))du
                  -\int_r^s u\,d(F(u)^n)
 \le(s-r)k_\alpha.
\]
To approach $(\lambda-r)k_\alpha$, put mass
$z_*=\min\{\alpha,(n-1)/n\}$ just below $r$, the remaining body mass
above $s$, and let $s\uparrow\lambda$. The CDF is $z_*$ on $[r,s]$.
At $r=0$, put the first mass in $(0,\varepsilon)$; the discarded
contribution is $O(\varepsilon)$. These are absolutely continuous completions.

For a tail competitor, the largest revenue is $R_G(r_{\rm T})$, whereas
$R_F(r)\ge T_\lambda+rp_\alpha$. This gives the third branch
$B_\lambda+(\lambda-r)p_\alpha$, attained when all body mass lies below
$r$; at zero, concentrate it arbitrarily close to zero.
Taking the largest of the three sharp bounds proves
\eqref{eq:unrestricted-risk}.

Finally, a lower reserve can beat $\lambda$ only when all bidders are
censored, so
\begin{equation}\label{eq:body-revenue-envelope}
 \sup_{F}\max_{s\le\lambda}R_F(s)=R_G(\lambda)+\lambda c_\alpha.
\end{equation}
For sharpness put the body in $(\lambda-\varepsilon,\lambda)$ and post
$\lambda-\varepsilon$. Hence at any tail reserve
\[
 W_{\rm all}(r)=\max\{B_\lambda,\lambda c_\alpha\}
                      +R_G(\lambda)-R_G(r).
\]
This proves the tail formula and the zero-regret frontier. If
$B_\lambda<\lambda c_\alpha$, the two crossings of $rc_\alpha$ with the decreasing
branches are exactly those in Corollary~\ref{cor:unrestricted-computation}.
Their larger crossing minimizes the body objective; comparison with
$r_{\rm T}$ finishes the solution. In particular, if $B_\lambda<\lambda c_\alpha$, the
minimum is $\min\{c_\alpha r_{\rm b},\lambda c_\alpha-B_\lambda\}>0$, so
$\Delta(\mathcal F(\lambda;G))>0$; this is the converse direction of
\eqref{eq:unrestricted-frontier}.
\end{proof}

\begin{remark}[An identifiable unrestricted instance]\label{rem:unrestricted-example}
For $n=2$, $\lambda=1$, and
$G=\frac12\mathrm{Unif}[0,1]+\frac12\mathrm{Unif}[2,3]$, every completion
has $R_F(2)=19/12$, while \eqref{eq:body-revenue-envelope} bounds every
body revenue by $4/3$. Thus all completions share an optimal tail reserve.
\end{remark}

\subsection{The zero-regret frontier}\label{app:proofs-ident}

\begin{lemma}[Tail domination]\label{lem:auction-tail-domination}
Let $\mathcal A\subseteq\mathcal F(\lambda;G)$ be nonempty.
Assume $R_F(r)\le R_F(\lambda)$ for every $F\in\mathcal A$ and every
$r\ge\lambda$. Then
\[
 \Delta(\mathcal A)=\inf_{r\in[0,\lambda]}
       \sup_{F\in\mathcal A}\{R_F^*-R_F(r)\}.
\]
\end{lemma}
\begin{proof}
Replacing any reserve above $\lambda$ by $\lambda$ weakly increases
revenue under every completion.
\end{proof}

\begin{remark}[Feasibility of an MHR body]\label{rem:left-feasible}
An MHR completion exists exactly when the observed cumulative hazard is
convex and $A\le\lambda h_0$. Necessity follows from
$A=\int_0^\lambda h_F\le\lambda h_0$; the chord
$H(x)=Ax/\lambda$ proves sufficiency. At equality the hazard must equal
$h_0$ almost everywhere, so the chord is the only body.
\end{remark}

\begin{lemma}[Feasible heights]\label{lem:left-pointwise-H}
Assume $\mathcal F_{\mathrm{MHR}}(\lambda;G)$ is nonempty and the
common right boundary hazard satisfies $0<h_0<\infty$.
Then, for every $F\in\mathcal F_{\mathrm{MHR}}(\lambda;G)$ and
every $r\in[0,\lambda]$,
\[
 (A-h_0(\lambda-r))_+\le H_F(r)\le Ar/\lambda.
\]
Each bound is attained by a completion in this full ambiguity set.
\end{lemma}
\begin{proof}
Convexity bounds $H_F$ by its chord. The remaining interval accumulates
at most $h_0(\lambda-r)$ hazard, giving the lower bound.
The chord and $H(x)=(A-h_0(\lambda-x))_+$ attain the two bounds.
The latter has a positive lower support endpoint, which our definition
of MHR permits. Exponentiation gives \eqref{eq:S-interval}.
\end{proof}

\begin{remark}[Measurability]\label{rem:measurability}
The supremum of the continuous functions $R_F^*-R_F(r)$ is lower
semicontinuous, hence Borel. Composing it with a measurable reserve gives
a random loss. Our empirical rules select the smallest maximizer from
specified finite candidate sets, so the reserves are measurable as well.
\end{remark}

\begin{lemma}[Revenue derivative]\label{lem:auction-revenue-derivative}
Let $F\in\mathcal G$ and $n\ge2$. Then $R_F$ is locally absolutely
continuous on $[0,\infty)$ and, for almost every $r\ge0$,
\begin{equation}\label{eq:auction-revenue-derivative}
 R_F'(r)=nF(r)^{n-1}\bigl(1-F(r)-rf(r)\bigr).
\end{equation}
Where $f(r)>0$, this equals $-nF(r)^{n-1}f(r)\phi_F(r)$.
Thus the sign of virtual value determines the direction of revenue.
\end{lemma}
\begin{proof}
The integrand in \eqref{eq:revenue} is bounded by $n(n-1)uf(u)$,
which is integrable. Differentiating its two terms on compact intervals
cancels the contributions containing $F^{n-2}(1-F)f$ and gives the formula.
An absolutely continuous function with nonnegative (nonpositive)
derivative almost everywhere is nondecreasing (nonincreasing).
\end{proof}

\begin{lemma}[Two-distribution lower bound]\label{lem:two-dist}
Let $\mathcal A\subseteq\mathcal F(\lambda;G)$ and let
$F^{(i)}\in\mathcal A$ have a unique optimal reserve
$r_i^*\in[0,\lambda]$ for $i\in\{1,2\}$.
If $r_1^*\ne r_2^*$, then $\Delta(\mathcal A)>0$.
\end{lemma}

\begin{proof}
The two continuous losses vanish at different reserves, so their maximum
is positive everywhere. Both revenues tend to zero at infinity, so this
maximum is bounded away from zero outside a compact interval. Its
minimum on that interval is positive as well. Nature's supremum dominates
this two-distribution loss, proving $\Delta(\mathcal A)>0$.
\end{proof}

\begin{proof}[Proof of Theorem~\ref{thm:frontier}]
If $h_0\le1/\lambda$, every body has $rh_F(r)\le1$, so revenue is
nondecreasing up to $\lambda$. Every completion therefore shares the tail
maximizers. If $A=\lambda h_0$, Remark~\ref{rem:left-feasible} gives a
uniquely determined completion. Both cases have zero regret.

In the remaining regime, tail revenue is nonincreasing, so it suffices
to exhibit two body completions with distinct unique optima. Choose
\begin{equation}\label{eq:choose-b}
 \max\{A/\lambda,1/\lambda\}<b_1<b_2<\min\{h_0,(A+1)/\lambda\}.
\end{equation}
This interval is nonempty. For each $i$, take
$\tau_i=\lambda-A/b_i+\varepsilon_i$, where
$0<\varepsilon_i<\min\{1/b_i-\lambda+A/b_i,1/b_i\}$, and set
$a_i=b_i\varepsilon_i/\tau_i$. Then
\begin{equation}\label{eq:twolevel-hazard}
 a_i\tau_i+b_i(\lambda-\tau_i)=A,
\end{equation}
with $0<a_i<b_i<h_0$, $\tau_i<1/b_i<\lambda$, and
\begin{equation}\label{eq:tau-a-invariant}
 a_i\tau_i=b_i\varepsilon_i<1.
\end{equation}
The hazard $a_i$ followed by $b_i$ pastes to the common tail and is MHR.
Revenue strictly increases to $1/b_i$ and strictly decreases afterwards.
These two unique optima differ, so Lemma~\ref{lem:two-dist} gives
$\Delta>0$.
\end{proof}

\begin{remark}\label{rem:hard-gloss}
Nonidentifiability concerns disagreement between completions, not the
location of the true optimum. For example, the hazard $1/2$ below
$\lambda=1$ and $2$ above it gives $r_G^*=1$, but $A=1/2<\lambda h_0=2$ and
$\phi_0>0$, so other compatible MHR bodies have different optima.
\end{remark}

\begin{proof}[Proof of the population claim in Proposition~\ref{prop:identifiable-iff}]
Let $W(r)=\sup_{F\in\mathcal A}\{R_F^*-R_F(r)\}$.
It is finite and lower semicontinuous. All tail revenues equal $R_G$,
which tends to zero, whereas $R_G(\lambda)>0$.
Thus some $K\ge\lambda$ makes reserve $\lambda$ dominate all reserves
above $K$, under every completion. The minimax infimum is attained on
$[0,K]$. It equals zero exactly when its minimizing reserve is optimal
for every completion. This proves the population equivalence.
\end{proof}

\subsection{Randomization and competition}\label{app:randomization-competition}

\begin{remark}[Randomized reserves]\label{rem:randomized}
A distribution $\mu$ over reserves defines an alternative population benchmark:
\[
 \Delta_{\mathrm{rand}}(\mathcal A)
 :=\inf_\mu\sup_{F\in\mathcal A}\mathbb E_{r\sim\mu}[R_F^*-R_F(r)].
\]
For every nonempty set $\mathcal A\subseteq\mathcal F(\lambda;G)$ of
regular distributions,
$\Delta(\mathcal A)/2\le\Delta_{\mathrm{rand}}(\mathcal A)\le\Delta(\mathcal A)$.
Here nature chooses a completion before the reserve is drawn. Our
learning criterion instead takes worst-case regret at the realized
reserve, then averages over the historical sample.
\end{remark}

\begin{proposition}[Randomization preserves the frontier]\label{prop:randomization-bound}
Let $\mathcal A\subseteq\mathcal F(\lambda;G)$ be nonempty and assume
every $F\in\mathcal A$ is regular. Then
\begin{equation}\label{eq:randomization-bound}
  \tfrac12\Delta(\mathcal A)
  \le\Delta_{\mathrm{rand}}(\mathcal A)\le\Delta(\mathcal A).
\end{equation}
Under the historical design of Section~\ref{subsec:censored-data}, every
possibly randomized data-driven rule satisfies, at every sample size,
\[
  \sup_{F\in\mathcal A}\mathbb E_F
  \bigl[R_F^*-R_F(r^\pi)\bigr]\ge\tfrac12\Delta(\mathcal A),
\]
where the expectation includes the historical data and the rule's
randomization; since every historical reserve is at least $\lambda$, the
censored record has the same law under every $F\in\mathcal A$, so
$\mathbb E_F$ may be read as $\mathbb E_G$ on the record. Thus the zero-regret conditions in Theorems~\ref{thm:frontier} and~\ref{thm:frontier-reg}
also hold for randomized reserves and for vanishing worst-case expected
regret of data-driven rules.
\end{proposition}

\begin{proof}[Proof of Proposition~\ref{prop:randomization-bound}]
For regular distributions, revenue is unimodal. For any reserve distribution $\mu$,
let $m_\mu$ be a median. For each completion, the half of $\mu$ lying on the
side of $m_\mu$ away from a maximizer has loss at least the loss at $m_\mu$.
Consequently
\[
 \mathbb E_{r\sim\mu}[R_F^*-R_F(r)]\ge\tfrac12(R_F^*-R_F(m_\mu)).
\]
Take the supremum over $F$ and the infimum over $\mu$ to get the lower
bound; point masses give the upper bound.
All compatible completions induce the same censored-data law, hence the
same reserve law $\mu_N$ for any rule, including its random seed.
Applying the inequality to $\mu_N$ proves the rule bound.
Conversely, the zero-regret rules of Proposition~\ref{prop:identifiable-iff}
have uniformly bounded loss,
$R_F^*\le n(\lambda+\mathbb E_G[(V-\lambda)_+])$.
Their convergence in probability therefore gives convergence in expectation.
\end{proof}

\begin{remark}[Which minimax criterion applies to a rule?]\label{rem:randomized-details}
Theorem~\ref{thm:finite-sample} controls $\mathbb E[\sup_F\{R_F^*-R_F(r^\pi)\}]$,
whose minimum with the distribution of censored records known is
$\Delta$. For the weaker criterion
$\sup_F\mathbb E[R_F^*-R_F(r^\pi)]$, the corresponding minimax regret is
$\Delta_{\rm rand}$: even a deterministic function of the random record
posts a random reserve. Explicit randomization selects that reserve law
directly. Proposition~\ref{prop:randomization-bound} bounds the potential
benefit by a factor of two.

Mixing is allowed over $[0,\infty)$, the same domain as deterministic
reserves. In the nonidentifiable regime, tail domination permits all mass
above $\lambda$ to be moved to $\lambda$. This restriction would be
wrong in a tail-optimal instance with its common optimum above $\lambda$.
\end{remark}

\begin{proposition}[An exponential bound in the number of bidders]\label{prop:competition}
Let $G\in\mathcal A\subseteq\mathcal F(\lambda;G)$, and assume every
$F\in\mathcal A$ has an optimal reserve in $[0,\lambda]$. Then
\begin{equation}\label{eq:competition-bound}
  \Delta(\mathcal A)
  \le\sup_{F\in\mathcal A}\{R_F^*-R_F(\lambda)\}
  \le\lambda\alpha^n,
  \qquad
  \frac{\Delta(\mathcal A)}{R_G^*}
  \le\frac{\alpha^n}{1-\alpha^n}.
\end{equation}
This applies to both shape classes in the nonidentifiable regime.
\end{proposition}

\begin{proof}[Proof of Proposition~\ref{prop:competition}]
Let $s_F\le\lambda$ be an optimum. Whenever some value clears
$\lambda$, posting $\lambda$ weakly dominates $s_F$. Otherwise the gain
is at most $\lambda$. That event has probability $\alpha^n$, proving
$\Delta(\mathcal A)\le\lambda\alpha^n$.
Also $R_G^*\ge R_G(\lambda)\ge\lambda(1-\alpha^n)$, which gives the
relative bound.
\end{proof}

\begin{remark}[Sharpness and an MHR refinement]\label{rem:competition-sharpness}
For an exponential tail of hazard $h\ge1/\lambda$, $B_\lambda=0$.
At fixed $\alpha<1$ and sufficiently large $n$, $k_\alpha=p_\alpha$,
so the unrestricted formula gives
\[
 \Delta(\mathcal F(\lambda;G))=\lambda\alpha^n
           \frac{n(1-\alpha)}{\alpha+n(1-\alpha)}\sim\lambda\alpha^n.
\]
Thus the exponential order is sharp for unrestricted completions.
For MHR in the nonidentifiable regime, every completion has an optimum at or below
$u=\min\{\lambda,\lambda-(A-1)/h_0\}$.
The same event argument at $u$, using $H_F(u)\le Au/\lambda$, gives
\begin{equation}\label{eq:competition-mhr}
 \Delta\le\sup_F\{R_F^*-R_F(u)\}
       \le u(1-e^{-Au/\lambda})^n.
\end{equation}
This improves the bound when $A>1$. Neither bound asserts monotonicity
of the exact minimax regret in $n$ or a rate when $\alpha$ also tends to one.
\end{remark}

\subsection{MHR geometry and the equalizer}\label{app:proofs-dro}

We justify the two-level replacement, establish stability, and then prove
the unique equalizer.

\begin{proof}[Proof of Proposition~\ref{thm:two-scalar}]
Since $h_0>1/\lambda$, every completion's revenue is nonincreasing above
$\lambda$ (Section~\ref{sec:two-scalar-red}), so its optimum and every
alternative reserve can be taken in $[0,\lambda]$. By \eqref{eq:mhr-cdf-gap} and the characterization of feasible bodies
in Section~\ref{sec:two-scalar-red},
both the revenue comparison and the feasible set of bodies depend on the
tail only through $(A,h_0)$.
\end{proof}

\begin{figure}[htbp]
 \centering
 \includegraphics[width=0.90\textwidth]{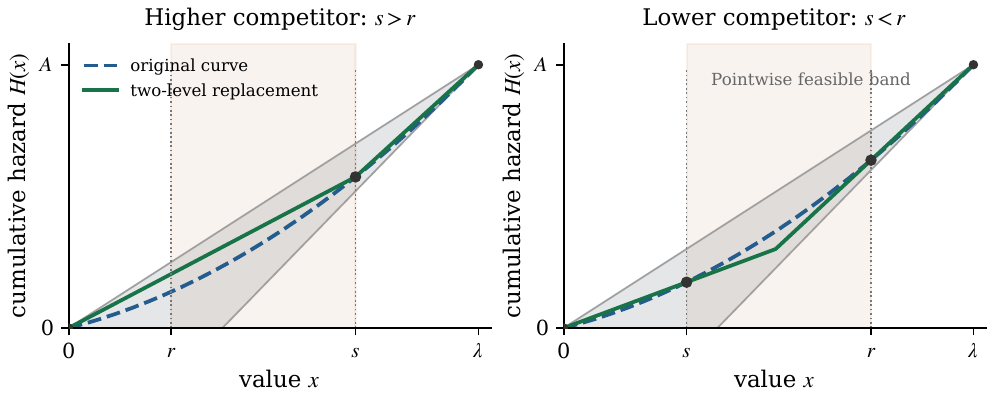}
 \caption{Replacing a convex cumulative hazard by a two-level body.
 A higher alternative reserve uses two chords to raise the curve between
 the reserves; a lower one uses the maximum of two extended chords to
 lower it. The preserved heights make each replacement improve nature's
 revenue comparison.}
 \label{fig:mhr-replacement}
\end{figure}

\begin{proof}[Proof of the two-level reduction (Theorem~\ref{thm:two-level})]
Fix the alternative reserve $s$ and a feasible body $H$.
If $s>r$, replace $H$ by the chords through $(0,0),(s,H(s))$ and
$(\lambda,A)$. The replacement is convex, has terminal slope at most
$h_0$, and raises $H$ on $[r,s]$ while preserving $H(s)$.
Thus it raises the CDF there, improving both the negative endpoint term
at $r$ and $-\int_r^s\psi(F)$ in \eqref{eq:mhr-cdf-gap}.

If $s<r$, take the maximum of the ray through $(0,0),(s,H(s))$ and
the line through $(r,H(r)),(\lambda,A)$. Convexity puts this maximum
below $H$ on $[s,r]$, preserving both endpoint heights. Its slopes are
ordered and bounded by $h_0$, and its break lies in $[s,r]$.
It lowers the CDF and increases $\int_s^r\psi(F)$.
At $s=0$ use the zero ray; at $r=\lambda$ use the backward line of
slope $h_0$. At $s=\lambda$ the first replacement is the whole-body chord.

Both replacements are feasible two-level bodies (Figure~\ref{fig:mhr-replacement}).
The reverse inequality
follows from inclusion. Continuity on the compact level rectangle
(Lemma~\ref{lem:body-functional}) gives attainment; the paragraph on
attainment in the two-level family below gives the details.
\end{proof}

For $t=H_F(r)$ and $\tilde t=H_F(\tilde r)$, the density form of the comparison is
\begin{equation}\label{eq:regret-decomp}
\begin{aligned}
 R_F(\tilde r)-R_F(r)
 ={}&n\tilde r(1-e^{-\tilde t})^{n-1}e^{-\tilde t}
       -nr(1-e^{-t})^{n-1}e^{-t}\\
 &+\int_{\tilde r}^r x\,n(n-1)F(x)^{n-2}(1-F(x))f(x)dx.
\end{aligned}
\end{equation}
Define
\begin{equation}\label{eq:beta-def}
 \beta(z)=n(n-1)(1-e^{-z})^{n-2}e^{-2z}.
\end{equation}
Its unknown integral becomes
\begin{equation}\label{eq:int-beta}
 I(F)=\int_{\tilde r}^r x\beta(H_F(x))h_F(x)dx.
\end{equation}
On each affine piece with positive slope, $z=H_F(x)$ makes $x$ affine in $z$;
Lemma~\ref{lem:beta-closed-form} evaluates the resulting integral.
A zero-slope piece contributes zero.

\begin{lemma}[Continuity of the body revenue difference]\label{lem:body-functional}
Let $n\ge2$, $\lambda>0$, and let $H,\bar H$ be nonnegative
cumulative-hazard bodies on $[0,\lambda]$. Define $D_{r,s}(H)$ by
\eqref{eq:mhr-cdf-gap}, with $F=1-e^{-H}$.
Then, for every $r,s\in[0,\lambda]$,
\begin{equation}\label{eq:two-level-lip}
 |D_{r,s}(H)-D_{r,s}(\bar H)|
 \le(r+s+2|r-s|)\|H-\bar H\|_\infty
 \le3\lambda\|H-\bar H\|_\infty.
\end{equation}
The bound does not require equal endpoint heights, and its constant is
independent of the bidder count.
\end{lemma}
\begin{proof}
In cumulative-hazard
coordinates put $p=1-e^{-z}$ and define
$\Psi_1(z)=1-p^n$, $\Psi_2(z)=\psi(p)$. Their derivatives satisfy
\[
 |\Psi_1'(z)|=np^{n-1}(1-p)\le1,\qquad
 |\Psi_2'(z)|=n(n-1)p^{n-2}(1-p)^2\le2.
\]
The two expressions are, respectively, a binomial probability and twice a
binomial probability. Applying these derivative bounds to the two endpoint terms and the
integral in \eqref{eq:mhr-cdf-gap} gives \eqref{eq:two-level-lip}.
\end{proof}

\paragraph{Attainment in the two-level family}
For the two-level parameterization,
\[
 \|H_{a,b}-H_{a',b'}\|_\infty
       \le\lambda\max\{|a-a'|,|b-b'|\}.
\]
Regret, including its maximization over $s$, is therefore continuous on the
compact level rectangle. For a fixed $s>r$, the family with break at $s$ is
closed: its levels satisfy $as=A-b(\lambda-s)$. For $s<r$, a break in $[s,r]$
is equivalently described by the two closed inequalities
$as\ge A-b(\lambda-s)$ and $ar\le A-b(\lambda-r)$. Constant bodies are
included in both descriptions. Thus the fixed-competitor suprema in
Theorem~\ref{thm:two-level} are attained.
Moreover, $(a,b,s)\mapsto D_{r,s}(H_{a,b})$ is jointly Lipschitz on the
compact set $[0,A/\lambda]\times[A/\lambda,h_0]\times[0,\lambda]$, with
constant $3\lambda^2$ in the levels by the bound above and
$\max\{1,\lambda h_0\}$ in $s$ by the revenue-derivative bound
$|R_F'|\le\max\{1,\lambda h_0\}$ for bodies with hazard at most $h_0$;
hence the supremum over bodies and alternative reserves in
Theorem~\ref{thm:two-level} is a maximum, and
$R_{F_{a,b}}^*-R_{F_{a,b}}(r)=\max_sD_{r,s}(H_{a,b})$ is continuous in
$(a,b)$.

\begin{proof}[Proof of global stability (Lemma~\ref{lem:stability})]
For any feasible body $H$, define
\begin{equation}\label{eq:Lambda-body}
 \Lambda_r(H)=\sup_{s\in[0,\lambda]}D_{r,s}(H),\qquad
 \mathsf W(r;A,h_0)=\sup_{H\in\mathcal H(A,h_0)}\Lambda_r(H),
\end{equation}
where $D_{r,s}$ is \eqref{eq:mhr-cdf-gap} with $F=1-e^{-H}$; the second
expression is the body objective \eqref{eq:W-body-def} of
Section~\ref{sec:dro}.
Lemma~\ref{lem:body-functional} makes $\Lambda_r$ $3\lambda$-Lipschitz
in $H$. We now match the feasible body sets.

Write $v=H'$ and $A_* =\min\{A,A'\}$. Scale $v$ to have integral
$A_*$; this moves its cumulative integral by at most $A-A_*$.
If $h_0'<h_0$, replace the scaled hazard by
\[
 v_\gamma(x)=\min\{h_0',\max\{\gamma,v(x)\}\},\qquad 0\le\gamma\le h_0'.
\]
Its integral ranges continuously from at most $A_*$ to
$\lambda h_0'\ge A'\ge A_*$. Choose $\gamma$ to preserve $A_*$.
The mass removed at the end equals that added at the beginning and is
at most $\lambda(h_0-h_0')$. Thus the cumulative integral moves by at
most this amount. If $h_0'\ge h_0$, no cap adjustment is needed.
Finally, interpolate with the constant hazard $h_0'$ until the integral
is $A'$. This adds mass $A'-A_*$ while preserving monotonicity and the cap.
The degenerate cases need no separate argument: if $A_*=0$ the scaled
hazard is zero and the floor $\gamma=0$ keeps it so; if $A'=\lambda h_0'$
the interpolation ends at the constant hazard $h_0'$, the only element of
$\mathcal H(A',h_0')$; if $A=\lambda h_0$ the source hazard is the
constant $h_0$. In each case every step above is defined as written.
The resulting $\bar H\in\mathcal H(A',h_0')$ satisfies
\[
 \|H-\bar H\|_\infty\le |A-A'|+\lambda|h_0-h_0'|.
\]
Repeat in the reverse direction and apply the functional bound to obtain
\eqref{eq:global-stability}.

The revenue derivative on the body also gives
\[
 |R_F'(r)|=nF(r)^{n-1}(1-F(r))|1-rh_F(r)|
       \le\max\{1,\lambda h_0\}=\mathrm{Lip}_r.
\]
This bound passes to the comparison suprema and proves reserve stability.
\end{proof}

\begin{proof}[Proof of the unique equalizer (Theorem~\ref{thm:mhr-equalizer})]
For a completion with optimum $r_F^*$, its directional losses are
\[
 \ell_F(r)=R_F(\min\{r,r_F^*\})-R_F(r),\qquad
 u_F(r)=R_F(\max\{r,r_F^*\})-R_F(r).
\]
Unimodality makes $\ell_F$ nondecreasing and $u_F$ nonincreasing.
Both are $\mathrm{Lip}_r$-Lipschitz, so their suprema $L,U$ are continuous
and have the same directions. The compact two-level family attains these
suprema, including their alternative reserves.

If $L(r_1)>0$, an attaining completion has already lost revenue before
$r_1$. Hence $F(r_1)>0$ and $r_1h_F(r_1^+)>1$.
Hazard monotonicity preserves both inequalities to the right, so its
revenue strictly decreases there. Retaining this completion proves
$L(r_2)>L(r_1)$ for $r_2>r_1$.

If $U(r_2)>0$, choose an attaining higher competitor $s>r_2$.
The two-level replacement has constant initial hazard $a$ up to $s$.
Positive gain requires $a>0$ and $ar_2<1$, so its revenue strictly
increases on $(0,r_2]$. Retaining it gives $U(r_1)>U(r_2)$ for $r_1<r_2$.

At $l=1/h_0$, every completion's revenue has been nondecreasing, so
$L(l)=0$. At $u=\min\{\lambda,\lambda-(A-1)/h_0\}$, convexity and
the backward tangent give $xh_F(x)\ge H_F(x)\ge1$ for $x\ge u$,
whenever $u<\lambda$. Thus $U(u)=0$; this is immediate if $u=\lambda$.
Since $\Delta>0$, the opposite directional losses at these endpoints are
positive. Continuity gives a positive crossing in $(l,u)$, and strictness
makes it unique. To its left $W=U$ strictly decreases; to its right
$W=L$ strictly increases.
\end{proof}

\begin{proof}[Proof of Proposition~\ref{prop:rdagger}]
The endpoints $l$ and $u$ above satisfy $L(l)=0<U(l)$ and $U(u)=0<L(u)$, so the
unique crossing lies strictly between them, which is \eqref{eq:rdagger-interval}.
Subtracting $1/h_0$ from the upper endpoint bounds the length by $\lambda-A/h_0$.
At fixed $A$, increasing $h_0$ enlarges $\mathcal H(A,h_0)$ and leaves the regret
functional unchanged, so $\mathsf W$ and its minimum do not decrease.
\end{proof}

\begin{lemma}[A sharp minimum]\label{lem:kink}
Let $n\ge2$, $\lambda>0$, $0<A<\lambda h_0$, and $h_0>1/\lambda$.
Write $r^\dagger=r^\dagger(A,h_0)$ for the minimax reserve.
There are $c_L,c_R>0$ such that, for every $r\in[0,\lambda]$,
\[
 \mathsf W(r;A,h_0)-\Delta(A,h_0)\ge
 \begin{cases}
 c_L(r^\dagger-r),&r\le r^\dagger,\\
 c_R(r-r^\dagger),&r\ge r^\dagger.
 \end{cases}
\]
These constants may deteriorate near the zero-regret frontier.
\end{lemma}
\begin{proof}
For the left side, retain an attaining higher competitor and its body
with initial hazard $a>0$, $ar^\dagger<1$. Its revenue derivative on
$[r^\dagger/2,r^\dagger]$ is at least
\[
 c_{\rm der}=n(1-e^{-ar^\dagger/2})^{n-1}
                    e^{-ar^\dagger}(1-ar^\dagger)>0.
\]
For $r$ in this half interval, integrate the bound. For smaller $r$,
monotonicity gives at least the loss over the upper half interval.
Thus $c_L=c_{\rm der}/2$ works on all $[0,r^\dagger]$.

For the right side, retain an attaining lower competitor.
Its positive loss at $r^\dagger$ implies $F(r^\dagger)>0$ and
$r^\dagger h_\dagger>1$, where $h_\dagger=h_F(r^\dagger{}^+)$.
For almost every $x\in[r^\dagger,\lambda]$,
\[
 -R_F'(x)\ge nF(r^\dagger)^{n-1}e^{-A}
                    (r^\dagger h_\dagger-1)=:c_R>0.
\]
Integration gives the claim.
\end{proof}

If two parameter pairs $\theta,\theta'$ have objectives within $\eta$ uniformly, optimality
of the perturbed minimizer gives
$\mathsf W(r^\dagger(\theta');\theta)-\Delta(\theta)\le2\eta$.
Thus, with $c=\min\{c_L,c_R\}$,
\[
 |r^\dagger(\theta')-r^\dagger(\theta)|\le2\eta/c.
\]
Together with global stability, this controls the reserve at each instance
in $\mathcal D$; it does not give a uniform positive $c$ up to the frontier.

\subsection{MHR numerical certificates}\label{app:mhr-computation}

\begin{lemma}[Closed-form integration]\label{lem:beta-closed-form}
Let $n\ge2$ be an integer and let $\mu,\nu,z_1,z_2\in\mathbb R$.
For $\beta$ in \eqref{eq:beta-def},
\begin{equation}\label{eq:beta-closed-form}
\begin{aligned}
 \int_{z_1}^{z_2}(\mu+\nu z)\beta(z)dz
 ={}&n(n-1)\sum_{k=0}^{n-2}\binom{n-2}{k}(-1)^k\\
 &\quad\cdot\left[-e^{-(k+2)z}
       \left(\frac{\mu+\nu z}{k+2}+\frac\nu{(k+2)^2}\right)\right]_{z_1}^{z_2}.
\end{aligned}
\end{equation}
\end{lemma}
\begin{proof}
Expand $(1-e^{-z})^{n-2}$ and integrate each exponential term.
On a positive-slope affine piece of $H$, the substitution $z=H(x)$
makes $x=\mu+\nu z$. A zero-slope piece contributes zero.
The extremal comparisons therefore require at most two such integrals.
\end{proof}

\subsubsection{The level-grid computation}\label{app:core-computation}

For $(A,h_0)$ with $0<A<\lambda h_0$ and $h_0>1/\lambda$,
grid the rectangle
$[0,A/\lambda]\times[A/\lambda,h_0]$, including endpoints.
Write $\mathsf W^{\mathcal T}(r)$ for the largest regret among these
bodies, using their exact body optima.

\begin{corollary}[Grid error]\label{cor:two-level-oracle}
Under Assumption~\ref{ass:mhr}, let $0<A<\lambda h_0$ and
$h_0>1/\lambda$. Use the endpoint-inclusive level grids defined above,
with maximum gaps $\varepsilon_a,\varepsilon_b$. Put
$\eta_{\mathcal T}=3\lambda^2\max\{\varepsilon_a,\varepsilon_b\}$ and
$\mathrm{Lip}_r=\max\{1,\lambda h_0\}$. Then
$0\le W_G(r)-\mathsf W^{\mathcal T}(r)\le\eta_{\mathcal T}$ for every
$r\in[0,\lambda]$.
For the minimum $\Delta^{\mathcal T}$ on an endpoint-inclusive posted
grid on $[0,\lambda]$ with maximum gap $\varepsilon_r$,
\begin{equation}\label{eq:two-level-bracket}
 \Delta^{\mathcal T}-\mathrm{Lip}_r\varepsilon_r
      \le\Delta\le\Delta^{\mathcal T}+\eta_{\mathcal T}.
\end{equation}
Maximizing each body's revenue on an endpoint-inclusive grid on
$[0,\lambda]$ with maximum gap $\varepsilon_x$
adds $\mathrm{Lip}_r\varepsilon_x$ to the upper error bounds.
By Proposition~\ref{thm:two-scalar}, $W_G=\mathsf W(\cdot;A,h_0)$ and
$\Delta=\Delta(A,h_0)$, so these bounds hold for every pair in $\mathcal D$,
in particular for the projected pair of Algorithm~\ref{alg:plug-in}.
\end{corollary}
\begin{proof}
Every grid body is feasible. Rounding the attaining levels changes
$H_{a,b}$ by at most $\lambda\max\{\varepsilon_a,\varepsilon_b\}$;
Lemma~\ref{lem:body-functional} gives the reverse bound with error
$\eta_{\mathcal T}$. A posted grid point near the true minimizer gives
$\Delta^{\mathcal T}\le\Delta+\mathrm{Lip}_r\varepsilon_r$.
At the minimizing grid point, $\Delta\le W_G\le\Delta^{\mathcal T}+\eta_{\mathcal T}$.
The price-grid error follows from the same reserve Lipschitz bound.
\end{proof}

For a two-level body with break $\tau=(b\lambda-A)/(b-a)$,
its hazard-weighted reserve crosses or jumps past one at the body optimum
\begin{equation}\label{eq:two-level-optimum}
 r_{a,b}^{\rm opt}
 =\min\left\{\lambda,
       \max\left\{\frac1b,\min\left\{\tau,\frac1a\right\}\right\}\right\}.
\end{equation}
Use $1/0=\infty$; for $a=b=A/\lambda$, take
$r_{a,b}^{\rm opt}=\min\{\lambda,1/a\}$ and $\tau=\lambda$.
The derivative formula proves this directly on the two segments.
If $a=0$ and $b\tau\ge1$, all reserves in $[0,\tau]$ are optimal,
and the formula chooses $\tau$.

\begin{corollary}[Closed-form worst-case regret]\label{cor:two-level-closed-form}
Let $n\ge2$, $\lambda>0$, $0\le A\le\lambda h_0$, and
$0\le a\le A/\lambda\le b\le h_0$. For $r\in[0,\lambda]$,
let $F_{a,b}$ be the two-level body joined to a common observed tail,
and put $z_r=H_{a,b}(r)$,
$z_\tau=a\tau$, and
$\mathsf B_{\mu,\nu}(z_1,z_2)=\int_{z_1}^{z_2}(\mu+\nu z)\beta(z)dz$,
evaluated by \eqref{eq:beta-closed-form}.
Let $T_\lambda$ denote the common second-highest-value contribution
above $\lambda$ in \eqref{eq:revenue}. Then
\begin{equation}\label{eq:two-level-closed-form}
\begin{aligned}
 R_{F_{a,b}}(r)-T_\lambda
 ={}&nr(1-e^{-z_r})^{n-1}e^{-z_r}\\
 &+\mathbf1\{a>0\}\mathsf B_{0,1/a}(z_r,\max\{z_r,z_\tau\})\\
 &+\mathsf B_{\lambda-A/b,1/b}(\max\{z_r,z_\tau\},A).
\end{aligned}
\end{equation}
The first integral is omitted when $a=0$. Under
Assumption~\ref{ass:mhr}, with $(A,h_0)\in\mathcal D$, differences
of this formula give
\[
 \mathsf W(r;A,h_0)=
 \max_{0\le a\le A/\lambda\le b\le h_0}
   \{R_{F_{a,b}}(r_{a,b}^{\rm opt})-R_{F_{a,b}}(r)\}.
\]
At $A=0$ the body is zero and all its comparison losses are zero.
Under Assumption~\ref{ass:mhr}, for every $r\ge\lambda$,
$W_G(r)=W_G(\lambda)+R_G(\lambda)-R_G(r)$.
\end{corollary}
\begin{proof}
In the body integral substitute $z=H(x)$. Its two inverse pieces are
$x=z/a$ and $x=\lambda-(A-z)/b$; a zero-slope segment contributes zero.
This proves the formula. The two-level reduction gives the maximum,
and common tail revenues give the last identity.
\end{proof}

\begin{corollary}[Reserve bisection]\label{cor:mhr-bisection}
Under the hypotheses and grid conventions of
Corollary~\ref{cor:two-level-oracle}, restrict the directional losses to
the level grid, obtaining $L_{\mathcal T},U_{\mathcal T}$.
Bisection on $L_{\mathcal T}-U_{\mathcal T}$ over $[0,\lambda]$ returns
an interval containing a minimizer of $\mathsf W^{\mathcal T}$.
If its length is at most $\varepsilon_r>0$, then any point
$\widehat r$ in it, with $v=\mathsf W^{\mathcal T}(\widehat r)$, satisfies
\[
 v-\mathrm{Lip}_r\varepsilon_r\le\Delta\le v+\eta_{\mathcal T},\qquad
 W_G(\widehat r)\le\Delta+\eta_{\mathcal T}+\mathrm{Lip}_r\varepsilon_r.
\]
If the grids have $M_a$ and $M_b$ hazard levels, the cost is
$O(M_aM_b[1+\log_+(\lambda/\varepsilon_r)])$ evaluations.
\end{corollary}
\begin{proof}
The finite directional losses are continuous and monotone in opposite
directions. Their difference is nonpositive at zero and nonnegative at
$\lambda$; every zero minimizes their maximum.
If that minimum is $d_{\mathcal T}$, then
$d_{\mathcal T}\le\Delta\le d_{\mathcal T}+\eta_{\mathcal T}$ and
$d_{\mathcal T}\le v\le d_{\mathcal T}+\mathrm{Lip}_r\varepsilon_r$.
Combining these bounds proves the certificate.
Each bisection step evaluates all bodies, and halves the interval.
\end{proof}

\section{Learning from Censored Samples}\label{app:learning}

We organize the proofs around the three settings in Section~\ref{sec:finite-data}:
a tail-optimal case, a nonidentifiable regime, and a fully identified case
with an optimum at or below
$\lambda$. The corresponding expected excess-regret bounds are
$O(N^{-1/2})$, $O(N^{-2/5})$, and $O(N^{-1})$, respectively. We then prove the
auto-selector guarantee for selecting between the tail and plug-in rules.

\subsection{Learning in the observed tail}\label{app:ident-policies}

When $h_0\le1/\lambda$, all compatible MHR distributions share an
optimal reserve in the observed tail. We learn it by maximizing empirical
revenue. The following identity includes bids equal to the reserve and
controls the contribution of an unbounded tail.

\begin{lemma}[Empirical revenue and uniform truncation]\label{lem:empirical-revenue}
Fix $n\ge2$. For any CDF $F$ on $[0,\infty)$ with finite first moment, auction revenue satisfies
\begin{equation}\label{eq:atom-safe-revenue}
  R_F(r)\;=\;r\bigl(1-F(r^-)^n\bigr)
  \;+\;\int_r^{\infty}\Bigl(1-nF(u)^{n-1}+(n-1)F(u)^n\Bigr)\,du .
\end{equation}
Let $G$ and $\widehat G_N$ be two such CDFs, continuous at a boundary
$\lambda>0$, and write $\widehat R_N=R_{\widehat G_N}$.
Then, for every $T\ge\lambda$, the following bound holds with $c_n=n^2$:
\begin{equation}\label{eq:uniform-tail-split}
  \sup_{r\ge\lambda}\bigl|\widehat R_N(r)-R_G(r)\bigr|
  \;\le\;
  c_n(1+T)\,\bigl\|\widehat G_N-G\bigr\|_{\infty,[\lambda,\infty)}
  \;+\;
  c_n\Bigl(\mathbb E_{\widehat G_N}\bigl[V\mathbf 1\{V>T\}\bigr]+\mathbb E_{G}\bigl[V\mathbf 1\{V>T\}\bigr]\Bigr).
\end{equation}
The same bound holds with $\lambda=0$, without a continuity assumption
at zero. Thus it also controls revenues uniformly over all reserves.
\end{lemma}

\begin{proof}
The payment identity
\[
 \mathrm{rev}(r,\mathbf V)
 =r\mathbf1\{V_{(1)}\ge r\}+(V_{(2)}-r)_+
\]
proves \eqref{eq:atom-safe-revenue}. The left limit includes bids equal
to the reserve. Put $\psi(p)=1-np^{n-1}+(n-1)p^n$ and
$\delta_G=\|\widehat G_N-G\|_{\infty,[\lambda,\infty)}$.
The bounds $|(p^n)'|\le n$, $|\psi'|\le n(n-1)$, and
$0\le\psi(p)\le n(n-1)(1-p)$ compare the two formulas on
$[r,T]$ and bound the remaining integrals by their tail first moments.
For $r\le T$ this gives
\[
 |\widehat R_N(r)-R_G(r)|\le n^2T\delta_G+
 n^2\sum_{F\in\{\widehat G_N,G\}}\mathbb E_F[V\mathbf1\{V>T\}].
\]
For $r>T$, each revenue is at most
$n\mathbb E_F[V\mathbf1\{V>T\}]$. Continuity at $\lambda$ extends
$\delta_G$ to the left limits there; at zero both left limits are zero.
This proves \eqref{eq:uniform-tail-split}.
\end{proof}

Suppose a lower bound $h_{\min}>0$ on $h_0$ is known. Set
\[
 \bar\lambda=\max\{\lambda,1/h_{\min}\}.
\]
For $x\ge\lambda$, with $\mathcal U_N$ the multiset of uncensored
bids, the observable empirical left CDF is
\[
 \widehat G_N^-(x)=\frac{N-m+\sum_{U\in\mathcal U_N}\mathbf1\{U<x\}}N.
\]
With $\psi(p)=1-np^{n-1}+(n-1)p^n$, define
\begin{equation}\label{eq:tail-empirical-gap}
 \widehat Q_N(r)=r\bigl(1-\widehat G_N^-(r)^n\bigr)
 -\lambda\bigl(1-\widehat G_N^-(\lambda)^n\bigr)
 -\int_\lambda^r\psi(\widehat G_N^-(x))\,dx,
 \qquad \lambda\le r\le\bar\lambda.
\end{equation}
Let $\hat r_N^{\rm tail}$ be its smallest maximizer over
$\{\lambda,\bar\lambda\}\cup(\mathcal U_N\cap[\lambda,\bar\lambda])$.
Using the left CDF treats a bid equal to the reserve as qualifying.
Between bids the objective is affine with nonnegative slope; at a bid it takes the upper
value before a downward jump. Thus this finite search attains the
supremum on the interval.

Subtracting revenue at $\lambda$ cancels payments from values beyond the
search interval; only the CDF on it needs to be estimated.

\begin{proof}[Proof of Proposition~\ref{prop:tail-rate}]
MHR and $h_0\ge h_{\min}$ make revenue nonincreasing beyond
$\bar\lambda$, so the tail optimum is in $[\lambda,\bar\lambda]$.
The population analogue $Q$ of \eqref{eq:tail-empirical-gap} equals
$R_G(r)-R_G(\lambda)$. Since $|(p^n)'|\le n$ and
$|\psi'(p)|=n(n-1)p^{n-2}(1-p)\le n$,
\[
 \sup_{[\lambda,\bar\lambda]}|\widehat Q_N-Q|
 \le2n\bar\lambda\sup_{x\ge\lambda}|\widehat G_N^-(x)-G(x)|.
\]
The Dvoretzky--Kiefer--Wolfowitz (DKW) inequality applies to the left empirical
CDF by taking left limits in its usual
bound. Empirical maximization costs twice this error, proving the result.
For $h_0\le1/\lambda$, all body revenues are nondecreasing, so every
completion shares the tail optimum. Integrating the DKW tail bound gives
$\mathbb E\|\widehat G_N^--G\|_{\infty,[\lambda,\infty)}=O(N^{-1/2})$,
and hence the stated expected-regret bound.
\end{proof}

\begin{proof}[Tail-optimum part of Proposition~\ref{prop:identifiable-iff}]
The unrestricted frontier and the tail-optimal cases of the two shape
classes give
\begin{equation}\label{eq:tail-common}
 R_F^*=\max_{r\ge\lambda}R_G(r)=:R^\star
 \qquad(F\in\mathcal A).
\end{equation}
For the shape classes this is the monotonicity of every completion's
revenue on $[0,\lambda]$ in the tail-optimal case
(Theorems~\ref{thm:frontier} and~\ref{thm:frontier-reg}); for the
unrestricted class it is the frontier condition $B_\lambda\ge\lambda\alpha^n$
of Theorem~\ref{thm:unrestricted} together with
\eqref{eq:body-revenue-envelope}, which bounds every completion's body
revenue by $R_G(\lambda)+\lambda\alpha^n\le R^\star$.
Assign every censored record to zero and form its empirical CDF
$\widehat G_N$. This recovers the empirical tail without completing the
hidden values for revenue comparisons above $\lambda$.
Post the smallest empirical-revenue maximizer $\hat r_N$ over $\{\lambda\}\cup\mathcal U_N$,
where $\mathcal U_N$ is the multiset of uncensored bids.
Between bids empirical revenue is affine and nondecreasing; above the
largest bid it is zero. The finite search thus attains the tail supremum,
and returns $\lambda$ when no bids are observed.

Glivenko--Cantelli gives uniform tail CDF convergence. For fixed
$T\ge\lambda$, the empirical tail first moment beyond $T$ converges by
the law of large numbers. First choosing $T$ large and then applying
Lemma~\ref{lem:empirical-revenue} gives
$\sup_{r\ge\lambda}|R_{\widehat G_N}(r)-R_G(r)|=o_P(1)$.
The maximizer inequality then yields
\[
 \sup_{F\in\mathcal A}\{R_F^*-R_F(\hat r_N)\}
 =R^\star-R_G(\hat r_N)
 \le2\sup_{r\ge\lambda}|R_{\widehat G_N}(r)-R_G(r)|=o_P(1).
\]
This needs no uniqueness of the tail optimum.
\end{proof}

\begin{remark}[The uncensored benchmark]\label{rem:baseline-literature}
Learning a common reserve for all bidders differs from learning the full
revenue-optimal auction mechanism. Jin et al.~\citep{jin2022learning} obtain
$\tilde\Theta(\varepsilon^{-2})$ sample complexity for bounded-support
and MHR values, without the bidder-count factor in the general-mechanism
bounds of Guo et al.~\citep{guo2019settling}.
These results use multiplicative revenue approximation and samples of
all bidders' values; here we use additive loss and individual i.i.d.\ bids.
For continuous MHR values, the best-known lower bound is only
$\tilde\Omega(\varepsilon^{-3/2})$; the matching
$\varepsilon^{-2}$ lower bound permits discrete distributions.
\end{remark}

\subsection{Learning in the nonidentifiable regime}\label{app:learning-proofs}

The projected plug-in rule estimates $A,h_0$ and solves the corresponding
minimax problem. We first bound the two estimation errors, then transfer
them to regret using projection and global stability. The matching lower
bound shows why estimating the boundary density determines the rate.

\subsubsection{Estimator and grid conventions}\label{app:learning-details}

For $m\ge1$, set $b_N=c_bN^{-1/5}$, $w=b_N/M$, $z_j=(j-1/2)/M$,
$K_j=\tfrac34(1-z_j^2)$, and $S_k=\sum_{j=1}^M K_jz_j^k$.
The histogram and local-linear weights are
\[
 \hat g_j=\frac{\#\{U\in\mathcal U_N:
       \lambda+(j-1)w\le U<\lambda+jw\}}{mw},\qquad
 \omega_j=\frac{K_j(S_2-S_1z_j)}{S_0S_2-S_1^2},\qquad
 \hat h_0=\sum_j\omega_j\hat g_j.
\]
The denominator is positive because the $M\ge2$ design points are
distinct and have positive weights. The weights depend only on $M$ and
satisfy $\sum_j\omega_j=1$, $\sum_j\omega_jz_j=0$.
Algorithm~\ref{alg:plug-in} divides each grid interval of length $\ell_{\mathrm g}$
into $\max\{1,\lceil\ell_{\mathrm g}/\varepsilon_N\rceil\}$ equal subintervals,
including endpoints; a zero-length interval is a single point.
Euclidean projection and the smallest-index tie-breaking make the plug-in
rule measurable. The bandwidth constant $c_b$ is fixed before sampling.

\begin{lemma}[Boundary-parameter concentration]\label{lem:est-rates}
Under Assumption~\ref{ass:boundary-smoothness}, fix $c_b>0$, $M\ge2$,
and $m_{\min}\ge1$. For every $\delta\in(0,1)$, there are finite constants
$\kappa_A,\kappa_h$ and $\underline N(\delta)$ such that, for
$N\ge \underline N(\delta)$, with probability at least $1-\delta$,
\begin{equation}\label{eq:est-conc}
 \begin{gathered}
 |\hat A_N-A|\le\kappa_A\sqrt{\log(6/\delta)/N},\qquad
 |\hat h_0-h_0|\le\kappa_h\sqrt{\log(6/\delta)}\,N^{-2/5},\\
 m\ge Nq_\lambda/2\ge m_{\min}.
 \end{gathered}
\end{equation}
The constants depend on $q_\lambda$, the bounds on $\tilde g$ and
$|\tilde g''|$ over $[\lambda,\lambda+\bar b]$, and the fixed choices
$c_b,M,m_{\min},\bar b$. Their explicit choices appear in the proof below.
\end{lemma}

\begin{proof}[Proof of Lemma~\ref{lem:est-rates}]
Write $q=q_\lambda$, $L_\delta=\log(6/\delta)$,
$\bar g=\sup_{[\lambda,\lambda+\bar b]}\tilde g$, and
$\bar g_2=\sup_{[\lambda,\lambda+\bar b]}|\tilde g''|$.
For $\hat\alpha_N=1-m/N$, Hoeffding gives
$|\hat\alpha_N-\alpha|\le\sqrt{L_\delta/(2N)}$ with failure probability
$\delta/3$. If $N\ge2L_\delta/q^2$, then $m/N\ge q/2$ on this event, and
the logarithm is $2/q$-Lipschitz between $q$ and $m/N$. Thus
$|\hat A_N-A|\le\kappa_A\sqrt{L_\delta/N}$ with $\kappa_A=\sqrt2/q$.
Also, for $\mathcal B_N=\{m\ge Nq/2\}$,
$\mathbb P(\mathcal B_N^c)\le e^{-Nq/8}$ by Chernoff.

Conditionally on $m$, the observed bids are independent with density
$\tilde g$. Put $\bar\omega=\max_j|\omega_j|$ and
\[
 \mu_2=\sum_j|\omega_j|z_j^2+\frac1{12M^2}\sum_j|\omega_j|.
\]
Cell averaging differs from the density at a midpoint by at most
$\bar g_2w^2/24$. Taylor expansion at $\lambda$ and
$\sum\omega_j=1$, $\sum\omega_jz_j=0$ therefore give
\[
 |\mathbb E[\hat h_0\mid m]-h_0|\le\tfrac12\mu_2\bar g_2b_N^2.
\]
For $\varpi(u)=\sum_j\omega_j\mathbf1\{u\text{ is in cell }j\}$,
$\hat h_0=(mw)^{-1}\sum_{i=1}^m\varpi(U_i)$,
$|\varpi-\mathbb E\varpi|\le2\bar\omega$, and
$\operatorname{Var}(\varpi)\le\bar\omega^2\bar g b_N$.
Bernstein consequently gives conditional failure probability at most
$\delta/3$ for
\[
 |\hat h_0-\mathbb E[\hat h_0\mid m]|
 \le\bar\omega M\sqrt{\frac{2\bar g L_\delta}{mb_N}}
       +\frac{4\bar\omega ML_\delta}{3mb_N}.
\]
On $\mathcal B_N$, $mb_N\ge q c_bN^{4/5}/2$.
The second term is at most $\sqrt{L_\delta}\,N^{-2/5}$ when
$N\ge(8\bar\omega M\sqrt{L_\delta}/(3qc_b))^{5/2}$.
A valid choice is therefore
\[
 \kappa_h=\tfrac12\mu_2\bar g_2c_b^2
       +\bar\omega M\sqrt{4\bar g/(qc_b)}+1,
\]
and, including the requirements $b_N\le\bar b$ and $Nq/2\ge m_{\min}$,
\begin{equation}\label{eq:N0-explicit}
 \underline N(\delta)=\max\left\{
 \frac{2\log(6/\delta)}{q^2},\frac{8\log(3/\delta)}q,
 \frac{2m_{\min}}q,\left(\frac{c_b}{\bar b}\right)^5,
 \left(\frac{8\bar\omega M\sqrt{\log(6/\delta)}}{3qc_b}\right)^{5/2}
 \right\}.
\end{equation}
A union bound proves \eqref{eq:est-conc}.

For the expectation and bounds near full identification, the same argument gives
\begin{equation}\label{eq:learning-moments}
 \begin{gathered}
 \mathbb E[|\hat A_N-A|\mathbf1_{\mathcal B_N}]
 \le2\sqrt{\alpha/(qN)},\qquad
 \mathbb E[(\hat A_N-A)^2\mathbf1_{\mathcal B_N}]
 \le4\alpha/(qN),\\
 \mathbb E[|\hat h_0-h_0|\mathbf1_{\mathcal B_N}]=O(N^{-2/5}).
 \end{gathered}
\end{equation}
The first two follow from the logarithm bound and
$\operatorname{Var}(\hat\alpha_N)=\alpha q/N$; the last follows from
the conditional bias and variance by Cauchy--Schwarz. The same moment
orders hold on any fixed compact rectangle of finite raw estimates:
outside $\mathcal B_N$ their errors are bounded and its probability is
exponentially small. The raw $\hat A_N$ has no unconditional finite
moment, since it is infinite when $m=0$.
\end{proof}

In Theorem~\ref{thm:finite-sample}, take
\[
 C_\star=6\sqrt2\lambda(\kappa_A+\lambda\kappa_h),\qquad
 C_{\rm num}=3\lambda^2+\max\{1,\lambda h_0\}.
\]
The statistical constant $C_\star$ bounds the effect of estimation
error after projection. The numerical constant $C_{\rm num}$ bounds
the errors from the hazard and reserve grids: the mesh $\varepsilon_N$
in Algorithm~\ref{alg:plug-in} is applied to each grid in its own units,
units of value for the reserve grid and of hazard for the two-level grids, and by Corollary~\ref{cor:two-level-oracle} a level mesh of
$\varepsilon_N$ moves the worst case by at most $3\lambda^2\varepsilon_N$,
the first term of $C_{\rm num}$.
The auxiliary Lipschitz constants used for numerical certificates are
\[
 \mathrm{Lip}_r=\max\{1,\lambda h_0\},\qquad
 \mathrm{Lip}_\Theta=3\lambda\max\{1,\lambda\},
\]
where the latter uses the unscaled parameter $\ell^1$ norm. Both are
independent of $n$.

\begin{corollary}[Expectation and sample complexity]
\label{cor:expected-regret}\label{cor:sample-complexity}
Under the conditions of Theorem~\ref{thm:finite-sample}, expected regret is
$\Delta(G)+O(N^{-2/5})$. For $\varepsilon>0$ and $\delta\in(0,1)$,
excess regret at most $\varepsilon$ with
probability $1-\delta$ is guaranteed when
\[
 N\ge\max\left\{\underline N(\delta),
 \left(\frac{2C_\star}{\varepsilon}\right)^{5/2}
       \bigl(\log(6/\delta)\bigr)^{5/4},
 \left(\frac{2c_{\mathrm g}C_{\rm num}}{\varepsilon}\right)^2\right\}.
\]
\end{corollary}
The sample bound follows by making each term of
\eqref{eq:finite-sample-hp} at most $\varepsilon/2$.
The expectation proof below uses moments, so it adds no logarithmic factor.

\paragraph{Data-driven bandwidth}
A wider window reduces sampling noise but can increase bias.
The rule compares the estimates from the fixed grid
$\{\lambda/4,\lambda/2,\lambda,2\lambda\}$ and widens the window
until its estimate no longer agrees with those from narrower windows.
The last accepted window is chosen. This is Lepski's rule~\citep{lepski1997optimal};
Appendix~\ref{app:exp-truths} reports its effect.
For each grid point $c$, compute $\hat h_0(c)$ and
the plug-in standard error
$s(c)=(\hat p\sum_j\omega_j^2/(mw_c))^{1/2}$, where $w_c=cN^{-1/5}/M$ and
$\hat p$ is the share of uncensored bids within $2\lambda N^{-1/5}$ of
$\lambda$ divided by that width. Starting with the narrowest grid point,
accept successive candidates while their estimates lie within $2s(c')$
of the estimate at every narrower point $c'$. Stop at the first failure
and return the last accepted point. Let $\mathcal C_b$ denote this four-point grid and
$\hat c$ the selected point. Write $\kappa_h(c)$ and $\underline N_c$ for the
constants in Lemma~\ref{lem:est-rates} at bandwidth constant $c$.
Applying that lemma at level $\delta/4$ to each candidate gives a
simultaneous event of probability at least $1-\delta$. Thus the selected
rule satisfies the high-probability bound of Theorem~\ref{thm:finite-sample}
with $\log(6/\delta)$ replaced by $\log(24/\delta)$, using
\[
 \kappa_h^{\rm sel}=\max_{c\in\mathcal C_b}\kappa_h(c),\qquad
 \underline N^{\rm sel}(\delta)=\max_{c\in\mathcal C_b}\underline N_c(\delta/4).
\]
The expectation bound in Corollary~\ref{cor:expected-regret} also holds:
the selected estimation error is bounded by the maximum of the four
candidate errors, whose moments give $O(N^{-2/5})$ without a logarithmic
factor, as shown below. The rate guarantee uses only simultaneous concentration over the four
fixed candidates. It therefore holds for any measurable selection from
this grid, even though the plug-in standard errors need not be calibrated
confidence bounds. Together with the moment computation at the end of the
proof of Theorem~\ref{thm:finite-sample} below, this paragraph proves the
following statement.
\begin{corollary}[Data-driven bandwidth]\label{cor:selected-bandwidth}
Let the bandwidth constant in Algorithm~\ref{alg:plug-in} be chosen from
the fixed grid $\mathcal C_b=\{\lambda/4,\lambda/2,\lambda,2\lambda\}$ by
the rule above, or by any measurable selection from this grid. Under the
conditions of Theorem~\ref{thm:finite-sample}, for every $\delta\in(0,1)$
and $N\ge \underline N^{\rm sel}(\delta)$, the bound \eqref{eq:finite-sample-hp}
holds with probability at least $1-\delta$ with $\log(6/\delta)$ replaced
by $\log(24/\delta)$, where $C_\star$ uses $\kappa_h^{\rm sel}$. The
expectation bound $\mathbb E[W_G(\hat r_N^\dagger)]\le\Delta(G)+O(N^{-2/5})$
is unchanged.
\end{corollary}

\begin{lemma}[Bounded regret]\label{lem:bounded-regret}
For every $F\in\mathcal F(\lambda;G)$ and $r\ge0$,
\[
 R_F^*-R_F(r)\le\overline W=
 n\bigl(\lambda+\mathbb E_G[(V-\lambda)_+]\bigr)<\infty.
\]
Under Assumption~\ref{ass:mhr}, if $h_0\ge1/\lambda$, then $W_G(r)\le\lambda$ for
$r\in[0,\lambda]$.
\end{lemma}
\begin{proof}
Revenue is at most the largest value, bounded in expectation by
$n\mathbb E_F[V]\le\overline W$. For the sharper bound, optima lie in
$[0,\lambda]$. Payments at any two reserves there agree if the second
highest value clears both; otherwise both payments are in $[0,\lambda]$.
\end{proof}

\begin{proof}[Proof of Theorem~\ref{thm:finite-sample}]
We first compare the selected reserve with a grid point near the true
minimax reserve. Stability controls the error from estimating the
parameters; the grid bounds control the numerical error.

Use scaled coordinates $\theta=(A,\lambda h_0)$,
$\hat\theta=(\hat A_N,\lambda\hat h_0)$, and
$\bar\theta=\operatorname{proj}_{\mathcal P_\lambda}\hat\theta$;
$\mathsf W(r;\theta)$ abbreviates $\mathsf W(r;A,h_0)$.
Nonexpansiveness implies
\begin{equation}\label{eq:projection-error}
 \|\bar\theta-\theta\|_1\le\sqrt2\|\hat\theta-\theta\|_1.
\end{equation}
When the fallback is inactive and $(\bar A_N,\bar h_N)\in\mathcal D$, let $r_{\mathrm g}$
be a grid reserve within $\varepsilon_N$ of a true minimizer.
The inner objective $\mathsf W^{\mathcal T}$ underestimates $\mathsf W$
by at most $3\lambda^2\varepsilon_N$. Grid optimality and stability give
\begin{align*}
 \mathsf W(\hat r_N^\dagger;\theta)
 &\le\mathsf W^{\mathcal T}(\hat r_N^\dagger;\bar\theta)
       +3\lambda^2\varepsilon_N+3\lambda\|\bar\theta-\theta\|_1\\
 &\le\mathsf W(r_{\mathrm g};\theta)+3\lambda^2\varepsilon_N
       +6\lambda\|\bar\theta-\theta\|_1\\
 &\le\Delta(G)+C_{\rm num}\varepsilon_N
       +6\sqrt2\lambda\|\hat\theta-\theta\|_1.
\end{align*}
Reserve rounding uses the Lipschitz constant at the true pair, so no bound
on the random projected hazard is needed. At a degenerate projected
pair the algorithm returns an exact minimizer: $\lambda$ for $A=0$ or
$\lambda h_0=1$, and the linear-completion reserve for $A=\lambda h_0$.
The same comparison holds with zero grid error.
Lemma~\ref{lem:est-rates} now proves the high-probability bound.

To obtain the expectation bound, we use moments of the same estimation
errors. For sufficiently large $N$, the fallback is inactive on $\mathcal B_N$.
Integrating the pathwise bound there and using
\eqref{eq:learning-moments} gives expected excess $O(N^{-2/5})$.
On $\mathcal B_N^c$ the rule stays in $[0,\lambda]$, so
Lemma~\ref{lem:bounded-regret} adds at most $\lambda e^{-Nq_\lambda/8}$.
This proves the expectation bound and Corollary~\ref{cor:expected-regret}.
The distribution enters $\kappa_A$, $\kappa_h$, $\underline N(\delta)$ and
the expectation bound only through $q_\lambda$ and the bounds on
$\tilde g,|\tilde g''|$ near $\lambda$. The remaining choices
$c_b,M,m_{\min}$ are fixed. The numerical constant $C_{\rm num}$
depends on $h_0=\tilde g(\lambda^+)$, covered by the same density bound.
Thus both regret bounds hold uniformly when $q_\lambda$ is bounded away
from zero and the density and smoothness bounds are common to the class.

For the finite-grid bandwidth selector, the simultaneous concentration
event in Appendix~\ref{app:learning-details} proves the probability bound.
For its expectation bound, on $\mathcal B_N$ we have
\[
 \mathbb E\!\left[|\hat h_0(\hat c)-h_0|\mathbf1_{\mathcal B_N}\right]
 \le\sum_{c\in\mathcal C_b}
       \mathbb E\!\left[|\hat h_0(c)-h_0|\mathbf1_{\mathcal B_N}\right]
 =O(N^{-2/5}),
\]
by \eqref{eq:learning-moments} for each of the four fixed candidates.
The constant can be taken as four times the largest candidate moment
constant. The pathwise regret comparison and the complement term are
unchanged, so no logarithmic factor is needed in the expected regret.
\end{proof}

\begin{proof}[Proof of Proposition~\ref{rem:parametric-limit}]
On $[0,\lambda]$, $W_G$ is Lipschitz with constant $\mathrm{Lip}_r$
(Lemma~\ref{lem:stability}). For $r\ge\lambda$, every completion
agrees with $G$ on $[\lambda,\infty)$, so
$W_G(r)=W_G(\lambda)+R_G(\lambda)-R_G(r)$, which is continuous. Since $W_G$ is
bounded (Lemma~\ref{lem:bounded-regret}), convergence in probability of the
reserve gives convergence of $\mathbb E[W_G(\hat r_N^{\rm par})]$ to
$W_G(r^{\rm par})$; Corollary~\ref{cor:expected-regret} gives the limit $\Delta(G)$
for the plug-in. By Theorem~\ref{thm:mhr-equalizer}, $W_G>\Delta$ on
$[0,\lambda]\setminus\{r^\dagger\}$. For $r>\lambda$, $G$ is MHR with
$h_G(r)\ge h_0>1/\lambda$, so its virtual value is positive and $R_G$ is
nonincreasing on $[\lambda,\infty)$ (Lemma~\ref{lem:auction-revenue-derivative});
hence $W_G(r)\ge W_G(\lambda)>\Delta$, because $r^\dagger<\lambda$
(Proposition~\ref{prop:rdagger}). Finally,
$G\in\mathcal F_{\mathrm{MHR}}(\lambda;G)$ gives
$R_G^*-R_G(r)\le W_G(r)$ for every $r$, and the realized-regret bound follows
from Corollary~\ref{cor:expected-regret}. For the last claim, let
$r^{\rm par}\ne r^\dagger$ and let $F^\star$ be a completion attaining
$W_G(\min\{r^{\rm par},\lambda\})$ (Theorem~\ref{thm:two-level}); for
$r^{\rm par}>\lambda$ it also attains $W_G(r^{\rm par})$, since all completions
agree above $\lambda$. Its censored records have the law of $G$'s, so the
rule has the same limit $r^{\rm par}$ when the data come from $F^\star$,
and $\Delta(F^\star)=\Delta(G)$ because both completion sets coincide.
Hence $R_{F^\star}^*-R_{F^\star}(r^{\rm par})=W_G(r^{\rm par})>\Delta(F^\star)$.
Table~\ref{tab:exp-truths} shows an instance: at $N=3000$ the Weibull MLE loses
$2.49\%$ under the exponential-body completion, against $\Delta=1.35\%$.
\end{proof}

\subsubsection{A matching interior lower bound}\label{app:lb-statistical-reconstruction}

The lower bound combines two facts: regret grows linearly with reserve
error, and the minimax reserve changes to first order with the boundary
hazard at some interior instance. We prove these facts before constructing
two smooth tails that are difficult to distinguish.

Normalize $\lambda=1$ and fix $n\ge2$. Write $W_{A,h}$ for the body
objective, $\Delta(A,h)$ for its minimum, and $r(A,h)$ for its unique
minimizer. For $0<A<h$ and $h>1$ this agrees with the full objective on
$[0,1]$.

\begin{lemma}[A uniform sharp minimum]\label{lem:lb-uniform-kink}
Fix $A>0$, $1<h^+<\infty$, and $\Delta_{\min}>0$. For pairs $(A,h)\in\mathcal D$ with $h\le h^+$ and
$\Delta(A,h)\ge\Delta_{\min}$, put
\[
 a_*=(\Delta_{\min}/n)^{1/(n-1)},\qquad
 c_L=\tfrac n2(1-e^{-a_*/(2h^+)})^{n-1}e^{-A}\Delta_{\min},
\]
\[
 c_R=e^{-A}\Delta_{\min},\qquad c=\min\{c_L,c_R\}>0.
\]
Then $W_{A,h}(t)-\Delta(A,h)\ge c|t-r(A,h)|$ for $t\in[0,1]$.
For the full objective of any $G$ satisfying Assumption~\ref{ass:mhr}
with this pair, replace $t$
on the right by $\min\{t,1\}$ for all $t\ge0$.
\end{lemma}
\begin{proof}
Let $r_*=r(A,h)$. An attaining upper-loss witness has initial hazard
$a>0$ constant up to its competitor $s>r_*$, with $ar_*<1$.
Bounding the positive revenue derivative on $[r_*,s]$ gives
\[
 \Delta_{\min}\le\Delta\le na^{n-1},\qquad \Delta_{\min}\le\Delta\le1-ar_*.
\]
Thus $a\ge a_*$, $1-ar_*\ge\Delta_{\min}$, and $r_*>1/h^+$.
On $[r_*/2,r_*]$,
$R'(x)\ge n(1-e^{-a_*/(2h^+)})^{n-1}e^{-A}\Delta_{\min}=2c_L$.
Retaining this witness and integrating proves the left bound; for
$t<r_*/2$, integration over the upper half gives at least
$c_Lr_*\ge c_L(r_*-t)$.

For an attaining lower-loss witness put $z=h_F(r_*^+)$.
Positive loss implies $r_*z>1$, and
\[
 \Delta_{\min}\le\Delta=\int_s^{r_*}-R'(x)\,dx
 \le nF(r_*)^{n-1}(r_*z-1).
\]
For $x\in[r_*,1]$, monotonicity and $1-F(x)\ge e^{-A}$ give
$-R'(x)\ge e^{-A}nF(r_*)^{n-1}(r_*z-1)\ge c_R$.
Integration proves the right bound. Tail revenue is nonincreasing,
so $W_G(t)\ge W_G(1)$ for $t\ge1$.
\end{proof}

\begin{lemma}[An interior sensitivity point]\label{lem:lb-existential-sensitivity}
For $A=1/2$, there are $h_*>1$, $k_*>0$, and $\varepsilon_*>0$ such that
\begin{equation}\label{eq:lb-S}
 |r(A,h_*+\varepsilon)-r(A,h_*)|\ge k_*\varepsilon,
 \qquad0<\varepsilon\le\varepsilon_*.
\end{equation}
\end{lemma}
\begin{proof}
For $\varphi(h)=r(1/2,h)$, Proposition~\ref{prop:rdagger} gives $1/h<\varphi(h)<1$.
Hence $\varphi(h)\to1$ as $h\downarrow1$, but $\varphi(2)<1$.
Choose $h_L\in(1,2)$ with $\varphi(h_L)>\varphi(2)$.
Stability makes $\Delta(1/2,h)$ continuous, so its positive minimum
on $[h_L,2]$ supplies a common $c$ in Lemma~\ref{lem:lb-uniform-kink}.
Optimality and stability give
\[
 W_{1/2,h}(\varphi(h'))-\Delta(1/2,h)\le6|h'-h|,
 \qquad |\varphi(h')-\varphi(h)|\le(6/c)|h'-h|.
\]
Thus $\varphi$ is absolutely continuous, and
$\int_{h_L}^2 \varphi'(h)\,dh=\varphi(2)-\varphi(h_L)<0$.
At some interior differentiability point $h_*$, $\varphi'(h_*)<0$.
Take $k_*=|\varphi'(h_*)|/2$ and shrink $\varepsilon_*$ so that
\eqref{eq:lb-S} holds and $h_*+\varepsilon_*<2$.
\end{proof}

\begin{proof}[Proof of Proposition~\ref{prop:interior-lower-bound}]
Take the sensitivity point above, $A=1/2$, $q=e^{-A}$, and a base
distribution with cumulative hazard
\[
 H_0(x)=Ax\ (0\le x\le1),\qquad
 H_0(1+u)=A+h_*u+u^2/2\ (u\ge0).
\]
Keep its body fixed and perturb its right hazard to
\[
 h_N(1+u)=h_*+u+\delta_N(1-u/w_N)_+^3,
 \qquad w_N=N^{-1/5},\quad\delta_N=\kappa w_N^2,
\]
where $\kappa>0$ is fixed with $q\kappa^2/(14h_*)\le1/8$.
For large $N$, $1-3\kappa w_N\ge1/2$, so the new hazard is increasing
and joins the body with an upward jump. It is $C^2$ on the right,
including at the cutoff. On $u\in[0,1]$, its height and first two
derivatives are bounded uniformly: $h_N\le h_*+1+\kappa$,
$1/2\le h_N'\le1$, $0\le h_N''\le6\kappa$.
The identity $g_N''=(h_N''-3h_Nh_N'+h_N^3)e^{-H_N}$ gives common
density bounds. Survival is at most the base Gaussian-tail survival,
so means are finite. Both distributions have height $A$, and their
boundary hazards differ by $\delta_N$. Call the base distribution $G_0$
and the perturbed one $G_N$, with densities $g_0$ and $g_N$.

Let $P_0,P_N$ be their one-record laws. Their no-bid atoms both have
mass $1-q$. Conditional on a bid, write
$\xi(u)=h_*+u$, $\vartheta(u)=\delta_N(1-u/w_N)_+^3$, and
$S_0(u)=e^{-h_*u-u^2/2}$. The conditional densities are $\xi S_0$ and
$(\xi+\vartheta)S_0e^{-\int_0^u\vartheta}$, so
\begin{align*}
 D_{\rm KL}(P_0\Vert P_N)
 &=q\int_0^\infty S_0(u)
    \left[\vartheta(u)-\xi(u)\log\left(1+\frac{\vartheta(u)}{\xi(u)}\right)\right]du\\
 &\le\frac{q}{2h_*}\int_0^{w_N}\vartheta(u)^2du
 =\frac{q\kappa^2w_N^5}{14h_*}.
\end{align*}
For the equality, Tonelli transforms the cumulative-hazard contribution
$\int\xi S_0\int_0^u\vartheta$ into $\int S_0\vartheta$, using the identity
\[
 \int_v^\infty\xi(u)S_0(u)\,du=S_0(v).
\]
The inequality uses
$z-\log(1+z)\le z^2/2$ for $z\ge0$.
For $N$ records, KL is at most $1/8$ and total variation at most $1/4$.
Thus the two records are difficult to distinguish, even though their
boundary hazards differ by order $N^{-2/5}$.

Write $r_0=r(A,h_*)$, $r_N=r(A,h_*+\delta_N)$.
For large $N$, $|r_N-r_0|\ge k_*\delta_N$ by
\eqref{eq:lb-S}, and stability gives
$\Delta(A,h_*+\delta_N)\ge\Delta(A,h_*)/2>0$.
Lemma~\ref{lem:lb-uniform-kink} supplies one $c>0$ for both losses.
For every $t\ge0$, with $\ell_i(t)=W_{G_i}(t)-\Delta(G_i)$,
\[
 \ell_0(t)+\ell_N(t)\ge
 c\bigl(|\min\{t,1\}-r_0|+|\min\{t,1\}-r_N|\bigr)
 \ge ck_*\delta_N.
\]
Let $K_D$ be the conditional reserve law of a possibly randomized
rule, and let $p_0,p_N$ be densities of its two record laws with
respect to a common measure. Nonnegativity gives
\[
 \mathbb E_{G_0}\ell_0(r^\pi)+\mathbb E_{G_N}\ell_N(r^\pi)
 \ge\int\min\{p_0,p_N\}\int(\ell_0+\ell_N)\,dK_D
 \ge\tfrac34 ck_*\kappa N^{-2/5}.
\]
The maximum risk is therefore at least $3ck_*\kappa N^{-2/5}/8$.
The fixed class $\{G_0\}\cup\{G_N:N\ge N_0\}$, with $N_0$ the threshold
beyond which $1-3\kappa w_N\ge1/2$ and $\delta_N\le\varepsilon_*$, has common smoothness
bounds and boundary hazards in $[h_*,h_*+\varepsilon_*]$, uniformly
away from both frontiers. Finally, multiplying values by $\lambda$
preserves information distance and multiplies regret by $\lambda$.
The density bounds rescale on $[\lambda,2\lambda]$.
\end{proof}

\subsection{Learning when the body is fully identified}\label{app:edge-pretest}\label{app:ident-body-policies}

Here the body is determined by $A$. When its optimum is at or below
$\lambda$, the censored fraction alone suffices to learn the optimal reserve.
We first establish consistency, including the case of an optimum in the
observed tail, then prove the fast rate and the bounds for nearby instances.

\begin{proof}[Fully identified part of Proposition~\ref{prop:identifiable-iff}]
When $A=\lambda h_0$, the body is $H(x)=Ax/\lambda$.
With $m=|\mathcal U_N|$, estimate $A$ by $\hat A_N=-\log(m/N)$ and paste
$1-e^{-\hat A_Nx/\lambda}$ below $\lambda$ to the empirical tail.
The result is a CDF $\widehat F_N$: both pieces have boundary mass
$1-e^{-\hat A_N}$. Its body revenue is maximized at
$r_c(\hat A_N)$, the linear-completion reserve in \eqref{eq:chord-reserve}, so maximize empirical revenue over
$\{\lambda,r_c(\hat A_N)\}\cup\mathcal U_N$.
When all observations are censored, post zero; this event has probability
$\alpha^N\to0$. Otherwise $\hat A_N\to_P A$, and, with $F$ the unique compatible
completion,
\[
 \|\widehat F_N-F\|_\infty
 \le|\hat A_N-A|+
           \|\widehat G_N-G\|_{\infty,[\lambda,\infty)}\to_P0.
\]
The tail moments are the same as before. The all-reserves form of
Lemma~\ref{lem:empirical-revenue} gives uniform revenue convergence, and
its maximizer inequality again gives vanishing regret.
When $A\ge1$, the population optimum is $\lambda/A\le\lambda$; the simpler
rule based on the censored fraction alone has the fast $O(N^{-1})$ expected
regret rate proved
in Proposition~\ref{prop:edge}.
\end{proof}

\begin{proof}[Proof of Proposition~\ref{prop:edge}]
All compatible optima lie in a short interval. We first bound revenue
loss quadratically in the distance to possible optima, then apply the
mean-square bound for $\hat A_N$.

Write $h=h_0$, $l=1/h$, and $u=(1+d)/h$.
On the body, $h_F\le h$ and convexity gives
$v h_F(v)\ge H_F(v)\ge hv-d$. Revenue thus rises up to $l$
and falls from $u$ onward; the common tail also has decreasing revenue.
Every completion has an optimum in $[l,\min\{\lambda,u\}]$.
Since $nF^{n-1}(1-F)\le k_1$,
\[
 (R_F'(v))_+\le k_1(1-hv+d)_+,
 \qquad(-R_F'(v))_+\le k_1(hv-1)_+.
\]
Integrating between a reserve and an optimum gives
\begin{equation}\label{eq:edge-envelope}
 W_G(r)\le\frac{k_1h}{2}
 \max\{(u-r)_+^2,(r-l)_+^2\},\qquad r\in[0,\lambda].
\end{equation}
At the midpoint, both distances are at most $d/(2h)$. If the midpoint
exceeds $\lambda$, posting $\lambda$ cannot under-reserve because every
optimum is at most $\lambda$; its remaining distance $\lambda-l$ is
less than $d/(2h)$. The linear-completion reserve is in $[l,\min\{\lambda,u\}]$:
it equals $\lambda$ for $A\le1$, and for $A>1$,
$hr_c(A)=1+d/A\le1+d$. Its two distances are at most $d/h$.
These observations prove \eqref{eq:quadratic-edge}.

The map $r_c$ is $\lambda$-Lipschitz on finite nonnegative inputs, so
\begin{equation}\label{eq:chord-pathwise}
 W_G(r_c(\hat A_N))\le\frac{k_1}{2h}
       \bigl(d+\lambda h|\hat A_N-A|\bigr)^2.
\end{equation}
On $\mathcal B_N$, apply $(x+y)^2\le2x^2+2y^2$ and the two moment bounds for
$\hat A_N$ in \eqref{eq:learning-moments}, which use only the logarithm bound and the variance of the censored
fraction, not Assumption~\ref{ass:boundary-smoothness}. The complement contributes at most
$\lambda e^{-Nq_\lambda/8}$ by bounded regret, including $m=0$.
This proves \eqref{eq:chord-expected}.
\end{proof}

\begin{proposition}[The statistical scale near full identification]\label{prop:lower-bound}
Fix $n\ge2$ and $\lambda>0$. There is $c>0$ such that, for every $N\ge1$,
there are two MHR distributions with smooth right-boundary densities and boundary
pairs in $\mathcal D$ satisfying
\[
 \max_{i=1,2}\mathbb E_{G_i}[W_{G_i}(r^\pi)-\Delta(G_i)]\ge c/N
\]
for every rule using $N$ independent records at $\lambda$.
Their distances from the fully identified frontier are $O(N^{-1/2})$,
and their boundary density bounds are uniform in $N$.
\end{proposition}
\begingroup \begin{proof}
We choose two nearby distributions whose optimal reserves are separated
by order $N^{-1/2}$. Their minimax regrets are small enough that the
quadratic revenue loss from confusing the two still forces order $N^{-1}$
excess regret.

Let $A_\vartheta=\lambda\vartheta-\zeta$ and use cumulative hazard
$(A_\vartheta/\lambda)x$ below $\lambda$, and
$A_\vartheta+\vartheta(x-\lambda)+(x-\lambda)^2$ above it.
Fix $\lambda\vartheta_1\ge2$, set
$\vartheta_2=\vartheta_1+c_\vartheta N^{-1/2}$, $\zeta=c_\zeta N^{-1/2}$,
with $0<c_\vartheta\le1/\lambda$, $0<c_\zeta\le1/2$.
The two distributions, $G_1$ and $G_2$ with densities $g_1$ and $g_2$,
are MHR with $A_i\in[3/2,\lambda\vartheta_1+1]$.
One record has an atom $p_\vartheta=1-e^{-A_\vartheta}$ and tail density
$g_\vartheta(u)=e^{-A_\vartheta}(\vartheta+2u)e^{-\vartheta u-u^2}$.
The atom is bounded away from zero and one, while
\[
 \partial_\vartheta\log g_\vartheta(u)=-\lambda+(\vartheta+2u)^{-1}-u.
\]
On this compact parameter range, $g_\vartheta/g_{\vartheta_1}$ is bounded and
the tail is Gaussian. Hence
$\chi^2(P_{\vartheta_2}\Vert P_{\vartheta_1})\le C_\chi(\vartheta_2-\vartheta_1)^2$,
with $C_\chi$ uniform in $\zeta$, by integrating the squared density derivative.
Choose $c_\vartheta$ small enough that the $N$-record total variation is
at most $1/2$.

The actual optima are $r_i=\lambda/A_i$, separated by at least
$c_rc_\vartheta N^{-1/2}$, where $c_r=\lambda^2/(\lambda\vartheta_1+1)^2$.
On $|x-r_i|\le r_i/2$, the prefactor in
$-R_{G_i}'(x)=nG_i(x)^{n-1}g_i(x)(x-r_i)$ has a uniform positive lower
bound. Integration and unimodality therefore give
$R_{G_i}^*-R_{G_i}(r)\ge c_q\min\{(r-r_i)^2,(r_i/2)^2\}$ for all $r$,
with $c_q>0$. Also Proposition~\ref{prop:edge} gives
$\Delta(G_i)\le c_\zeta^2/(8\vartheta_1N)$.
Since $r_i\ge\lambda/(\lambda\vartheta_1+1)$, choosing $c_\vartheta$ small
ensures that $\sigma_N=c_rc_\vartheta N^{-1/2}/2$ is at most $r_i/2$
for every $N$. Every reserve $r$ is at least $\sigma_N$ from one of the
optima, say $r_i$. Since $G_i$ is a compatible completion of its own record,
$W_{G_i}(r)\ge R_{G_i}^*-R_{G_i}(r)$. Hence
\[
 W_{G_i}(r)-\Delta(G_i)
 \ge c_q\sigma_N^2-\frac{c_\zeta^2}{8\vartheta_1N}
 =\frac1N\left(\frac{c_qc_r^2c_\vartheta^2}{4}
                -\frac{c_\zeta^2}{8\vartheta_1}\right)
 \ge\frac{c_{\mathrm{lb}}}{N},\qquad
 c_{\mathrm{lb}}:=\frac{c_qc_r^2c_\vartheta^2}{8},
\]
provided $c_\zeta^2\le\vartheta_1c_qc_r^2c_\vartheta^2$.
This choice is independent of $N$.
Integrating against the overlap of the two record laws gives a sum of
expected risks at least $c_{\mathrm{lb}}/(2N)$, and hence the claimed bound.
\end{proof} \endgroup

\subsection{Selecting between the tail and plug-in rules}\label{app:pretest-proof}

The auto-selector compares the estimated boundary hazard with $1/\lambda$.
Away from this threshold, it selects the correct rule with high probability.
Near the threshold, selecting the other rule has a controlled cost, as
shown below.

Fix bounds $0<h_{\min}<1/\lambda<h_{\max}<\infty$ and $q_{\min}>0$.
Let $\mathcal G^\sharp$ be the class of distributions satisfying
Assumptions~\ref{ass:mhr} and~\ref{ass:boundary-smoothness} with
$h_{\min}\le h_0\le h_{\max}$, $q_\lambda\ge q_{\min}$, and common bounds
$\bar g,\bar g_2$ for $\tilde g,|\tilde g''|$ on
$[\lambda,\lambda+\bar b]$. Set $\bar\lambda=\max\{\lambda,1/h_{\min}\}$.
Denote by $r^{\rm sel}$ the auto-selector reserve of
Algorithm~\ref{alg:pretest}: on $m<m_{\min}$ it posts $r_c(\hat A_N)$;
otherwise it posts $\hat r_N^{\rm tail}$ when the raw estimate
$\hat h_0\le1/\lambda$, and $\hat r_N^\dagger$ when it exceeds this
threshold.

\begin{proposition}[Uniform auto-selector guarantee]\label{prop:pretest-uniform}
For the class $\mathcal G^\sharp$ defined above and every $\delta\in(0,1)$,
there are $K_A,K_h$ and $N_\sharp(\delta)$ depending only on these class
bounds and the fixed tuning constants. Put
\[
 \begin{gathered}
 t_N=(K_A+\lambda K_h)\sqrt{\log(12/\delta)}N^{-2/5},\qquad
 s_N=\sqrt{\log(4/\delta)/(2N)},\\
 e_N^{\rm num}=c_{\mathrm g}\bigl(3\lambda^2+\max\{1,\lambda h_{\max}\}\bigr)N^{-1/2}.
 \end{gathered}
\]
For $N\ge N_\sharp(\delta)$ and every $G\in\mathcal G^\sharp$,
with probability at least $1-\delta$,
\[
 W_G(r^{\rm sel})-\Delta(G)
 \le6\lambda(2\sqrt2+1)t_N+\frac{t_N^2}{2h_{\min}}
     +e_N^{\rm num}+4n\bar\lambda s_N.
\]
The excess is therefore uniformly $O_P(N^{-2/5})$, including at both
zero-regret frontiers.
\end{proposition}

\begin{proof}[Proof of Proposition~\ref{prop:pretest-uniform}]
Choose $K_A,K_h$ from the concentration constants using $q_{\min}$ and
the common density bounds. Take $N_\sharp(\delta)$ from
\eqref{eq:N0-explicit} with these bounds and $\delta/2$.
Together with DKW, with probability $1-\delta$ we have $m\ge m_{\min}$,
\[
 \varsigma_N=|\hat A_N-A|+\lambda|\hat h_0-h_0|\le t_N,
 \qquad\sup_{x\ge\lambda}|\widehat G_N^-(x)-G(x)|\le s_N.
\]
Write $\theta=(A,\lambda h_0)$ and $\hat\theta=(\hat A_N,\lambda\hat h_0)$. Correct selection of the body rule costs at most
$6\sqrt2\lambda \varsigma_N+e_N^{\rm num}$; correct tail selection costs at most
$4n\bar\lambda s_N$.

If $\lambda h_0\ge1\ge\lambda\hat h_0$, then $\lambda h_0-1\le \varsigma_N$. The feasible pair
$\theta_0=(\min\{A,1\},1)$ has $\|\theta-\theta_0\|_1\le2\varsigma_N$ and
$\mathsf W(\lambda;\theta_0)=0$. Stability gives
$\mathsf W(\lambda;\theta)\le6\lambda \varsigma_N$; for a tail reserve $r$,
\begin{equation}\label{eq:tail-branch-regret}
 W_G(r)\le\mathsf W(\lambda;\theta)
       +\sup_{s\ge\lambda}R_G(s)-R_G(r).
\end{equation}
The wrong tail selection therefore costs at most
$6\lambda \varsigma_N+4n\bar\lambda s_N$.

If $\lambda h_0<1<\lambda\hat h_0$, put $\theta_0=(A,1)\in\mathcal P_\lambda$.
\[
 \|\operatorname{proj}_{\mathcal P_\lambda}\hat\theta-\theta\|_1
 \le\sqrt2\|\hat\theta-\theta_0\|_1+\|\theta_0-\theta\|_1
 \le(2\sqrt2+1)\varsigma_N.
\]
Comparing the selected body reserve with the grid endpoint $\lambda$
gives $\mathsf W(r^{\rm sel};\theta)\le
6\lambda(2\sqrt2+1)\varsigma_N+e_N^{\rm num}$, since
$\mathsf W(\lambda;\theta)=0$. The additional tail gain satisfies
\begin{equation}\label{eq:tail-gain}
 B_\lambda=\sup_{s\ge\lambda}R_G(s)-R_G(\lambda)
 \le\frac{(1-\lambda h_0)_+^2}{2h_0}\le\frac{\varsigma_N^2}{2h_{\min}}.
\end{equation}
Indeed, at a tail reserve $v\ge\lambda$ the positive revenue derivative
is at most $(1-vh_0)_+$.
For a body reserve $W_G(r)\le\mathsf W(r;\theta)+B_\lambda$, and here
$\Delta(G)=0$. These bounds cover both correct selections and both possible mistakes,
which proves the guarantee.

The same comparison applies if $\hat h_0$ is selected from the four
bandwidth candidates in Appendix~\ref{app:learning-details}. Apply
Lemma~\ref{lem:est-rates} to each candidate at level $\delta/8$, and
DKW at level $\delta/2$. Using the largest class constants and thresholds
over the grid replaces $\log(12/\delta)$ in $t_N$ by $\log(48/\delta)$;
the remaining terms are unchanged. This selected-bandwidth auto-selector
therefore retains the uniform $O_P(N^{-2/5})$ rate.
\end{proof}

\subsection{Participation counts and adaptive historical reserves}\label{app:design-extensions}

\begin{remark}[An incorrect bidder count]\label{rem:count-error}
If the seller uses $\tilde N=(1+\nu)N\ge m$, with $\nu>-1$, then
\begin{equation}\label{eq:count-bias}
 \tilde A_N=-\log(m/\tilde N)=\hat A_N+\log(1+\nu).
\end{equation}
A fixed multiplicative count error therefore causes a persistent height
bias. The bandwidth constant changes by $(1+\nu)^{-1/5}$, preserving
the density-estimation rate. The projected plug-in argument adds at most
$6\sqrt2\lambda|\log(1+\nu)|$ to its excess-regret bound, besides the
sampling and numerical terms. This is an error bound, not a way to
infer unrecorded participation.
\end{remark}

\begin{remark}[Adaptive historical reserves]\label{rem:endogenous}
Our finite-sample results use a design fixed before values are drawn.
If a reserve is chosen from earlier records, current values can still be
independent draws from $G$. At a fixed level posted infinitely often,
centered censoring and tail indicators form martingale differences, so
population identification at that level survives by the martingale strong law.
The finite-sample proof would need martingale concentration for a random
boundary count and control of the data-dependent minimum reserve; it is
not established here. If the reserve depends on the very values it censors,
or on a common contemporaneous demand shock, the censoring probability
conditional on the reserve need not be $G(\lambda)$ at all.
\end{remark}

\section{Extensions to Regular Distributions}\label{app:regular}\label{sec:regular}

Regular distributions have nondecreasing virtual value.
We now solve the reserve problem over
\begin{equation}\label{eq:regular_ambiguity_set}
    \mathcal F_{\mathrm{reg}}(\lambda;G)=\{F\in\mathcal F(\lambda;G):F\text{ is regular}\},
\end{equation}
which contains $\mathcal F_{\mathrm{MHR}}(\lambda;G)$ because every MHR
distribution is regular. Enlarging the compatible set changes the worst case, but retains a tractable
structure.
In the nonidentifiable regime, the worst-case problem still reduces to
two one-parameter families.
A body with constant virtual value gives the largest loss from setting
the reserve too high; a body whose virtual value changes from negative
to positive as the bidder's value increases gives the largest loss from
setting it too low. As under MHR, the unique minimax reserve equates these
losses. The resulting plug-in rule achieves expected excess regret
$O(N^{-2/5})$ under local density smoothness, when supplied with a
parameter rectangle containing $(A,h_0)$ in its interior
(Proposition~\ref{prop:reg-finite}).

\subsection{Regular completions and the zero-regret frontier}\label{sec:auction-regular}

For the regular-completion analysis, assume $G\in\mathcal G$ is regular,
with positive density on the interior of its effective support. Assume
its right density at $\lambda$ exists and
$h_0=g(\lambda^+)/(1-G(\lambda))\in(0,\infty)$.
We retain $\lambda>0$, $n\ge2$, and $0<G(\lambda)<1$ from the model.
We use its right-continuous virtual value, with
value $-\infty$ below the support. Its observed boundary value is
$\phi_0=\lambda-1/h_0$.

To describe regular distributions geometrically, use a single bidder's acceptance probability
$q=1-F(v)$ at price $v$. This differs from the auction sale probability
$1-F(v)^n$. Increasing $q$ moves toward lower values. With
$v(q):=F^{-1}(1-q)$, define the \emph{single-bidder revenue curve}
\begin{equation}\label{eq:revcurve}
  \Phi_F(q):=q\,v(q),\qquad 0<q<1,
\end{equation}
which is single-bidder posted-price revenue, not the $n$-bidder auction revenue
$R_F$. Virtual value governs the derivatives of both revenue functions. Extend $\Phi_F$
continuously to the endpoints. Finite mean gives $\Phi_F(0)=0$,
while $\Phi_F(1)$ is the lower endpoint of the support and may be positive.
We drop the subscript when the distribution is clear. For almost every
$q\in(0,1)$, $\Phi'(q)=\phi_F(v(q))$, so $F$ is regular exactly when $\Phi_F$ is concave
(Lemma~\ref{lem:reg-concave}, Appendix~\ref{app:regular-geometry}). Recall the boundary
quantile $q_\lambda=1-\alpha=e^{-A}$ of \eqref{eq:qlam} and write
\begin{equation}\label{eq:c0-def}
  c_0:=\lambda q_\lambda
\end{equation}
for the revenue curve's height there.
The observed tail fixes $\Phi$ on $[0,q_\lambda]$. A regular completion
extends this curve to $[q_\lambda,1]$, preserving concavity and starting
at $(q_\lambda,c_0)$ with slope at most $\phi_0$. It therefore lies below
the line of slope $\phi_0$ through that point and above the chord to $(1,0)$.

If $\phi_0\le0$, monotonicity of virtual value gives $\phi_F\le0$ below
$\lambda$ for every completion. Auction revenue is then nondecreasing
on $[0,\lambda]$, so an optimal reserve lies in the observed tail. If $\phi_0>0$, the observed tail no longer
determines the sign of virtual value below $\lambda$. Compatible
completions can have disjoint optimal reserve sets, giving positive
minimax regret.

\begin{theorem}[Zero-regret frontier for regular distributions]\label{thm:frontier-reg}
Suppose $G\in\mathcal G$ is regular, has positive density on the interior
of its effective support, and has right density at $\lambda$ with
$h_0=g(\lambda^+)/(1-G(\lambda))\in(0,\infty)$. Then
\[
\Delta\bigl(\mathcal F_{\mathrm{reg}}(\lambda;G)\bigr)=0
\qquad\Longleftrightarrow\qquad
\phi_G(\lambda)\le 0.
\]
In that case, every maximizer of $R_G$ on $[\lambda,\infty)$ is optimal for all $F\in\mathcal F_{\mathrm{reg}}(\lambda;G)$.
\end{theorem}

For regular distributions, the problem is identifiable exactly in the tail-optimal
case $\phi_0\le0$; the smaller MHR set has one additional zero-regret
case: the fully identified body $A=h_0\lambda$ of Theorem~\ref{thm:frontier}.

\begin{proof}
If $\phi_0\le0$, every compatible distribution has nondecreasing virtual
value, so $\phi_F\le0$ below $\lambda$. Thus $R_F$ is nondecreasing there, and any maximizer of the
common tail revenue is optimal for every completion.

If $\phi_0>0$, the tail revenue is nonincreasing, so reserves above
$\lambda$ are dominated by $\lambda$. Consider two body revenue curves:
the line of slope $\phi_0$ through $(q_\lambda,c_0)$, and the chord from
that point to $(1,0)$. Both paste to the observed regular tail
(Lemma~\ref{lem:C-realized}). The first has constant positive virtual
value on its support, so its optimal reserves form
$[0,c_0+\phi_0(1-q_\lambda)]$; the second has negative virtual value below
$\lambda$, so its optimal set lies in $[\lambda,\infty)$. These optimal sets
are disjoint because $c_0+\phi_0(1-q_\lambda)<\lambda$. Their two loss
functions are continuous on $[0,\lambda]$, so their pointwise maximum
has a strictly positive minimum. Hence
$\Delta(\mathcal F_{\mathrm{reg}}(\lambda;G))>0$.
\end{proof}

\subsection{Revenue curves and regular completions}\label{sec:regular-oracle}

With $\rho=1-F(r)$, changing variables in
\eqref{eq:revenue} gives (Lemma~\ref{lem:affine})
\begin{equation}\label{eq:affine}
 R_F(r)=R_\Phi(\rho)
 :=n(1-\rho)^{n-1}\Phi(\rho)
   +n(n-1)\int_0^\rho\Phi(q)(1-q)^{n-2}\,dq.
\end{equation}
The objective is affine in the curve when the quantile is fixed. A posted
price does not fix that quantile. Above the lower support endpoint, its
quantile is the intersection with the ray $\Phi(\rho)=r\rho$; at or below
that endpoint, we set the quantile to $1$.

The tail determines $\Phi$ on $[0,q_\lambda]$, with
$q_\lambda=e^{-A}$ and $c_0=\lambda q_\lambda$. Concavity across the boundary
caps the first unknown slope by $\phi_0=\lambda-1/h_0$. For the
regular-completion reduction and learning rule below, take $A>0$ and
$\phi_0>0$, the regular nonidentifiable regime. The possible bodies are
\begin{equation}\label{eq:Cset}
 \mathcal C=\mathcal C(A,h_0):=
 \bigl\{\Phi\in C([q_\lambda,1]):\ \Phi\text{ concave},\ \Phi\ge0,
       \ \Phi(q_\lambda)=c_0,\ \Phi'(q_\lambda^+)\le\phi_0\bigr\}.
\end{equation}
This geometric class describes exactly the regular completions.

\begin{lemma}[Realization of the regular revenue curves]\label{lem:C-realized}
Suppose $G\in\mathcal G$ is regular, its right density at $\lambda$
exists, and $h_0=g(\lambda^+)/(1-G(\lambda))>1/\lambda$ is finite.
Every $\Phi\in\mathcal C(A,h_0)$ is the body revenue curve of exactly one
$F\in\mathcal F_{\mathrm{reg}}(\lambda;G)$. Its price curve
$v_\Phi(q)=\Phi(q)/q$ is strictly decreasing, and its inverse, extended by
$\rho_\Phi(r)=1$ for $r\le\Phi(1)$, is $h_0e^A$-Lipschitz on $[0,\lambda]$.
The completion is $F(x)=1-\rho_\Phi(x)$ below $\lambda$, pasted to $G$
above it. It is absolutely continuous, has density at most $h_0e^A$ on the
body, and has no atom. Thus $F\mapsto\Phi_F|_{[q_\lambda,1]}$ is a bijection.
\end{lemma}

The slope cap makes $v_\Phi$ strictly decreasing, so its inverse constructs
the unique completion. Lemma~\ref{lem:price-slope} gives the quantitative
inverse bound, which will control the effect of changing a posted price.

The common observed tail cancels from revenue differences. In particular, with
\begin{equation}\label{eq:U-telescope}
 U_\Phi(q):=n(1-q)^{n-1}\Phi(q)
       +n(n-1)\int_{q_\lambda}^q\Phi(u)(1-u)^{n-2}\,du,
\end{equation}
the worst-case regret is
\begin{equation}\label{eq:W-reg}
 \mathsf W_{\mathrm{reg}}(r;A,h_0)
 :=\sup_{\Phi\in\mathcal C}
 \left\{\max_{q\in[q_\lambda,1]}U_\Phi(q)
              -U_\Phi\bigl(\rho_\Phi(r)\bigr)\right\},
 \qquad r\in[0,\lambda].
\end{equation}
Every completion's optimum lies in $[0,\lambda]$, since its virtual value
above $\lambda$ is at least $\phi_0>0$. Hence \eqref{eq:W-reg} is the
original worst-case regret, and it depends on the observed tail only through
$(A,h_0)$. Write
$\Delta_{\mathrm{reg}}(A,h_0):=\min_{r\in[0,\lambda]}
\mathsf W_{\mathrm{reg}}(r;A,h_0)$ for its minimum.

\subsection{Two families and a unique minimax reserve}

Both families begin at the observed boundary point
$(q_\lambda,c_0)$ and respect the boundary slope cap $\phi_0$.
Write $\ell(t)=c_0+\phi_0(t-q_\lambda)$ for the cap line. The families are
\begin{equation}\label{eq:reg-families}
 S_s(u)=c_0+s(u-q_\lambda),\quad 0\le s\le\phi_0;
 \qquad
 E_t(u)=\begin{cases}
 \ell(u),&q_\lambda\le u\le t,\\
 \ell(t)\dfrac{1-u}{1-t},&t\le u\le1,
 \end{cases}\quad q_\lambda\le t<1,
\end{equation}
and $E_1=\ell$. The straight curve $S_s$ has constant virtual value $s$ on
the body. Every reserve below its lower support endpoint
$b_s=c_0+\alpha s$ is optimal (recall $q_\lambda=1-\alpha$).
For $t<1$, $E_t$ has virtual value $\phi_0$ at values above its kink and a
negative virtual value below it. In increasing quantile order its slope
therefore changes from positive to negative. Its optimal reserve is
$r^*_{E_t}=\ell(t)/t$.

Figure~\ref{fig:regular-families} in Appendix~\ref{app:regular-scalar} shows both families and how a posted
reserve determines the seller's quantile.

As in the MHR problem, $\mathsf L_{\mathrm{reg}}$ compares with lower
alternative reserves, and $\mathsf U_{\mathrm{reg}}$ with higher ones.
For a reserve that is too high, the regret of $S_s$ is
\begin{equation}\label{eq:reg-upper-family}
 u_s(r):=\begin{cases}
 0,&r\le b_s,\\
 s\left(\dfrac{r-c_0-\alpha s}{r-s}\right)^n,&r>b_s.
 \end{cases}
 \qquad
 \mathsf L_{\mathrm{reg}}(r):=\max_{0\le s\le\phi_0}u_s(r).
\end{equation}
For $t<1$, put $\sigma_t=\ell(t)/(1-t)$. The loss from posting below the optimum
of $E_t$ is
\begin{equation}\label{eq:reg-lower-family}
 d_t(r):=\begin{cases}
 \sigma_t\left[(1-t)^n-\left(\dfrac{r}{\sigma_t+r}\right)^n\right],&r\le r^*_{E_t},\\
 0,&r>r^*_{E_t},
 \end{cases}
 \quad d_1(r):=0,
 \qquad
 \mathsf U_{\mathrm{reg}}(r):=\max_{q_\lambda\le t\le1}d_t(r).
\end{equation}
Setting $d_t$ to zero above $r^*_{E_t}$ retains only the loss from a reserve that
is too low. The other direction is already covered by
\eqref{eq:reg-upper-family}.

\begin{theorem}[Exact regular reduction and the minimax reserve]\label{thm:regular-reduction}
Suppose $G\in\mathcal G$ is regular, its right density at $\lambda$
exists, and $h_0=g(\lambda^+)/(1-G(\lambda))>1/\lambda$ is finite. Then
\begin{equation}\label{eq:regular-reduction}
 \mathsf W_{\mathrm{reg}}(r;A,h_0)
 =\max\{\mathsf L_{\mathrm{reg}}(r),\mathsf U_{\mathrm{reg}}(r)\},
 \qquad 0\le r\le\lambda.
\end{equation}
A worst-case completion is attained in one of the two families in
\eqref{eq:reg-families}. The function $\mathsf U_{\mathrm{reg}}$ is continuous
and strictly decreasing on $[0,\lambda]$; $\mathsf L_{\mathrm{reg}}$ is
continuous, vanishes on $[0,c_0]$, and is strictly increasing on $[c_0,\lambda]$.
There is a unique minimax reserve $r^\dagger_{\mathrm{reg}}\in(c_0,\lambda)$,
characterized by
\begin{equation}\label{eq:regular-equalizer}
 \mathsf L_{\mathrm{reg}}(r^\dagger_{\mathrm{reg}})
 =\mathsf U_{\mathrm{reg}}(r^\dagger_{\mathrm{reg}})
 =\Delta\bigl(\mathcal F_{\mathrm{reg}}(\lambda;G)\bigr).
\end{equation}
\end{theorem}

Together with Theorem~\ref{thm:frontier-reg}, this theorem proves
Proposition~\ref{prop:regular-summary}.
The proof fixes the seller's and competitor's quantiles. Raising the
competitor's feasible height yields $S_s$ for lower competing prices and
$E_t$ for higher ones.

The equalizer balances the worst losses from posting too high and too low.
The posted price fixes a ray, rather than a quantile, so both families
are needed to solve nature's problem.

For computation, the straight-family loss has the explicit quadratic
formula \eqref{eq:reg-quadratic}; only the kink location requires a grid.
A quantile grid with maximum gap $\varepsilon$, followed by reserve
bisection to tolerance $\eta$, returns a reserve with worst-case regret at
most $\Delta_{\mathrm{reg}}+n\lambda\varepsilon+n\lambda h_0e^A\eta$.
Appendix~\ref{app:regular-computation} gives the certificate and
Algorithm~\ref{alg:regular-scalar}, used in the experiments.

\begin{remark}[What regular distributions permit]\label{rem:reg-scope}
On a linear revenue-curve piece, virtual value is constant, say
$\varphi$, and the hazard $h_F(v)=1/(v-\varphi)$ decreases. Thus MHR
excludes both families in \eqref{eq:reg-families}, despite their continuous
distributions. Table~\ref{tab:exp-shape} compares the resulting ambiguity costs at the
same boundary scalars, with the reserves in Appendix~\ref{app:exp-regular-more};
neither tests a shape assumption.
\end{remark}

\subsection{Regular completion geometry}\label{app:regular-geometry}

We verify the correspondence between concave revenue curves and
regular completions before proving the two-family reduction.

\begin{lemma}[Regular distributions and concave revenue curves]\label{lem:reg-concave}
Let $F\in\mathcal G$ have positive density on the interior of its effective
support, and let $v(q)=F^{-1}(1-q)$. Then
$\Phi_F'(q)=\phi_F(v(q))$ for almost every $q\in(0,1)$.
Consequently, $F$ is regular if and only if $\Phi_F$ is concave.
\end{lemma}

\begin{lemma}[Auction revenue in quantile coordinates]\label{lem:affine}
Let $F\in\mathcal G$ have positive density on the interior of its effective
support, and let $n\ge2$. For every reserve $r\ge0$, with survival quantile
$\rho=1-F(r)$, $R_F(r)=R_{\Phi_F}(\rho)$, where $R_{\Phi_F}$ is the
functional in \eqref{eq:affine}.
\end{lemma}

\begin{lemma}[The price curve is uniformly steep]\label{lem:price-slope}
Let $\lambda>0$, $A>0$, $h_0>1/\lambda$, and
$\Phi\in\mathcal C(A,h_0)$. The price curve
$v_\Phi(\rho)=\Phi(\rho)/\rho$ satisfies, for almost every
$\rho\in(q_\lambda,1)$,
\begin{equation}\label{eq:price-slope}
 v_\Phi'(\rho)\le-\frac{q_\lambda}{h_0\rho^2}
 \le-\frac{e^{-A}}{h_0}.
\end{equation}
It is strictly decreasing on $[q_\lambda,1]$. Its inverse, extended by
$\rho_\Phi(r)=1$ below the lower support endpoint $\Phi(1)$, is
$h_0e^A$-Lipschitz on $[0,\lambda]$.
\end{lemma}

\begin{proof}[Proof of Lemma~\ref{lem:reg-concave}]
Fix $q$ in the interior with $f(v(q))>0$. From $q=1-F(v)$ we get $dv/dq=-1/f(v)$,
so differentiating $\Phi(q)=q\,v(q)$,
\[
  \Phi'(q)=v(q)+q\,v'(q)=v-\frac{1-F(v)}{f(v)}=\phi_F(v(q)).
\]
Since $v(\cdot)$ is strictly decreasing on the interior of the effective support,
$\phi_F$ is nondecreasing in $v$ if and only if $\Phi'$ is nonincreasing in
$q$. The identity holds almost everywhere; the revenue curve is locally absolutely continuous, so a nonincreasing version of its derivative is equivalent to concavity. The canonical endpoint limits preserve that concavity on the closed interval.
\end{proof}

\begin{proof}[Proof of Lemma~\ref{lem:affine}]
If $\rho=0$, both sides are zero.
If $r$ is below the lower support endpoint, $\rho=1$, the boundary term
vanishes because $n\ge2$, and both sides equal the expected second-highest
value. Otherwise substitute $q=1-F(x)$, $x=v(q)$ and $f(x)\,dx=-dq$ in \eqref{eq:revenue}, noting
that $x=r$ corresponds to $q=\rho$ and $x\to\infty$ to $q\to0$. The boundary term
becomes
$n\,r\,F(r)^{n-1}(1-F(r))=n\,v(\rho)\,(1-\rho)^{n-1}\rho=n(1-\rho)^{n-1}\Phi(\rho)$,
using $\rho\,v(\rho)=\Phi(\rho)$. The integral becomes
\[
\begin{aligned}
  n(n-1)\!\int_r^\infty\! x\,F(x)^{n-2}(1-F(x))f(x)\,dx
  &=n(n-1)\!\int_0^{\rho}\! v(q)\,(1-q)^{n-2}\,q\,dq\\
  &=n(n-1)\!\int_0^{\rho}\!\Phi(q)(1-q)^{n-2}dq,
\end{aligned}
\]
again by $q\,v(q)=\Phi(q)$. Adding the two gives \eqref{eq:affine}.
\end{proof}

\begin{proof}[Proof of Lemma~\ref{lem:price-slope}]
Concavity gives $\Phi'(\rho)\le\bigl(\Phi(\rho)-c_0\bigr)/(\rho-q_\lambda)$, the
chord slope from the pasting point, so
\[
  \Phi'(\rho)\rho-\Phi(\rho)\;\le\;\frac{q_\lambda\bigl(\Phi(\rho)-\lambda\rho\bigr)}{\rho-q_\lambda},
\]
using $c_0=\lambda q_\lambda$. The slope cap gives
$\Phi(\rho)\le c_0+\phi_0(\rho-q_\lambda)$, and $\phi_0-\lambda=-1/h_0$, so
$\Phi(\rho)-\lambda\rho\le-(\rho-q_\lambda)/h_0$. Substituting and dividing by
$\rho^{2}$ yields \eqref{eq:price-slope}. The bound is attained at $\rho=1$ by
the completion that leaves $(q_\lambda,c_0)$ at the cap, so it cannot be
improved.
\end{proof}

\begin{proof}[Proof of Lemma~\ref{lem:C-realized}]
Write $v(q)=\Phi(q)/q$ and $b=\Phi(1)$. The price-slope bound makes $v$
a continuous strictly decreasing bijection from $[q_\lambda,1]$ onto
$[b,\lambda]$, with $h_0e^A$-Lipschitz inverse $\rho$. Define $F=0$ below
$b$, $F(x)=1-\rho(x)$ on $[b,\lambda]$, and $F=G$ above $\lambda$.
The endpoint values match, and the inverse bound makes $F$ absolutely
continuous with density at most $h_0e^A$. On the interior, finite negative
one-sided derivatives of $v$ give positive density; at a kink we use the
corresponding one-sided version. The completion has finite mean because
its changed values are bounded by $\lambda$.

Its body virtual value is $\Phi'(q)$ at $x=v(q)$. Concavity and the reversed
quantile order make this nondecreasing in $x$, and the slope cap
$\Phi'(q_\lambda^+)\le\phi_0$ preserves nondecreasing virtual value at the join. Conversely,
any regular completion has such a concave revenue curve. The quantile
function determines its distribution uniquely, proving the bijection.
The construction also preserves revenue at reserves below $b$: all
bidders qualify there, and the convention $\rho_\Phi(r)=1$ gives the
same second-highest-value revenue.
\end{proof}

\subsection{The regular scalar computation}\label{app:regular-scalar}

\begin{figure}[htbp]
 \centering
 \includegraphics[width=\textwidth]{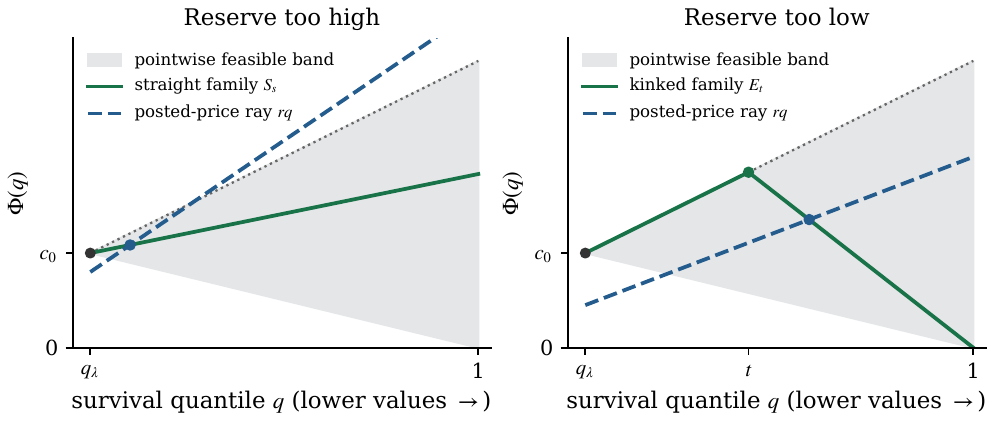}
 \caption{The regular families share boundary point
 $(q_\lambda,c_0)$, with $c_0=\lambda q_\lambda$, and slope cap
 $\phi_0=\lambda-1/h_0$. Gray shading gives the pointwise feasible range;
 curves must also be concave. $S_s$ gives the worst loss against lower
 alternative reserves, and $E_t$ against higher ones. The price ray $\Phi=rq$
 intersects each displayed curve at the seller's quantile; a reserve below
a curve's lower support endpoint has quantile $1$. Moving right, toward
 larger survival quantiles, means moving to lower bidder values.}
 \label{fig:regular-families}
\end{figure}

\begin{proof}[Proof of Theorem~\ref{thm:regular-reduction}]
Write $q=q_\lambda$, $p=1-q$, $c=c_0$ and $\phi=\phi_0$.
On an affine piece of slope $\sigma$, differentiation of
\eqref{eq:affine} gives
\begin{equation}\label{eq:reg-affine-difference}
 R_\Phi(v)-R_\Phi(u)=\sigma[(1-u)^n-(1-v)^n].
\end{equation}
Fix the seller's price $r$, its quantile $\rho$, and a competitor quantile
$t$. Usually $\Phi(\rho)=r\rho$. If $\rho=1$ and $\Phi(1)>r$, replace the
curve after $t<1$ by the chord to $(1,r)$. This preserves concavity and
the slope cap, lowers the curve, and increases the competitor's revenue
gap because the terminal boundary term vanishes for $n\ge2$. Thus we may
assume the seller anchor is $(\rho,r\rho)$ in every case.

\emph{Lower competing prices ($t>\rho$).}
For $\rho>q$, set $s=(r\rho-c)/(\rho-q)$. Concavity puts the curve below
the line $S_s$ on $[\rho,t]$. Raising it to this line increases both
the competitor boundary term and the positively weighted integral in
\eqref{eq:affine}. The gap is therefore at most
$s[(1-\rho)^n-(1-t)^n]$. Nonpositive $s$ gives no positive gap; otherwise
$s\le\phi$ and the bound is at most $s(1-\rho)^n$, attained by $S_s$
with competitor quantile $1$. At $\rho=q$ the same argument uses the cap
line $s=\phi$. Solving $S_s(\rho)=r\rho$ gives
\[
 1-\rho=\frac{r-c-ps}{r-s}\qquad(r>b_s),
\]
and hence exactly $u_s(r)$.

\emph{Higher competing prices ($t<\rho$).}
Let $y=\Phi(t)$. Concavity puts $\Phi$ above the chord between the two
anchors; the integral now has a negative sign. Consequently the gap is
at most
\begin{equation}\label{eq:reg-height-monotone}
 \frac{y-r\rho}{\rho-t}[(1-t)^n-(1-\rho)^n],
 \qquad
 y\le\min\left\{\ell(t),\frac{r\rho(1-t)}{1-\rho}\right\},
\end{equation}
where the second upper bound follows from $\Phi(1)\ge0$ and is omitted
at $\rho=1$. The gap increases with $y$.

If the terminal bound is active, put $\gamma=r\rho/(1-\rho)$.
The lines $\ell(u)$ and $\gamma(1-u)$ meet at some $\tau\in[q,t]$:
the latter is above $c$ at $q$ by concavity and below $\ell(t)$ at $t$.
Thus $E_\tau$ has seller quantile $\rho$. Moving the competitor from $t$
to its optimum $\tau$ only increases the gap, giving $d_\tau(r)$.

If the cap bound is active, put $\upsilon=\rho-t$, $w=1-t$, and
$C=\ell(t)-rt$. Then $y-r\rho=C-r\upsilon$, and the bound becomes
\[
 \left(\frac C\upsilon-r\right)[w^n-(w-\upsilon)^n]
 =C\,\frac{w^n-(w-\upsilon)^n}{\upsilon}-r\,[w^n-(w-\upsilon)^n].
\]
A positive gap implies $C>0$. The first term is $C$ times the average of
the decreasing function $z\mapsto n(w-z)^{n-1}$ over $[0,\upsilon]$, so it
is nonincreasing in $\upsilon$; the second is $-r$ times a nondecreasing
function of $\upsilon$. The bound therefore decreases with $\upsilon$, that
is, with $\rho$. Since the cap bound is active,
$(1-\rho)\ell(t)\le r\rho(1-t)$, so
$\rho\ge\rho_0=\ell(t)/(\ell(t)+r(1-t))$. The inequality $\rho_0>t$ is
$\ell(t)(1-t)>rt(1-t)$, which is $C>0$; it also gives $r<\ell(t)/t=r^*_{E_t}$.
At $\rho=\rho_0$ the line through the anchors $(t,\ell(t))$ and
$(\rho_0,r\rho_0)$ passes through $(1,0)$, because
$\rho_0(\ell(t)+r(1-t))=\ell(t)$. The bounding curve is therefore $E_t$,
with slope $-\sigma_t$ on $[t,1]$ and seller quantile $\rho_0=\sigma_t/(\sigma_t+r)$, and
the bound equals $\sigma_t[(1-t)^n-(r/(\sigma_t+r))^n]=d_t(r)$. For $r=0$,
$\rho_0=1$ and the same formula
applies. Thus every positive higher-competitor gap is bounded by
$\max_t d_t(r)$.

Conversely both families are feasible, and their optimal reserves give
$u_s(r)$ and $d_t(r)$. This proves the exact reduction. Both maxima are
attained: $u_s(r)$ is continuous on $[0,\phi]$, and
$0\le d_t(r)\le\ell(t)(1-t)^{n-1}\to0$ as $t\uparrow1$.

For the closed expression for $\mathsf L_{\rm reg}$, assume $c<r<\lambda$
and write $\varrho=r-c$. The derivative of $s[(\varrho-ps)/(r-s)]^n$ has the sign of
\[
 ps^2-\xi_rs+r\varrho,\qquad
 \xi_r=\varrho+pr+nq(\lambda-r).
\]
On $[0,\varrho/p]$ this polynomial decreases strictly from a positive to a
negative value: its derivative is at most
$2\varrho-\xi_r=-(n+1)q(\lambda-r)<0$.
Its smaller root, clipped at $\phi$, gives \eqref{eq:reg-quadratic}.
The endpoints give $\mathsf L_{\rm reg}=0$ for $r\le c$ and
$\mathsf L_{\rm reg}(\lambda)=\phi p^n$.

Each $d_t(r)$ decreases strictly while positive, and $E_q$ has optimum
$\lambda$, so their maximum is strictly decreasing for $r<\lambda$.
Each positive $u_s(r)$ increases strictly, so its maximum strictly
increases above $c$. Joint continuity and the compact parameter
intervals give continuity. Since
\[
 \mathsf L_{\rm reg}(c)=0<\mathsf U_{\rm reg}(c),\qquad
 \mathsf U_{\rm reg}(\lambda)=0<\mathsf L_{\rm reg}(\lambda),
\]
the two losses cross once in $(c,\lambda)$. Their maximum decreases
before that crossing and increases after it, proving uniqueness.
\end{proof}

\subsubsection{Computing the regular minimax reserve}\label{app:regular-computation}

The exact reduction separates computation into a closed expression for
$\mathsf L_{\mathrm{reg}}$, the loss against lower alternative reserves,
and a one-dimensional maximization for $\mathsf U_{\mathrm{reg}}$, the
loss against higher ones. Throughout, $A>0$, $h_0>1/\lambda$, and $n\ge2$;
$q_\lambda=e^{-A}$, $c_0=\lambda q_\lambda$, $\alpha=1-q_\lambda$, and
$\phi_0=\lambda-1/h_0$.

The straight-family loss is explicit. For $c_0<r<\lambda$, let
\begin{equation}\label{eq:reg-quadratic}
 \varrho=r-c_0,\qquad \xi_r=\varrho+\alpha r+nq_\lambda(\lambda-r),\qquad
 s_r=\min\left\{\phi_0,
 \frac{2r\varrho}{\xi_r+\sqrt{\xi_r^2-4\alpha r\varrho}}\right\}.
\end{equation}
Then $\mathsf L_{\mathrm{reg}}(r)=u_{s_r}(r)$;
$\mathsf L_{\mathrm{reg}}(r)=0$ for $r\le c_0$, and
$\mathsf L_{\mathrm{reg}}(\lambda)=\phi_0\alpha^n$.

\begin{corollary}[Certified scalar search]\label{cor:regular-scalar-grid}
Let $\lambda>0$, $A>0$, $h_0>1/\lambda$, and $n\ge2$.
Let $\mathcal Q\subset[q_\lambda,1]$ be a finite grid containing both
endpoints, with maximum successive gap $\varepsilon>0$, and put
$\mathsf U_{\mathcal Q}(r)=\max_{t\in\mathcal Q}d_t(r)$.
For every $r\in[0,\lambda]$,
\[
 0\le\mathsf U_{\mathrm{reg}}(r)-\mathsf U_{\mathcal Q}(r)
       \le n\lambda\varepsilon.
\]
Let $r_{\mathcal Q}$ be the unique solution of
$\mathsf L_{\mathrm{reg}}(r)=\mathsf U_{\mathcal Q}(r)$ and
$\Delta_{\mathcal Q}=\mathsf L_{\mathrm{reg}}(r_{\mathcal Q})$. Then
\[
 0\le\Delta_{\mathrm{reg}}-\Delta_{\mathcal Q}\le n\lambda\varepsilon,
 \qquad
 \mathsf W_{\mathrm{reg}}(r_{\mathcal Q})\le\Delta_{\mathrm{reg}}+n\lambda\varepsilon.
\]
For every $\eta>0$ and $r\in[0,\lambda]$ with
$|r-r_{\mathcal Q}|\le\eta$, the last bound increases by at most
$n\lambda h_0e^A\eta$.
\end{corollary}

The derivative bound $|\partial d_t(r)/\partial t|\le n\lambda$ controls
how much a quantile grid can underestimate the higher-competitor loss.
Algorithm~\ref{alg:regular-scalar} combines this grid with bisection of the
equalizer and requires no integration. Table~\ref{tab:exp-shape} and Appendix~\ref{app:exp-regular-more}
report the computed values.

\begin{proof}[Proof of Corollary~\ref{cor:regular-scalar-grid}]
Write $q=q_\lambda$, $c=c_0$, and $\phi=\phi_0$.
We bound the sensitivity to the kink location. On the active interval
$r\le r^*_{E_t}$, put
\[
 w=1-t,\qquad \Gamma=c+\phi(1-q)\le\lambda,\qquad
 \sigma=\sigma_t=\Gamma/w-\phi,\qquad z=r/(\sigma+r)\le w.
\]
Since $\sigma'=\Gamma/w^2$, differentiation gives
\[
 \frac{\partial d_t(r)}{\partial t}
 =\frac \Gamma{w^2}\left[w^n-z^n+\frac{n\sigma z^n}{\sigma+r}\right]
       -n\sigma w^{n-1}.
\]
The bracket is between $0$ and $nw^n$, because the coefficient of $z^n$
is between $-1$ and $n-1$ and $z\le w$. Thus the first term lies in
$[0,n\Gamma w^{n-2}]$ and the second in $[0,n\ell(t)w^{n-2}]$. Both are in
$[0,n\lambda]$, so the derivative has absolute value at most $n\lambda$.
The function meets zero continuously at the boundary of its active interval
and extends continuously at $t=1$. It is consequently $n\lambda$-Lipschitz
in $t$ on the whole interval. Rounding an attaining kink to a grid point
proves $0\le\mathsf U_{\mathrm{reg}}-\mathsf U_{\mathcal Q}
\le n\lambda\varepsilon$ uniformly in $r$.

The finite maximum $\mathsf U_{\mathcal Q}$ is continuous. Because
$q\in\mathcal Q$, it is positive and strictly decreasing for $r<\lambda$,
by the same maximizing-curve argument as in the theorem. It therefore has
a unique crossing $r_{\mathcal Q}\in(c,\lambda)$ with
$\mathsf L_{\mathrm{reg}}$, and the minimum of
$\max\{\mathsf L_{\mathrm{reg}},\mathsf U_{\mathcal Q}\}$ is
$\Delta_{\mathcal Q}$. The pointwise grid bound implies
$\Delta_{\mathcal Q}\le\Delta_{\mathrm{reg}}\le\Delta_{\mathcal Q}+n\lambda\varepsilon$.
Also $\mathsf U_{\mathcal Q}\le\mathsf U_{\mathrm{reg}}$ gives
$r_{\mathcal Q}\le r^\dagger_{\mathrm{reg}}$. Hence
\[
 \mathsf W_{\mathrm{reg}}(r_{\mathcal Q})
 =\mathsf U_{\mathrm{reg}}(r_{\mathcal Q})
 \le\mathsf L_{\mathrm{reg}}(r_{\mathcal Q})+n\lambda\varepsilon
 \le\Delta_{\mathrm{reg}}+n\lambda\varepsilon.
\]
Finally, for every $\Phi\in\mathcal C$, $U_\Phi$ has derivative
$n(1-x)^{n-1}\Phi'(x)$ at quantile $x$, of absolute value at most
$n\lambda$.
Lemma~\ref{lem:price-slope} bounds its inverse-price map by $h_0e^A$.
Their composition, and therefore the supremum regret, is
$n\lambda h_0e^A$-Lipschitz in the reserve. This gives the bisection error.
\end{proof}

\paragraph{Cost of the scalar algorithm}
Algorithm~\ref{alg:regular-scalar} implements the theoretical reduction of
Theorem~\ref{thm:regular-reduction}.

\begin{algorithm}[!htbp]
\caption{Certified scalar search for the regular minimax reserve}
\label{alg:regular-scalar}
\small
\begin{algorithmic}[1]
\REQUIRE $n\ge2$, $\lambda>0$, $A>0$, $h_0>1/\lambda$; quantile mesh $\varepsilon>0$; reserve tolerance $\eta>0$
\STATE $q_\lambda\leftarrow e^{-A}$, $c_0\leftarrow\lambda q_\lambda$, $\phi_0\leftarrow\lambda-1/h_0$
\STATE form a grid $\mathcal Q\subset[q_\lambda,1]$ containing both endpoints, with maximum gap at most $\varepsilon$
\STATE evaluate $\mathsf L_{\mathrm{reg}}$ by \eqref{eq:reg-quadratic} and its endpoint conventions; set $\mathsf U_{\mathcal Q}(r)=\max_{t\in\mathcal Q}d_t(r)$ using \eqref{eq:reg-lower-family}
\STATE $a\leftarrow c_0$; $b\leftarrow\lambda$
\WHILE{$b-a>\eta$}
 \STATE $r\leftarrow(a+b)/2$
 \IF{$\mathsf L_{\mathrm{reg}}(r)<\mathsf U_{\mathcal Q}(r)$}
  \STATE $a\leftarrow r$
 \ELSE
  \STATE $b\leftarrow r$
 \ENDIF
\ENDWHILE
\STATE \textbf{return} reserve $(a+b)/2$ and the interval
$[\mathsf L_{\mathrm{reg}}(a),\ \mathsf L_{\mathrm{reg}}(b)+n\lambda\varepsilon]$ containing the minimax regret
\end{algorithmic}
\end{algorithm}

Each bisection step costs $O(|\mathcal Q|)$ scalar evaluations. A uniform grid
therefore costs $O(\varepsilon^{-1}\log(1+\lambda/\eta))$ evaluations for fixed
$(n,\lambda)$, without quadrature. The returned interval contains the true
minimax regret by Corollary~\ref{cor:regular-scalar-grid} and monotonicity of
$\mathsf L_{\mathrm{reg}}$.

\subsection{Stability and learning for regular distributions}\label{app:regular-stability}\label{sec:reg-finite}

The rule estimates the same two boundary quantities as under MHR.
Here we supply a rectangle containing the true pair; for regular
distributions, the MHR constraint $A\le\lambda h_0$ need not hold.

\begin{definition}[Regular plug-in rule]\label{def:plug-in-reg}
Fix $\mathcal K_{\mathrm{reg}}=[A_-,A_+]\times[h_-,h_+]$ with
$0<A_-\le A_+<\infty$ and $1/\lambda<h_-\le h_+<\infty$.
Compute the raw boundary estimates of Section~\ref{sec:est-boundary}
and clip them coordinatewise into $\mathcal K_{\mathrm{reg}}$. At the clipped
pair, run Algorithm~\ref{alg:regular-scalar} with quantile mesh $\varepsilon_N^{\mathrm q}\asymp N^{-1/2}$ and bisection tolerance $\eta_N\asymp N^{-1/2}$, and post its reserve
$\hat r_N^{\dagger,\mathrm{reg}}$. If fewer than the fixed threshold
$m_{\min}$ bids are uncensored, post $\lambda$ instead.
\end{definition}

\begin{proposition}[Stability and regret of the regular rule]\label{prop:reg-finite}
Fix $\mathcal K=\mathcal K_{\mathrm{reg}}$ as in
Definition~\ref{def:plug-in-reg}.
\begin{itemize}
\item[(i)] For all $(A,h_0),(A',h_0')\in\mathcal K$,
\begin{equation}\label{eq:L-reg}
\begin{gathered}
 \sup_{r\in[0,\lambda]}\left|\mathsf W_{\mathrm{reg}}(r;A',h_0')
          -\mathsf W_{\mathrm{reg}}(r;A,h_0)\right|
 \le C_{\mathrm{reg}}\bigl(|A'-A|+\lambda|h_0'-h_0|\bigr),\\
 C_{\mathrm{reg}}:=n\lambda(4n+4+3\lambda h_+e^{2A_+}).
\end{gathered}
\end{equation}
For each $(A,h_0)\in\mathcal K$, the function
$r\mapsto\mathsf W_{\mathrm{reg}}(r;A,h_0)$ is Lipschitz on $[0,\lambda]$
with constant $\mathrm{Lip}_r^{\mathrm{reg}}\le n\lambda h_+e^{A_+}$.
\item[(ii)] Suppose $G\in\mathcal G$ is regular, $(A,h_0)$ is in the
interior of $\mathcal K$, and Assumption~\ref{ass:boundary-smoothness} holds.
Then the rule of Definition~\ref{def:plug-in-reg} satisfies
\[
 \mathbb E\!\left[\sup_{F\in\mathcal F_{\mathrm{reg}}(\lambda;G)}
 \{R_F^*-R_F(\hat r_N^{\dagger,\mathrm{reg}})\}\right]
 \le\Delta\bigl(\mathcal F_{\mathrm{reg}}(\lambda;G)\bigr)+O(N^{-2/5}).
\]
\end{itemize}
\end{proposition}

The expectation is over the boundary sample. The proof transports
concave revenue curves between the two quantile intervals, then controls
the quantile at which a fixed posted price meets each curve.

\begingroup \begin{proof}
For a feasible curve define the revenue primitive
\[
 V_\Phi(q)=n(1-q)^{n-1}\Phi(q)
 -n(n-1)\int_q^1\Phi(u)(1-u)^{n-2}\,du.
\]
It differs from $U_\Phi$ by a constant, so its completion regret is
$J_\Phi(r)=\max V_\Phi-V_\Phi(\rho_\Phi(r))$.
The cap gives $0\le\Phi\le\lambda$. Concavity gives
$-\Phi(q)/(1-q)\le\Phi'(q)\le\phi_0$, and hence
$|V_\Phi'|=|n(1-q)^{n-1}\Phi'(q)|\le n\lambda$.
Together with Lemma~\ref{lem:price-slope}, this proves the stated
reserve Lipschitz bound.

From here on, write $\theta=(A,h_0)$, $\theta'=(A',h_0')$, $q=e^{-A}$, $p=1-q$,
and use primes for the second pair. Put
$a=|A'-A|$, $b=\lambda|h_0'-h_0|$, and $\varsigma=a+b$.
Map the new interval onto the old one by
$\chi(u)=1-(p/p')(1-u)$, an affine bijection of $[q',1]$ onto $[q,1]$,
and transport any $\Phi\in\mathcal C(\theta)$ to
\[
 \widehat\Phi(u)=\min\{\ell'(u),(c_0'/c_0)\Phi(\chi(u))\}.
\]
This is concave as a minimum of concave functions and nonnegative,
since $\ell'\ge c_0'>0$ on $[q',1]$. Both branches equal $c_0'$ at $q'$,
because $\chi(q')=q$, and the first caps the right slope at $q'$ by
$\phi_0'$. Thus $\widehat\Phi\in\mathcal C(\theta')$.

\emph{Step 1: three elementary bounds.} On $[q',1]$ we have
$1-\chi(u)=(p/p')(1-u)$ and $1-u\le p'$. Hence
\begin{gather*}
 |\chi(u)-u|=\frac{(1-u)|p'-p|}{p'}\le|p'-p|\le a,\\
 \Bigl|\frac{c_0'}{c_0}-1\Bigr|=|e^{A-A'}-1|\le e^aa,\qquad
 |\phi_0'-\phi_0|=\frac{|h_0'-h_0|}{h_0h_0'}\le\lambda b,
\end{gather*}
using $|e^{-A}-e^{-A'}|\le|A-A'|$ and $h_0,h_0'>1/\lambda$.

\emph{Step 2: distance between the curves.} Put
$\Psi=(c_0'/c_0)\,\Phi\circ\chi$, so that $\widehat\Phi=\Psi-(\Psi-\ell')_+$
and
\[
 \widehat\Phi-\Phi\circ\chi=\Bigl(\frac{c_0'}{c_0}-1\Bigr)\Phi\circ\chi-(\Psi-\ell')_+.
\]
The cap gives $0\le\Phi\le\ell\le\lambda$, so the first term is at most
$\lambda e^aa$ in absolute value. For the second, $\Phi\le\ell$ and
$\chi(u)-q=(p/p')(u-q')$ give
\[
 \Psi(u)-\ell'(u)\le\frac{c_0'}{c_0}\,\ell(\chi(u))-\ell'(u)
 =\frac{u-q'}{p'}\Bigl[\phi_0\Bigl(\frac{c_0'}{c_0}p-p'\Bigr)+p'(\phi_0-\phi_0')\Bigr],
\]
and $(c_0'/c_0)p-p'=c_0'/c_0-1$. With $0\le u-q'\le p'\le1$ and
$\phi_0\le\lambda$, the excess $(\Psi-\ell')_+$ is at most
$\lambda e^aa+\lambda b$. Together,
\[
 z:=\|\widehat\Phi-\Phi\circ\chi\|_\infty\le\lambda(2e^aa+b).
\]

\emph{Step 3: distance between the revenue primitives.} The substitution
$v=\chi(w)$, with $dv=(p/p')\,dw$ and $1-v=(p/p')(1-w)$, gives
$V_\Phi\circ\chi=(p/p')^{n-1}V_{\Phi\circ\chi}$, where $V_{\Phi\circ\chi}$ is
the primitive of $\Phi\circ\chi$ on $[q',1]$. For any curve $\Psi$ on
$[q',1]$, $n(n-1)\int_u^1(1-v)^{n-2}dv=n(1-u)^{n-1}$ gives
$|V_\Psi(u)|\le2n(1-u)^{n-1}\|\Psi\|_\infty$. Since $V$ is linear in the
curve,
\[
 V_{\widehat\Phi}-V_\Phi\circ\chi
 =V_{\widehat\Phi-\Phi\circ\chi}+\bigl[1-(p/p')^{n-1}\bigr]V_{\Phi\circ\chi}.
\]
The first term is at most $2nz$. The second is at most
\[
 2n\lambda(1-u)^{n-1}\bigl|1-(p/p')^{n-1}\bigr|
 =2n\lambda\Bigl(\frac{1-u}{p'}\Bigr)^{n-1}|p'^{\,n-1}-p^{n-1}|
 \le2n(n-1)\lambda a,
\]
using $1-u\le p'$ and $|p'^{\,n-1}-p^{n-1}|\le(n-1)|p'-p|\le(n-1)a$. Hence
\[
 z_V:=\|V_{\widehat\Phi}-V_\Phi\circ\chi\|_\infty
 \le2n[z+(n-1)\lambda a].
\]
Because $\chi$ maps $[q',1]$ onto $[q,1]$,
$\max V_\Phi\circ\chi=\max V_\Phi$, so the two maxima also differ by at
most $z_V$.

\emph{Step 4: the seller's quantile.} Let $\rho=\rho_\Phi(r)$,
$\hat\rho=\rho_{\widehat\Phi}(r)$, and $\check\rho=\chi(\hat\rho)\in[q,1]$.
If $\hat\rho<1$, then $\widehat\Phi(\hat\rho)=r\hat\rho$, and with
$|\Phi(\check\rho)-\widehat\Phi(\hat\rho)|\le z$, $|\hat\rho-\check\rho|\le a$,
$r\le\lambda$ and $\check\rho\ge q=e^{-A}$,
\[
 |v_\Phi(\check\rho)-r|
 \le\frac{|\Phi(\check\rho)-\widehat\Phi(\hat\rho)|}{\check\rho}
   +r\hat\rho\,\Bigl|\frac1{\check\rho}-\frac1{\hat\rho}\Bigr|
 \le\frac{z+r|\hat\rho-\check\rho|}{\check\rho}
 \le e^A(z+\lambda a).
\]
Since $\rho=\rho_\Phi(r)$ and $\check\rho=\rho_\Phi(v_\Phi(\check\rho))$,
the $h_0e^A$-Lipschitz inverse of Lemma~\ref{lem:price-slope} gives
$|\check\rho-\rho|\le h_0e^{2A}(z+\lambda a)$. If $\hat\rho=1$, then
$\check\rho=1$ and $r\le\widehat\Phi(1)\le\Phi(1)+z$: either
$r\le\Phi(1)$ and $\rho=1$, or the clipped inverse gives
$|1-\rho|\le h_0e^A(r-\Phi(1))\le h_0e^Az$. In both cases
$|\check\rho-\rho|\le h_0e^{2A}(z+\lambda a)$.

\emph{Step 5: the completion regret.} Using Step 3 twice and
$|V_\Phi'|\le n\lambda$,
\[
 J_{\widehat\Phi}(r)=\max V_{\widehat\Phi}-V_{\widehat\Phi}(\hat\rho)
 \ge\bigl(\max V_\Phi-z_V\bigr)-\bigl(V_\Phi(\check\rho)+z_V\bigr)
 \ge J_\Phi(r)-2z_V-n\lambda h_0e^{2A}(z+\lambda a).
\]
Taking the supremum over $\Phi\in\mathcal C(\theta)$ bounds
$\mathsf W_{\rm reg}(r;\theta)-\mathsf W_{\rm reg}(r;\theta')$ by the error
term, and the transport from $\theta'$ to $\theta$ gives the reverse
inequality with the primed parameters, which lie in the same rectangle.

\emph{Step 6: the constant.} Inserting Steps 2 and 3, the error term is at
most $4n[z+(n-1)\lambda a]+n\lambda h_0e^{2A}(z+\lambda a)$. Uniformly on
the rectangle, $e^a=1+O(a)$, so $z=\lambda(2a+b)+O(\varsigma^2)$ and the
error is at most
\[
 n\lambda\bigl[(4n+4)a+4b+\lambda h_+e^{2A_+}(3a+b)\bigr]+O(\varsigma^2)
 \le n\lambda(4n+4+3\lambda h_+e^{2A_+})\varsigma+O(\varsigma^2).
\]
The rectangle is convex, so the segment from $\theta$ to $\theta'$ stays in
it; divide the segment into $k$ equal parts. The linear errors add to
$C_{\rm reg}\varsigma$, whereas the quadratic errors, whose constant is
uniform on the rectangle, add to $O(\varsigma^2/k)$. Letting
$k\to\infty$ proves \eqref{eq:L-reg}.

For the learning claim, on the event that $m\ge m_{\min}$
and the raw estimate lies in the rectangle, put
$e_N=|\hat A_N-A|+\lambda|\hat h_0-h_0|$.
Stability, optimality at the estimated pair, and the scalar certificate give
\[
 \mathsf W_{\rm reg}(\hat r_N^{\dagger,\mathrm{reg}};\theta)
 \le\Delta_{\mathrm{reg}}(\theta)+2C_{\rm reg}e_N
 +n\lambda\varepsilon_N^{\mathrm q}+n\lambda h_+e^{A_+}\eta_N.
\]
Lemma~\ref{lem:est-rates} bounds the expected estimation term by
$O(N^{-2/5})$. Because the true pair is interior to the rectangle,
its concentration bounds make the complementary event $O(N^{-2})$;
the low-bid-count event is exponentially small. Regret is bounded for
reserves in $[0,\lambda]$, including the fallback. These events therefore
add $O(N^{-2})$, and the numerical tolerances add $O(N^{-1/2})$.
\end{proof} \endgroup

\subsection{An arbitrary observed tail}\label{app:local-tail-extension}

An arbitrary observed tail contributes an additional observable quantity:
its best revenue gain over the boundary reserve. We first separate this
gain from body comparisons for any compatible class, then specialize to
shape-restricted bodies with separately imposed boundary slope caps.

For a nonempty class $\mathcal A\subseteq\mathcal F(\lambda;G)$, write
\[
 B_\lambda=\max_{s\ge\lambda}R_G(s)-R_G(\lambda),\qquad
 r_{\mathrm T}\in\argmax_{s\ge\lambda}R_G(s).
\]
The maximum exists by continuity and finite mean. For body reserves put
\[
\begin{aligned}
 W_{\rm body}(r)&=\sup_{F\in\mathcal A}\max_{s\in[0,\lambda]}
                   \{R_F(s)-R_F(r)\},\\
 V_{\rm body}(r)&=\sup_{F\in\mathcal A}\{R_F(\lambda)-R_F(r)\},
 &J_{\mathrm{body}}&=W_{\rm body}(\lambda).
\end{aligned}
\]
Here $J_{\mathrm{body}}$ is the largest gain that can hide below the boundary. All three
body quantities are independent of the common tail revenue.

\begin{proposition}[Body and tail gains]
\label{prop:local-tail-extension}
Let $\mathcal A\subseteq\mathcal F(\lambda;G)$ be nonempty, with
$B_\lambda,r_{\mathrm T},W_{\rm body},V_{\rm body},J_{\mathrm{body}}$ defined above.
Then, for every $r\in[0,\lambda]$,
\begin{equation}\label{eq:local-tail-body}
 \sup_{F\in\mathcal A}\{R_F^*-R_F(r)\}
 =\max\{W_{\rm body}(r),B_\lambda+V_{\rm body}(r)\}.
\end{equation}
The best tail reserve has worst-case regret
\begin{equation}\label{eq:local-tail-reserve}
 \sup_{F\in\mathcal A}\{R_F^*-R_F(r_{\mathrm T})\}
 =(J_{\mathrm{body}}-B_\lambda)_+.
\end{equation}
Consequently
\begin{equation}\label{eq:local-tail-minimax}
 \Delta(\mathcal A)=\min\left\{
 \inf_{r\in[0,\lambda]}\max\{W_{\rm body}(r),
 B_\lambda+V_{\rm body}(r)\},\ (J_{\mathrm{body}}-B_\lambda)_+\right\}.
\end{equation}
\end{proposition}

\begin{proof}
Put $d_F(r)=R_F(r)-R_G(\lambda)$ and
$m_F=\max_{s\in[0,\lambda]}d_F(s)$. Since tail revenue is common,
$R_F^*=R_G(\lambda)+\max\{m_F,B_\lambda\}$. Subtracting $R_F(r)$
and taking the supremum proves \eqref{eq:local-tail-body}.
At $r_{\mathrm T}$ the loss is $(m_F-B_\lambda)_+$, proving
\eqref{eq:local-tail-reserve}. This reserve dominates every other
choice above $\lambda$, which proves the minimax formula.
\end{proof}

For this specialization, let $\mathcal A$ be the full class of completions
obtained by pasting to $G$ a body satisfying the following constraints.
For MHR bodies, use convex nondecreasing cumulative hazards with $H(0)=0$,
$H(\lambda)=A$, and terminal slope at most $h_0$; for regular bodies,
use the concave revenue curves in \eqref{eq:Cset}, including the cap
$\Phi'(q_\lambda^+)\le\phi_0$.
Here the boundary caps are additional assumptions: body shape alone does
not constrain the join to an arbitrary tail. The pasted tail need not
share the body's shape. The existing body computations give
$W_{\rm body}$ and $J_{\mathrm{body}}$. The remaining loss $V_{\rm body}$ is attained
by the chord $H(x)=Ax/\lambda$ under MHR, and by the chord from
$(q_\lambda,c_0)$ to $(1,0)$ for regular bodies. Each chord gives the
stochastically smallest compatible values, and fixed-reserve auction
revenue is nondecreasing in each bidder's value.

For regular bodies with $\phi_0>0$, writing $q=e^{-A}$ and $p=1-q$, this gives
\[
 V_{\rm body}(r)=\frac{\lambda q}{p}
 \left[p^n-\left(\frac{rp}{\lambda q+rp}\right)^n\right],
 \qquad J_{\mathrm{body}}=\phi_0p^n.
\]
The nonidentifiable regime for the body problem is $0<A<\lambda h_0$ with
$h_0>1/\lambda$ for MHR bodies, and $A>0$ with $\phi_0>0$ for regular bodies.
For the corresponding full classes, $\inf_rW_{\rm body}(r)>0$.
The minimax formula therefore gives the exact frontier
\[
 \Delta(\mathcal A)=0\quad\Longleftrightarrow\quad
 B_\lambda\ge J_{\mathrm{body}},
\]
which is $B_\lambda\ge\phi_0p^n$ for regular bodies.
A sufficiently profitable tail makes learning the body unnecessary.
The minimax value depends on $(A,h_0,B_\lambda)$; the maximizing tail
reserve itself still comes from the full observed revenue curve.

\section{Experimental Details and Additional Results}\label{app:experiments}

This appendix gives the rule definitions, experimental settings and
numerical checks behind Section~\ref{sec:experiments}.

\subsection{Implementation and regret evaluation}\label{app:exp-methods}

The projected plug-in uses $M=12$, $m_{\min}=20$ and $c_{\mathrm g}=0.25$,
in units where $\lambda=1$. Its boundary-density window has width
$b_N=c_bN^{-1/5}$. In Section~\ref{sec:experiments} and
Appendices~\ref{app:exp-truths} and~\ref{app:exp-real} we choose $c_b$ by the data-driven
rule of Corollary~\ref{cor:selected-bandwidth}; in the frontier and non-MHR studies below
we fix $c_b=\lambda/2$. The auto-selector uses
$h_{\min}=0.1/\lambda$ to bound its tail search at $10\lambda$. Neither
rule receives the true distribution parameters. Table~\ref{tab:exp-policies} defines
the comparators.

\begin{table}[htbp]
\centering\small
\caption{Reserve rules compared in Section~\ref{sec:experiments} and Appendix~\ref{app:experiments}.}
\label{tab:exp-policies}
\begin{tabular}{@{}p{0.17\textwidth}p{0.76\textwidth}@{}}
\toprule
Rule & How the reserve is chosen\\
\midrule
Full-info & The full-information reserve of the held-out eBay study:
the exact revenue maximizer on simulated auctions among the uncensored
training bidders. It sees the values that censoring hides from every
other rule and serves as the oracle row.\\
Plug-in & Algorithm~\ref{alg:plug-in}, at the projected boundary estimates.\\
Auto-selector & Algorithm~\ref{alg:pretest}: empirical tail rule on $[\lambda,10\lambda]$
(on $[\lambda,r_{\max}]$ with $r_{\max}$ the larger of $4\lambda$ and twice the largest observed record in the
opening-bid study of Section~\ref{sec:exp-real}) when the raw
$\hat h_0\le1/\lambda$; plug-in otherwise.\\
Plug-in (Weibull hazard) & Algorithm~\ref{alg:plug-in} with the boundary hazard
replaced by the censored Weibull fit's hazard at $\lambda$; the count
estimate $\hat A_N$ is kept.\\
Censored MLE & Maximum likelihood fit in a Weibull, gamma (shape at least
one) or lognormal family. Observed bids contribute their fitted density;
no-bid records contribute the fitted probability below $\lambda$.
Post the fitted distribution's optimal reserve. It is an
oracle MLE when the full distribution belongs to the fitted family.\\
Linear completion & The linear-completion reserve $r_c(\hat A_N)$ of \eqref{eq:chord-reserve}: it uses only
the censored fraction and is optimal when the body hazard is constant below $\lambda$.\\
Boundary & Post the historical reserve $\lambda$.\\
Drop & Maximize exact empirical revenue from observed bids only;
choose the largest maximizer.\\
\bottomrule
\end{tabular}
\end{table}

To approximate worst-case regret under MHR, we search a $400\times400$
grid of two-level bodies and refine it near those giving the largest
regret, with $H(\lambda)=A$ imposed exactly. For
$r>\lambda$, it uses
$W_G(r)=W_G(\lambda)+R_G(\lambda)-R_G(r)$. In identifiable instances the
comparison also includes the observed tail's optimal revenue. Each completed
distribution's revenue and optimal reserve are in closed form. Searching only finitely
many bodies can underestimate $W_G$ at a fixed reserve; searching only
finitely many reserves can raise the computed minimax regret.
Appendix~\ref{app:exp-oracles} distinguishes the proved error bound based
on grid spacing from the observed change under refinement.

Empirical revenues are evaluated as finite sums, without interpolation.
Revenues of the Rayleigh and Weibull
distributions and of every two-level completion have closed forms. The gamma
and lognormal fits use numerical integration.
Standard errors measure uncertainty in the reported sample averages,
with the distribution fixed. Comparisons are paired: each rule receives
the same sample in a replication.
Each study runs as numbered tasks, one per cell; task $k$ uses NumPy \texttt{default\_rng} with seed $20260929+1000k$.
Different studies reuse these task seeds, so their tables are not
independent.\label{app:exp-seeds}
All computations use Python 3.13 with NumPy, SciPy and pandas.

At $\lambda=0.6$ in Figure~\ref{fig:exp-syn-lambda}, mean worst-case regret
is $1.5\%$ of $R_G^*$ for the plug-in and $0.07\%$ for the auto-selector.
The plug-in's search below $\lambda$ misses the optimum in the observed tail.

\subsection{Four compatible distributions and the two-level family}\label{app:exp-truths}

\paragraph{Instance selection}
We selected the instance by a deterministic scan over Weibull shapes
$k\in\{1.5,2,2.5,3,4\}$, scale $1$, and censored fraction
$\alpha\in\{0.3,0.4,\ldots,0.9\}$, with $\lambda=(-\log(1-\alpha))^{1/k}$.
The numerical gap $W_G(r_G^*)-\Delta$ ranges from $0.1$ to $1.4$
percentage points of $R_G^*$ over the nonidentifiable cells.
We chose $k=2.5$ and $\alpha=1/2$, where the gap is $1.14$ points.

\begin{figure}[htbp]
\centering\includegraphics[width=0.6\textwidth]{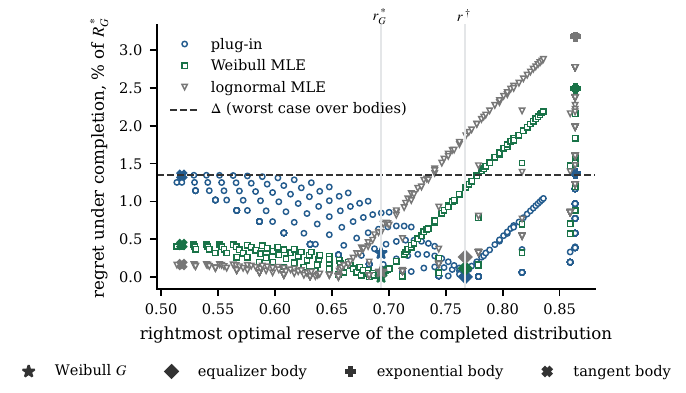}
\caption{Mean actual regret under $441$ two-level completions and the equalizer
completion of Weibull$(2.5,1)$, censored at $\lambda=(\log2)^{2/5}$
($n=3$, $N=3000$, $200$ paired replications).
Each completion is compared with its own optimal reserve; percentages
divide by $R_G^*$. Actual regret can fall below $\Delta$.}
\label{fig:exp-truths-bodies}
\end{figure}

\paragraph{The two-level family}
With $c=A/\lambda$, the worst case of every reserve is attained at a
two-level completion $H(x)=\max\{ax,\,A-b(\lambda-x)\}$, $a\in[0,c]$,
$b\in[c,h_0]$ (Theorem~\ref{thm:two-level}). We evaluate every posted
reserve under a $21\times21$ grid of them, $441$ bodies including the
exponential body ($a=b=c$) and the tangent ($a=0$, $b=h_0$), and under the
equalizer body, which has $a=0$ and $b=1/r^\dagger$. Each completion keeps $G$'s
hazard above $\lambda$; its revenue, optimal reserve and regret are in closed
form (Appendix~\ref{app:exp-methods}), and under $G$ itself the Weibull
closed form applies. Figure~\ref{fig:exp-truths-bodies} plots the mean actual regret of the
plug-in and of the Weibull and lognormal MLEs under every completion as a function of its rightmost optimal
reserve, at $N=3000$. A zero-hazard initial segment contains no bidder
values and gives flat revenue on
$[0,\tau]$; when this interval contains optimal reserves, we plot its
rightmost endpoint $\tau$.

For the four distributions of Table~\ref{tab:exp-truths}, $A=\log2$ and
$h_0=2.5\lambda^{3/2}\approx2.01$. The numerical minimax reserve is
$r^\dagger\approx0.767$, whereas $G$'s own optimal reserve is
$r_G^*=2.5^{-2/5}\approx0.693$. The exponential body has hazard
$A/\lambda$ throughout $[0,\lambda]$ and optimal reserve $\lambda$
since $A<1$. The equalizer body has hazard $0$ on $[0,\tau)$ and
$b=1/r^\dagger$ on $[\tau,\lambda]$, with
$\tau=\lambda-A/b\approx0.33$, making $r^\dagger$ optimal. The tangent
body has hazard $0$ below $\lambda-A/h_0\approx0.52$ and $h_0$ up to
$\lambda$. All three retain $G$'s hazard above $\lambda$.

The plug-in's mean actual regret is at most $1.36\%$ on the tested body grid.
Its larger worst-case regret, $1.46\%$ in Table~\ref{tab:exp-truths}, averages
each sample's worst case over a finer grid; the maximizing body need
not be the same across samples.

\paragraph{Smoothness of the hidden body}
The exponential body's hazard jumps from $c\approx0.80$ to
$h_0\approx2.01$ at $\lambda$, and the equalizer body has no values
below $\tau\approx0.33$. Both are MHR. Our learning assumption requires
smoothness only of the observed boundary density, not the hidden hazard.
Imposing a smoothness bound on the hidden hazard would shrink the
compatible class and could reduce minimax regret. Continuity alone
still permits steep approximations to a jump with the same limiting regret.

\paragraph{Rate and bandwidth on this instance}
\begin{figure}[htbp]
  \centering\includegraphics[width=\textwidth]{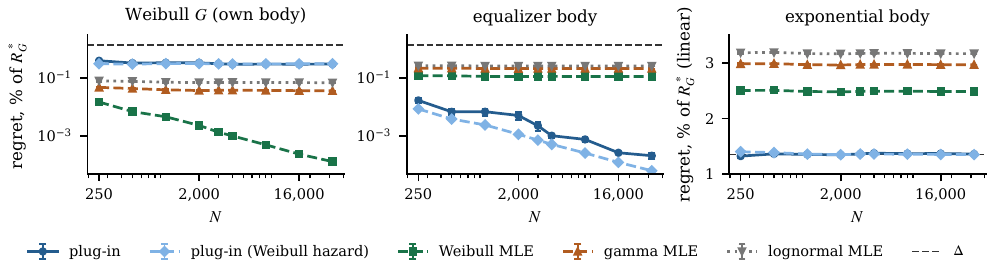}
  \caption{Mean actual regret under three compatible completions of Weibull$(2.5,1)$,
  censored at its median ($n=3$, $200$ paired replications per $N$).
  Percentages use $R_G^*$. Each rule posts the same reserves across panels.
  Plug-in (Weibull hazard) is the plug-in with a fitted Weibull boundary hazard. The right panel has a linear axis. Bars show one Monte-Carlo standard error.
  Actual regret can fall below $\Delta$.}
  \label{fig:exp-truths-rate}
\end{figure}

The plug-in's mean excess worst-case regret $W_G(r^\pi)-\Delta$ falls from $0.33$ to $0.04$
percentage points between $N=250$ and $32{,}000$; with fixed
$c_b=\lambda/2$, it falls from $0.66$ to $0.07$ points.
At $N=3000$, the data-driven rule selects $c_b=2\lambda$ in $94\%$
of samples and gives $W_G=1.46\%$, compared with $1.59\%$ for the
fixed bandwidth. Under each of the four distributions of Table~\ref{tab:exp-truths}, their mean actual regrets
differ by at most $0.14$ points.

Figure~\ref{fig:exp-truths-rate} shows that increasing $N$ does not
resolve the hidden-body ambiguity. The Weibull MLE's mean actual regret falls below
$0.001\%$ under $G$, whereas the parametric fits level off at
$2.49$--$3.17\%$ under the exponential completion. The censored
likelihoods and fitted reserves are the same across these distributions
(Proposition~\ref{rem:parametric-limit}).
The equalizer body favors the plug-in by construction, just as $G$
favors the oracle MLE.

\subsection{Experiments on eBay bids: data and further results}\label{app:exp-real}

\begingroup \paragraph{Data}
The data of Jank and Shmueli~\citep{jank2010modeling} record $10{,}681$ bids in $628$
eBay auctions of three items: $343$ auctions of Palm Pilot M515
handhelds, $149$ of Xbox game consoles and $136$ of Cartier
wristwatches. Each bid carries the bidder's identity, the auction's
opening bid and its closing price. A bidder's record in an auction is
their highest proxy bid; there are $3{,}022$, $1{,}229$ and $922$ such
records. The opening bid is the lowest acceptable bid, so bidders with
lower values never appear in the history: $211$, $107$ and $96$ auctions
have opening bids above one dollar and form the naturally censored samples
of the opening-bid study in Section~\ref{sec:exp-real}. The remaining $132$, $42$ and $40$ auctions
have opening bids of at most one dollar and reveal every participating
bidder's record; they contain $1{,}652$, $460$ and $392$ records,
$9.8$ to $12.5$ per auction. eBay records the winner at the closing price,
the second-highest record plus one increment, which is a lower bound on
the winner's value. We replace each of these $132$, $42$ and $40$ records
by a draw from the Kaplan--Meier conditional distribution of the item's
records above its closing price: one draw with a fixed seed in the
held-out study, and $50$ draws in the opening-bid study. Between $75\%$ and
$85\%$ of the records are whole dollars. \endgroup

\begingroup \paragraph{The refit}
For each item we fit a Gaussian kernel density with Silverman's
bandwidth, $15.7$, $15.2$ and $182$ dollars. On $4000$ equally spaced
points of $[0,1.6\max_i b_i]$ we project the kernel hazard onto
nondecreasing sequences by pool-adjacent-violators with weights equal to
the survival probability. The projected hazard is constant between grid
points and held at its last value beyond the grid, so its cumulative
hazard is piecewise linear and its distribution function, revenue and
inverse sampling rule are in closed form. This gives a continuous MHR
distribution $G$ with finite mean; it imposes MHR rather than testing it.
The boundaries are the whole-dollar quantiles $0.5$, $0.7$ and $0.8$ of
$G$. Table~\ref{tab:exp-real-cells-def} lists the nine cells. On the Xbox
refit the projected hazard is constant between $111$ and $141$ dollars,
so the two nonidentifiable cells share $h_0$; numerically they also have
the same minimax regret and, up to numerical error, the same minimax reserve. On the Cartier refit the hazard is
constant from below $555$ dollars, so $A$ is within $0.02$ of
$\lambda h_0$ at $\lambda=956$ and $1274$: these cells lie close to the fully
identified frontier and their minimax regret is below $0.01\%$.

\begin{table}[htbp]
\centering\small
\caption{The nine eBay cells ($n=3$). $G(\lambda)$ is the censored
fraction, $r_G^*$ the refit's optimal reserve, $\Delta$ the numerical
minimax regret in percent of $R_G^*$ and $r^\dagger$ the minimax
reserve; $\lambda$, $R_G^*$, $r_G^*$ and $r^\dagger$ are in dollars.}
\label{tab:exp-real-cells-def}
\begin{tabular}{lrcccclrrcr}
\toprule
Item & $\lambda$ & $G(\lambda)$ & $A$ & $h_0$ & $\lambda h_0$ & Regime & $R_G^*$ & $r_G^*$ & $\Delta$ & $r^\dagger$ \\
\midrule
Palm Pilot & $155$ & $0.50$ & $0.692$ & $0.0080$ & $1.23$ & nonident. & $156.74$ & $145.19$ & $0.45\%$ & $149.02$ \\
Palm Pilot & $199$ & $0.70$ & $1.202$ & $0.0169$ & $3.35$ & nonident. & $156.74$ & $145.19$ & $2.14\%$ & $143.38$ \\
Palm Pilot & $219$ & $0.80$ & $1.619$ & $0.0255$ & $5.59$ & nonident. & $156.74$ & $145.19$ & $3.36\%$ & $142.37$ \\
Xbox & $74$ & $0.50$ & $0.699$ & $0.0134$ & $0.99$ & ident. & $85.73$ & $74.19$ & $0$ & $74.19$ \\
Xbox & $111$ & $0.70$ & $1.204$ & $0.0137$ & $1.52$ & nonident. & $85.73$ & $74.19$ & $0.39\%$ & $84.18$ \\
Xbox & $141$ & $0.80$ & $1.614$ & $0.0137$ & $1.93$ & nonident. & $85.73$ & $74.19$ & $0.39\%$ & $84.17$ \\
Cartier & $555$ & $0.50$ & $0.694$ & $0.0013$ & $0.71$ & ident. & $717.26$ & $784.90$ & $0$ & $784.90$ \\
Cartier & $956$ & $0.70$ & $1.204$ & $0.0013$ & $1.22$ & nonident. & $717.26$ & $784.90$ & $<0.01\%$ & $789.30$ \\
Cartier & $1274$ & $0.80$ & $1.610$ & $0.0013$ & $1.62$ & nonident. & $717.26$ & $784.90$ & $<0.01\%$ & $790.18$ \\
\bottomrule
\end{tabular}
\end{table} \endgroup

\begingroup \paragraph{The shape assumption on the three items}
For each item we fit the unconstrained Gaussian kernel density at
bandwidths $0.5$, $1$ and $2$ times Silverman's, compute its hazard
$h=f/S$ on $400$ points, and resample the item's bidders with
replacement $2000$ times, keeping all of a bidder's records. On the
region where the full-sample survival lies between $0.05$ and $0.95$ we
form pointwise $90\%$ percentile bands and a simultaneous sup-$t$ band;
``significant decrease'' means some $x<y$ with the upper band at $y$ below
the lower band at $x$, and only the simultaneous band gives a
$10\%$-level test of a nondecreasing smoothed hazard. The prespecified
contrast $C_\lambda$ is the mean hazard on the far half of
$[\lambda,x_S]$ minus that on the near half, with $x_S$ the point where
the unscaled fit has survival $0.05$; a negative value is a decrease.
Table~\ref{tab:exp-real-hazard} reports the diagnostic and
Table~\ref{tab:exp-real-hazard-floors} the boundary scalars and minimax
regret of the kernel fit next to those of the refit.

On Palm Pilot no bandwidth shows a decrease, and every contrast is
positive with an interval excluding zero. On Xbox every bandwidth shows a
significant decrease: the hazard peaks near $115$ dollars and falls
beyond, and the contrast at $\lambda=74$ is negative in at least $99\%$ of
the draws at every bandwidth. On Cartier the two narrower bandwidths show a pointwise decrease
and the kernel fit has $A>\lambda h_0$ at every boundary, so no MHR
completion of the unconstrained fit exists. The refit imposes the shape
on these two items against their data; on Palm Pilot the kernel fit's
minimax regret is within $0.06$ points of the refit's.

\begin{table}[htbp]
\centering\scriptsize
\setlength{\tabcolsep}{2.5pt}
\caption{Kernel-hazard diagnostic on the reference records ($2000$
bidder-bootstrap draws). ``bw$\times$'' is the multiple of Silverman's
bandwidth; ``Peak'' is the location of the full-sample hazard's maximum
with its $90\%$ bootstrap interval; the contrasts are at the item's three
boundaries of Table~\ref{tab:exp-real-cells-def}.}
\label{tab:exp-real-hazard}
\begin{tabular}{lcccl*{3}{l}}
  \toprule
  & & \multicolumn{2}{c}{Significant decrease} & & \multicolumn{3}{c}{Contrast $10^3C_\lambda$ (90\% interval)} \\
  \cmidrule(lr){3-4}\cmidrule(lr){6-8}
  Item & bw$\times$ & pointwise & sup-$t$ & Peak (90\%) & $\lambda_1$ & $\lambda_2$ & $\lambda_3$ \\
  \midrule
    Palm Pilot & $0.5$ & no & no & $252$ {\scriptsize[247, 255]} & $24.2$ {\scriptsize[20.8, 28.0]} & $21.0$ {\scriptsize[15.6, 26.5]} & $13.6$ {\scriptsize[7.5, 19.6]} \\
    Palm Pilot & $1$ & no & no & $255$ {\scriptsize[252, 259]} & $20.4$ {\scriptsize[18.0, 23.0]} & $16.4$ {\scriptsize[13.6, 19.4]} & $11.5$ {\scriptsize[9.0, 14.2]} \\
    Palm Pilot & $2$ & no & no & $266$ {\scriptsize[264, 269]} & $12.9$ {\scriptsize[12.0, 13.8]} & $9.3$ {\scriptsize[8.6, 10.0]} & $6.6$ {\scriptsize[6.1, 7.1]} \\
    Xbox & $0.5$ & yes & yes & $120$ {\scriptsize[103, 205]} & $-5.1$ {\scriptsize[$-$8.6, $-$1.4]} & $0.5$ {\scriptsize[$-$3.5, 4.9]} & $4.0$ {\scriptsize[0.5, 8.2]} \\
    Xbox & $1$ & yes & yes & $113$ {\scriptsize[106, 119]} & $-5.4$ {\scriptsize[$-$8.7, $-$2.0]} & $-1.3$ {\scriptsize[$-$4.8, 2.6]} & $2.1$ {\scriptsize[$-$1.0, 5.6]} \\
    Xbox & $2$ & yes & yes & $115$ {\scriptsize[108, 122]} & $-5.1$ {\scriptsize[$-$7.7, $-$2.4]} & $-3.6$ {\scriptsize[$-$6.4, $-$0.7]} & $-1.2$ {\scriptsize[$-$3.6, 1.5]} \\
    Cartier & $0.5$ & yes & yes & $1874$ {\scriptsize[1555, 1944]} & $1.2$ {\scriptsize[0.7, 1.7]} & $1.4$ {\scriptsize[0.9, 2.1]} & $0.8$ {\scriptsize[0.1, 1.6]} \\
    Cartier & $1$ & yes & no & $1843$ {\scriptsize[1672, 1929]} & $1.0$ {\scriptsize[0.6, 1.4]} & $1.1$ {\scriptsize[0.7, 1.6]} & $0.8$ {\scriptsize[0.4, 1.3]} \\
    Cartier & $2$ & no & no & $1929$ {\scriptsize[1796, 2022]} & $0.5$ {\scriptsize[0.2, 0.7]} & $0.5$ {\scriptsize[0.3, 0.7]} & $0.4$ {\scriptsize[0.2, 0.6]} \\
  \bottomrule
\end{tabular}
\end{table}

\begin{table}[htbp]
\centering\small
\caption{Boundary scalars, regime and minimax regret (percent of the
fit's own $R_G^*$) of the unconstrained kernel fit and of the isotonic
refit at the nine boundaries. ``Infeasible'' means $A>\lambda h_0$, so no
MHR completion exists.}
\label{tab:exp-real-hazard-floors}
\begin{tabular}{lr*{2}{cccc}}
  \toprule
  & & \multicolumn{4}{c}{Kernel fit} & \multicolumn{4}{c}{Isotonic MHR refit} \\
  \cmidrule(lr){3-6}\cmidrule(lr){7-10}
  Item & $\lambda$ & $A$ & $h_0$ & Regime & $\Delta$ & $A$ & $h_0$ & Regime & $\Delta$ \\
  \midrule
    Palm Pilot & $155$ & $0.72$ & $0.0080$ & nonident. & $0.46\%$ & $0.69$ & $0.0080$ & nonident. & $0.45\%$ \\
    Palm Pilot & $199$ & $1.23$ & $0.0169$ & nonident. & $2.16\%$ & $1.20$ & $0.0169$ & nonident. & $2.14\%$ \\
    Palm Pilot & $219$ & $1.64$ & $0.0255$ & nonident. & $3.42\%$ & $1.62$ & $0.0255$ & nonident. & $3.36\%$ \\
    Xbox & $74$ & $0.75$ & $0.0134$ & ident. & $0$ & $0.70$ & $0.0134$ & ident. & $0$ \\
    Xbox & $111$ & $1.40$ & $0.0209$ & nonident. & $1.51\%$ & $1.20$ & $0.0137$ & nonident. & $0.39\%$ \\
    Xbox & $141$ & $1.93$ & $0.0111$ & infeasible & --- & $1.61$ & $0.0137$ & nonident. & $0.39\%$ \\
    Cartier & $555$ & $0.92$ & $0.0012$ & infeasible & --- & $0.69$ & $0.0013$ & ident. & $0$ \\
    Cartier & $956$ & $1.34$ & $0.0010$ & infeasible & --- & $1.20$ & $0.0013$ & nonident. & $<0.01\%$ \\
    Cartier & $1274$ & $1.68$ & $0.0012$ & infeasible & --- & $1.61$ & $0.0013$ & nonident. & $<0.01\%$ \\
  \bottomrule
\end{tabular}
\end{table} \endgroup

\begingroup \paragraph{The synthetic design on the refits}\label{app:exp-real-mirror}
The synthetic experiments of Sections~\ref{sec:exp-shape}
to~\ref{sec:exp-truths} need a known distribution, since worst-case
regret, the minimax regret $\Delta$ and a convergence rate are defined
relative to it. To repeat them on the eBay data we take the refit $G$ of
an item as the truth, draw training samples from $G$ censored at
$\lambda$, and score every rule against $G$ and its compatible
completions, exactly as in the synthetic design; $N$ is then a simulation
size rather than a count of recorded bids, and MHR is imposed on the
fitted law rather than tested. Table~\ref{tab:exp-real-shape-all} repeats
Table~\ref{tab:exp-shape} on the nine cells, Figure~\ref{fig:exp-real}
repeats Figure~\ref{fig:exp-syn} on the Palm Pilot refit, and
Table~\ref{tab:exp-real-truths} repeats Table~\ref{tab:exp-truths} on the
Palm Pilot cell at $\lambda=199$.

On the Palm Pilot cells, which lie far from full identification
($\lambda h_0-A$ is $0.54$, $2.15$ and $3.98$), the MHR minimax regret is
$0.45$--$3.36\%$ of $R_G^*$ and the minimax reserve's actual regret under
$G$ is at most $0.04\%$; under regular distributions the minimax regret is
$1.59$--$7.46\%$, and $9$--$33\%$ without a shape restriction, the ordering
of Table~\ref{tab:exp-shape}. The regular reduction of
Theorem~\ref{thm:regular-reduction} requires $\phi_0>0$, so it does
not apply on the two identifiable cells, where $\lambda h_0\le1$; there
the observed tail contains a shared optimum, so regular minimax regret is
zero (Proposition~\ref{prop:regular-summary}), as is MHR minimax regret
(Theorem~\ref{thm:frontier}). On the Xbox cell at $111$, $G$'s optimal reserve $74.19$ lies
well below $\lambda$ and the fitted hazard is nearly constant above it,
so the MHR minimax reserve $84.18$ pays almost its full minimax regret
under $G$ itself ($0.38\%$ of $0.39\%$). The Cartier cells at $956$ and
$1274$ lie close to the fully identified frontier, with minimax regret below
$0.01\%$.

\begin{table}[htbp]
\centering\small
\caption{Regret of the minimax reserves under three distribution classes
on the nine eBay cells ($n=3$), in percent of $R_G^*$, with exact censored
fraction and observed tail; the layout of Table~\ref{tab:exp-shape}.
Worst-case regret ranges over the compatible distributions in the
indicated class; actual regret uses the refit $G$. On the two identifiable
cells the observed tail contains a shared optimum, so regular minimax
regret is zero (Proposition~\ref{prop:regular-summary}).}
\label{tab:exp-real-shape-all}
\begin{tabular}{lrc*{3}{rr}}
\toprule
& & & \multicolumn{2}{c}{MHR} & \multicolumn{2}{c}{Regular} & \multicolumn{2}{c}{Unrestricted} \\
\cmidrule(lr){4-5}\cmidrule(lr){6-7}\cmidrule(lr){8-9}
Item & $\lambda$ & Regime & Worst-case & Actual & Worst-case & Actual & Worst-case & Actual \\
\midrule
Palm Pilot & $155$ & nonident. & $0.45\%$ & $0.04\%$ & $1.59\%$ & $0.03\%$ & $9.24\%$ & $1.22\%$ \\
Palm Pilot & $199$ & nonident. & $2.14\%$ & $0.01\%$ & $5.08\%$ & $<0.01\%$ & $24.54\%$ & $1.46\%$ \\
Palm Pilot & $219$ & nonident. & $3.36\%$ & $0.02\%$ & $7.46\%$ & $0.21\%$ & $33.35\%$ & $2.18\%$ \\
Xbox & $74$ & ident. & $0$ & $0$ & $0$ & $0$ & $8.21\%$ & $1.49\%$ \\
Xbox & $111$ & nonident. & $0.39\%$ & $0.38\%$ & $5.19\%$ & $0.18\%$ & $25.07\%$ & $0.66\%$ \\
Xbox & $141$ & nonident. & $0.39\%$ & $0.38\%$ & $6.91\%$ & $0.25\%$ & $39.20\%$ & $0.41\%$ \\
Cartier & $555$ & ident. & $0$ & $0$ & $0$ & $0$ & $7.74\%$ & $3.84\%$ \\
Cartier & $956$ & nonident. & $<0.01\%$ & $<0.01\%$ & $4.61\%$ & $0.12\%$ & $25.81\%$ & $2.14\%$ \\
Cartier & $1274$ & nonident. & $<0.01\%$ & $<0.01\%$ & $7.47\%$ & $0.06\%$ & $42.26\%$ & $1.36\%$ \\
\bottomrule
\end{tabular}
\end{table}

Figure~\ref{fig:exp-real-rate} shows excess worst-case regret against
$N$ on the three Palm Pilot cells. The plug-in's falls on every cell,
with fitted log-log slopes of $-0.41$, $-0.49$ and $-0.42$ (regression standard errors at most $0.03$), against the
$-0.4$ of Theorem~\ref{thm:finite-sample}; the censored Weibull MLE's and
the linear-completion rule's stay flat, since neither converges to the
minimax reserve, and by $N=2000$ the plug-in lies below both on every
cell. Figure~\ref{fig:exp-real-lambda} varies the boundary across the
Palm Pilot refit; the frontier is at $\lambda=r_G^*\approx145$ dollars,
and the picture matches Figure~\ref{fig:exp-syn-lambda}. The
auto-selector stays below $0.1\%$ on the identifiable side, where it
takes its tail branch in every sample at $\lambda\le128$ and in $58$ of
$80$ samples at $\lambda=142$, and tracks $\Delta$ within $0.33$ points
on the other side, where it takes the plug-in branch in all but one
sample. The Weibull MLE loses $1.0$--$1.7\%$ on the identifiable side and
sits $0.2$--$2.0$ points above $\Delta$ on the other; the
linear-completion rule nearly coincides with the plug-in on the
identifiable side, where both post reserves at or near $\lambda$, and peaks at $12.7\%$ at $\lambda=189$.

\begin{figure}[htbp]
  \centering
  \begin{subfigure}[b]{0.66\textwidth}
    \centering\includegraphics[width=\textwidth]{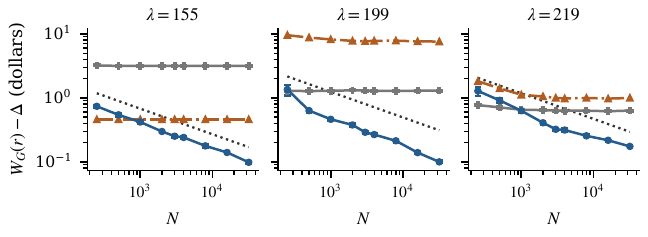}
    \caption{Excess worst-case regret against sample size, one panel per cell.}\label{fig:exp-real-rate}
  \end{subfigure}\hfill
  \begin{subfigure}[b]{0.32\textwidth}
    \centering\includegraphics[width=\textwidth]{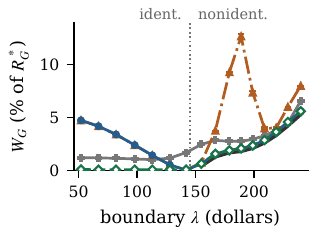}
    \caption{Regret across the frontier.}\label{fig:exp-real-lambda}
  \end{subfigure}
  \par\smallskip
  \includegraphics[width=0.92\textwidth]{plots/exp_syn_legend.pdf}
  \caption{\textbf{Learning the robust reserve on the Palm Pilot refit ($n=3$);
  the layout of Figure~\ref{fig:exp-syn}.}
  (a) Mean excess worst-case regret $W_G-\Delta$ in dollars over $200$
  replications on the three Palm Pilot cells: the plug-in, the censored Weibull MLE and the linear-completion rule.
  (b) Mean worst-case regret in percent of $R_G^*$ with $N=2000$ and $80$
  replications per $\lambda$, on a linear axis; the
  frontier is at $\lambda=r_G^*\approx145$ dollars. Bars show one
  Monte-Carlo standard error; some are smaller than the markers.}
  \label{fig:exp-real}
\end{figure}

Table~\ref{tab:exp-real-truths} repeats Table~\ref{tab:exp-truths} on
the Palm Pilot cell at $\lambda=199$, with the three completions of
Section~\ref{sec:exp-truths} built on $G$'s censored fraction and
observed tail. Here no MLE is an oracle: the refit is nonparametric, so
each family is misspecified under every completion. The plug-in's
worst-case regret is $2.32\%$ against the minimax regret $2.14\%$, and
its actual regret under $G$ is $0.01\%$. The three MLEs have worst-case
regret $2.70$--$2.96\%$ and actual regret $0.16$--$0.27\%$ under $G$.
The linear-completion reserve, which is optimal for the exponential
completion, loses $1.32\%$ under $G$ and $7.03\%$ in the worst case.
Posting the boundary loses $13.4\%$ under $G$, since $G$'s optimal
reserve $145$ lies well below $\lambda=199$.

\begin{table}[htbp]
  \centering\small
  \caption{Four distributions with the censored records of the Palm Pilot
  refit at $\lambda=199$; the layout of Table~\ref{tab:exp-truths}.
  Mean worst-case and actual regret in percent of $R_G^*$ ($N=3000$,
  $200$ paired replications). Actual regret compares each distribution
  with its own optimal revenue. Numerical minimax regret is
  $\Delta=2.14\%$; Monte-Carlo standard errors are at most $0.02$
  percentage points except in the Linear completion row ($0.12$). Rules
  that post the same reserves share a row.}
  \label{tab:exp-real-truths}
  \begin{tabular}{lccccc}
    \toprule
    & Worst-case regret & \multicolumn{4}{c}{Actual regret under each distribution} \\
    \cmidrule(lr){3-6}
    Rule & $W_G$ & Refit $G$ & Equalizer & Exponential & Tangent \\
    \midrule
    Plug-in, Auto-selector & $2.32$ & $0.01$ & $0.00$ & $0.39$ & $0.72$ \\
    Plug-in (Weibull hazard) & $2.46$ & $0.01$ & $0.01$ & $0.34$ & $0.92$ \\
    \addlinespace
    Weibull MLE & $2.96$ & $0.27$ & $0.09$ & $0.84$ & $0.04$ \\
    Gamma MLE & $2.85$ & $0.22$ & $0.07$ & $0.77$ & $0.08$ \\
    Lognormal MLE & $2.70$ & $0.16$ & $0.04$ & $0.69$ & $0.14$ \\
    \addlinespace
    Linear completion & $7.03$ & $1.32$ & $0.48$ & $0.01$ & $6.38$ \\
    Boundary & $25.02$ & $13.42$ & $3.04$ & $0.95$ & $25.02$ \\
    Drop & $25.13$ & $13.54$ & $3.15$ & $1.06$ & $25.13$ \\
    \bottomrule
  \end{tabular}
\end{table}

Table~\ref{tab:exp-real-cells} gives the worst-case regret and the
actual regret under $G$ of the plug-in and the three MLEs on the five
nonidentifiable Palm Pilot and Xbox cells at $N=3000$ and $N=32{,}000$.
The plug-in's worst-case regret falls on every cell; each MLE's stays
where it was, since the fit converges to the optimum of its own family
(Proposition~\ref{rem:parametric-limit}). On the Palm Pilot cells the
plug-in also has the lowest actual regret under $G$ at $\lambda=155$ and
$199$, and is within $0.02$ points of the best at $219$. The Xbox cells, where
the diagnostic above rejects MHR, are less favorable: at $N=3000$ the
Weibull MLE's worst-case regret is $0.52\%$ against the plug-in's
$0.69\%$ at $\lambda=111$, and its actual regret under $G$ is lower,
$0.27\%$ against $0.57\%$. The plug-in's worst-case regret keeps falling,
to $0.46\%$ at $N=32{,}000$ against $0.52\%$ for the MLE, while its actual
regret under $G$ settles at $0.40\%$, close to the $0.38\%$ cost of $r^\dagger$ itself
under $G$: on this cell the refit happens to favor a reserve near the
MLE's limit, another compatible body would reverse the comparison, and
the censored records cannot distinguish them.

\begin{table}[!tb]
\centering\small
\caption{Worst-case regret and actual regret under the refit $G$ on the
five nonidentifiable Palm Pilot and Xbox cells, in percent of $R_G^*$
($n=3$, mean over $200$ paired replications). Monte-Carlo standard
errors are at most $0.04$ points.}
\label{tab:exp-real-cells}
\begin{tabular}{lclcccc}
  \toprule
  Cell & $\Delta$ & Rule & \multicolumn{2}{c}{$N=3{,}000$} & \multicolumn{2}{c}{$N=32{,}000$} \\
  \cmidrule(lr){4-5}\cmidrule(lr){6-7}
  & & & $W_G$ & under $G$ & $W_G$ & under $G$ \\
  \midrule
    Palm Pilot, $\lambda=155$ & $0.45$ & Plug-in & $0.61$ & $0.06$ & $0.52$ & $0.04$ \\
     &  & Weibull MLE & $2.48$ & $0.70$ & $2.47$ & $0.70$ \\
     &  & Gamma MLE & $2.95$ & $1.02$ & $2.94$ & $1.01$ \\
     &  & Lognormal MLE & $3.08$ & $1.11$ & $3.07$ & $1.10$ \\
    \addlinespace
    Palm Pilot, $\lambda=199$ & $2.14$ & Plug-in & $2.32$ & $0.01$ & $2.20$ & $0.01$ \\
     &  & Weibull MLE & $2.96$ & $0.27$ & $2.96$ & $0.27$ \\
     &  & Gamma MLE & $2.85$ & $0.22$ & $2.85$ & $0.21$ \\
     &  & Lognormal MLE & $2.70$ & $0.16$ & $2.70$ & $0.15$ \\
    \addlinespace
    Palm Pilot, $\lambda=219$ & $3.36$ & Plug-in & $3.56$ & $0.02$ & $3.47$ & $0.01$ \\
     &  & Weibull MLE & $3.76$ & $0.12$ & $3.76$ & $0.11$ \\
     &  & Gamma MLE & $3.68$ & $0.01$ & $3.65$ & $<0.01$ \\
     &  & Lognormal MLE & $4.20$ & $0.01$ & $4.19$ & $0.01$ \\
    \addlinespace
    Xbox, $\lambda=111$ & $0.39$ & Plug-in & $0.69$ & $0.57$ & $0.46$ & $0.40$ \\
     &  & Weibull MLE & $0.52$ & $0.27$ & $0.52$ & $0.26$ \\
     &  & Gamma MLE & $0.71$ & $0.14$ & $0.72$ & $0.13$ \\
     &  & Lognormal MLE & $1.95$ & $0.18$ & $1.96$ & $0.18$ \\
    \addlinespace
    Xbox, $\lambda=141$ & $0.39$ & Plug-in & $0.58$ & $0.51$ & $0.44$ & $0.41$ \\
     &  & Weibull MLE & $0.54$ & $0.26$ & $0.53$ & $0.25$ \\
     &  & Gamma MLE & $0.72$ & $0.14$ & $0.72$ & $0.13$ \\
     &  & Lognormal MLE & $1.78$ & $0.11$ & $1.79$ & $0.11$ \\
  \bottomrule
\end{tabular}
\end{table} \endgroup

\begingroup \paragraph{Whole-dollar price points and the estimation window}
In the held-out study of Section~\ref{sec:exp-real}, $75\%$ of the Palm
Pilot records are whole dollars. The data-driven bandwidth rule selects
windows of $45$ to $69$ dollars holding $190$ to $380$ training records
at $\lambda=155$ and $219$, but a twelve-dollar window holding about
$100$ records at $\lambda=199$, where the MLEs are ahead. Jittering the training records by up to half a dollar, on the first $200$ splits, lowers the regret at $199$ from $1.02$ to $0.83\%$ for the fixed window and from $1.66$ to $1.44\%$ for the data-driven window (standard errors $0.07$--$0.08$), leaving their gap at about $0.6$ points, so the gap comes from the narrow window
rather than from the price points. \endgroup

\begingroup \paragraph{Held-out regret on Xbox and Cartier}
Table~\ref{tab:exp-ebay-holdout-other} repeats the held-out study of
Section~\ref{sec:exp-real} on the other two items, with $321$ or $322$ and $274$
training records. On Xbox the plug-in loses $1.8\%$ at $\lambda=74$, against $1.9$--$3.1\%$ for the MLEs, and $2.2\%$ at $111$, against $2.0\%$ for the gamma MLE and $2.3$ and $3.3\%$ for the Weibull and lognormal MLEs; the linear-completion reserve loses $1.3\%$ and $1.8\%$. At $\lambda=141$ the plug-in loses $22.6\%$
and the auto-selector $25\%$. The kernel hazard of the full sample is already
infeasible there (Table~\ref{tab:exp-real-hazard-floors}:
$\lambda h_0=1.57<A=1.93$), since the Xbox hazard peaks near $115$ dollars
and falls beyond; with only $43$ training records above $\lambda$, $23$ of
them in the selected $89$-dollar window, the estimated boundary hazard is
lower still. The projected estimates land on the face $\bar A_N=\lambda\bar h_N$ in every split, and the linear-completion reserve at the projected pair posts
$116$ dollars on average, while the auto-selector's tail search posts $194$.
The linear-completion reserve at the raw count, which needs no hazard estimate,
loses $2.1\%$. On Cartier every rule loses between $4.6\%$ and $14.6\%$ and
the full-information reserve (Full-info in Table~\ref{tab:exp-policies})
$8\%$: with $274$ records spread up to $3{,}103$
dollars, the test optimum is itself noisy. The auto-selector's $5.6\%$ at
$\lambda=555$ is below the full-information row, but the full-information
rule is itself estimated from a small training sample, so it need not have
the lowest held-out regret.

\begin{table}[htbp]
\centering\footnotesize
\setlength{\tabcolsep}{3pt}
\caption{Held-out regret on the recorded Xbox and Cartier bids, in
percent of each split's test optimum; the layout and experimental settings of
Table~\ref{tab:exp-ebay-holdout}. Standard errors at most $0.37$
points.}
\label{tab:exp-ebay-holdout-other}
\begin{tabular}{lrrccccccccc}
  \toprule
  & & & & & & \multicolumn{3}{c}{Censored MLE} & & & \\
  \cmidrule(lr){7-9}
  Item & $\lambda$ & $N$ & Full-info & Plug-in & Auto-selector & Weibull & Gamma & Lognormal & Linear compl. & Boundary & Drop \\
  \midrule
    Cartier & $555$ & $274$ & $8.16$ & $14.46$ & $5.60$ & $10.90$ & $11.85$ & $13.52$ & $14.60$ & $14.46$ & $7.53$ \\
    Cartier & $956$ & $274$ & $8.15$ & $9.19$ & $9.86$ & $11.11$ & $11.47$ & $14.04$ & $9.98$ & $4.56$ & $7.04$ \\
    Cartier & $1274$ & $274$ & $8.32$ & $9.47$ & $9.75$ & $10.49$ & $11.05$ & $12.27$ & $9.06$ & $6.38$ & $8.11$ \\
    Xbox & $74$ & $322$ & $1.09$ & $1.82$ & $1.88$ & $2.37$ & $1.93$ & $3.12$ & $1.33$ & $1.33$ & $1.83$ \\
    Xbox & $111$ & $321$ & $0.99$ & $2.22$ & $2.63$ & $2.31$ & $1.96$ & $3.25$ & $1.79$ & $17.84$ & $24.14$ \\
    Xbox & $141$ & $322$ & $1.07$ & $22.57$ & $25.24$ & $2.73$ & $1.86$ & $3.29$ & $2.05$ & $31.12$ & $23.86$ \\
  \bottomrule
\end{tabular}
\end{table} \endgroup

\begingroup \paragraph{Sellers' own reserves on Xbox and Cartier}
Table~\ref{tab:exp-real-natural-other} repeats the opening-bid study on
the other two items, with reference rates of $11.0$ and $9.8$ records per
auction. On Xbox the participation-based fraction is within $0.05$ of
$G(\lambda)$ at $50$ and $100$ and overstated at $10$; the auto-selector with
the data-driven bandwidth misclassifies the cell at $10$, where the
auto-selector with the fixed bandwidth takes the tail branch and loses
$0.7\%$ held out, and the cell at $100$ has $13$ records, below
$m_{\min}$, so the fallback posts the linear-completion reserve. On Cartier the
fraction is overstated at $10$ and $100$, the cell at $500$ is tail-optimal and lies close to the fully identified
frontier, where the auto-selector takes the
nonidentifiable branch (worst-case regret $2.80\%$ against $0.03\%$ for
the Weibull MLE), and the Weibull MLE has the lower regret in
two of the three cells.

\begin{table}[htbp]
\centering\scriptsize
\setlength{\tabcolsep}{2.5pt}
\caption{Sellers' own opening bids on Xbox and Cartier; the layout of
Table~\ref{tab:exp-real-natural}, with $\hat N$ from the item's own
reference rate. ``fallback'' marks fewer than $m_{\min}$ records.}
\label{tab:exp-real-natural-other}
\begin{tabular}{lrrrrcclcc*{2}{c}*{3}{c}}
  \toprule
  & & & & & & & & & & \multicolumn{2}{c}{Worst-case} & \multicolumn{3}{c}{Held-out} \\
  \cmidrule(lr){11-12}\cmidrule(lr){13-15}
  Item & $\lambda$ & $K$ & $m$ & $\hat N$ & $\hat\alpha$ & $G(\lambda)$ & Regime & Detected & $\Delta$ & Auto-selector & Weibull & Auto-selector & Weibull & Boundary \\
  \midrule
    Cartier & $10$ & $12$ & $75$ & $118$ & $0.36$ & $0.01$ & ident. & ident. & $0$ & $5.76$ & $0.48$ & $10.18$ & $6.81$ & $23.66$ \\
    Cartier & $100$ & $8$ & $39$ & $78$ & $0.50$ & $0.11$ & ident. & ident. & $0$ & $1.06$ & $3.89$ & $6.94$ & $14.18$ & $22.62$ \\
    Cartier & $500$ & $9$ & $49$ & $88$ & $0.44$ & $0.46$ & ident. & nonident. & $0$ & $2.80$ & $0.03$ & $12.01$ & $5.35$ & $12.01$ \\
    Xbox & $10$ & $13$ & $124$ & $142$ & $0.13$ & $0.05$ & ident. & nonident. & $0$ & $5.18$ & $0.05$ & $8.66$ & $1.24$ & $8.66$ \\
    Xbox & $50$ & $15$ & $112$ & $164$ & $0.32$ & $0.34$ & ident. & ident. & $0$ & $0.00$ & $0.11$ & $0.45$ & $1.39$ & $3.59$ \\
    Xbox & $100$ & $3$ & $13$ & $33$ & $0.61$ & $0.65$ & nonident. & fallback & $0.39$ & $2.47$ & $0.75$ & $5.66$ & $2.40$ & $5.66$ \\
  \bottomrule
\end{tabular}
\end{table} \endgroup

\begingroup \paragraph{The imputed bidder count}
Table~\ref{tab:exp-real-natural-count} reruns the Palm Pilot cells of
Table~\ref{tab:exp-real-natural} with the reference rate of $12.5$
records per auction replaced by $8$, $10$, $15$ and $20$, with the same
seeds, rules and replications; the rows at $12.5$ reproduce
Table~\ref{tab:exp-real-natural}. The auto-selector's
branch is unchanged at every rate in every cell except $\lambda=150$ at
rate $8$, where $12\%$ of the replications take the tail branch. A higher
rate overstates censoring and lowers the reserve, as
Remark~\ref{rem:count-error} predicts: at $\lambda=150$ from $149$
dollars at rate $8$ to $122$ at rate $20$, two to four dollars per record
in every nonidentifiable cell. Both the auto-selector and the Weibull fit use
the imputed count, and how their held-out regret moves with the rate
differs across cells: at $\lambda=150$ both do best at rate $8$, while at
$\lambda=240$ the Weibull fit's regret falls from $2.98\%$ at rate $8$ to
$1.04\%$ at rate $20$. The auto-selector stays below the Weibull fit in seven of the eight cells at
$12.5$ and $15$, in six at $10$, in five at $20$ and in four at $8$; at $50$ it
loses to the Weibull fit at every rate.
Rates below $10.8$ are ruled out by the data: at $\lambda=10$ the
opening-bid auctions themselves hold $10.8$ records each, so the count at
rates $8$ and $10$ is clipped to $m+1$, as is rate $8$ at $\lambda=50$.

\begin{table}[htbp]
\centering\small
\setlength{\tabcolsep}{4pt}
\caption{Sensitivity of the Palm Pilot opening-bid study to the imputed
bidder count: the reference rate (records per auction) replaced by $8$,
$10$, $15$ and $20$, with $12.5$ the rate of Table~\ref{tab:exp-real-natural}
($n=3$, $50$ draws of the winner imputation). Detected is the auto-selector's
majority branch, Reserve its mean posted reserve in dollars, and Auto-selector
and Weibull the held-out regret of the auto-selector and the censored Weibull
MLE in percent of the test optimum. $^\dagger$: $K$ times the rate is
below the $m$ observed records, so $\hat N=m+1$.}
\label{tab:exp-real-natural-count}
\begin{tabular}{rrrrcrrr}
\toprule
$\lambda$ & Rate & $\hat N$ & $\hat\alpha$ & Detected & Reserve & Auto-selector & Weibull \\
\midrule
    10 & $8$ & $348^\dagger$ & $0.00$ & ident. & $149$ & $0.15$ & $2.58$ \\
    10 & $10$ & $348^\dagger$ & $0.00$ & ident. & $149$ & $0.15$ & $2.58$ \\
    10 & $12.5$ & $400$ & $0.13$ & ident. & $149$ & $0.15$ & $2.52$ \\
    10 & $15$ & $480$ & $0.28$ & ident. & $149$ & $0.15$ & $1.25$ \\
    10 & $20$ & $640$ & $0.46$ & ident. & $149$ & $0.15$ & $21.05$ \\
  \addlinespace
    50 & $8$ & $74^\dagger$ & $0.01$ & nonident. & $50$ & $6.84$ & $1.58$ \\
    50 & $10$ & $90$ & $0.19$ & nonident. & $50$ & $6.84$ & $2.44$ \\
    50 & $12.5$ & $113$ & $0.35$ & nonident. & $49$ & $6.88$ & $3.39$ \\
    50 & $15$ & $135$ & $0.46$ & nonident. & $47$ & $6.95$ & $3.57$ \\
    50 & $20$ & $180$ & $0.59$ & nonident. & $44$ & $7.05$ & $3.79$ \\
  \addlinespace
    100 & $8$ & $72$ & $0.14$ & ident. & $150$ & $0.04$ & $2.14$ \\
    100 & $10$ & $90$ & $0.31$ & ident. & $150$ & $0.04$ & $3.14$ \\
    100 & $12.5$ & $113$ & $0.45$ & ident. & $150$ & $0.04$ & $4.18$ \\
    100 & $15$ & $135$ & $0.54$ & ident. & $150$ & $0.04$ & $4.22$ \\
    100 & $20$ & $180$ & $0.66$ & ident. & $150$ & $0.04$ & $5.15$ \\
  \addlinespace
    150 & $8$ & $112$ & $0.20$ & nonident. & $149$ & $0.73$ & $0.35$ \\
    150 & $10$ & $140$ & $0.36$ & nonident. & $144$ & $0.82$ & $1.65$ \\
    150 & $12.5$ & $175$ & $0.49$ & nonident. & $139$ & $1.17$ & $2.32$ \\
    150 & $15$ & $210$ & $0.57$ & nonident. & $132$ & $1.95$ & $3.10$ \\
    150 & $20$ & $280$ & $0.68$ & nonident. & $122$ & $2.27$ & $4.18$ \\
  \addlinespace
    175 & $8$ & $312$ & $0.54$ & nonident. & $150$ & $1.86$ & $1.94$ \\
    175 & $10$ & $390$ & $0.63$ & nonident. & $137$ & $1.43$ & $2.09$ \\
    175 & $12.5$ & $488$ & $0.70$ & nonident. & $126$ & $2.52$ & $2.95$ \\
    175 & $15$ & $585$ & $0.75$ & nonident. & $119$ & $2.49$ & $3.33$ \\
    175 & $20$ & $780$ & $0.81$ & nonident. & $107$ & $3.69$ & $4.08$ \\
  \addlinespace
    200 & $8$ & $64$ & $0.58$ & nonident. & $162$ & $2.51$ & $0.25$ \\
    200 & $10$ & $80$ & $0.66$ & nonident. & $150$ & $0.05$ & $1.15$ \\
    200 & $12.5$ & $100$ & $0.73$ & nonident. & $140$ & $0.94$ & $1.96$ \\
    200 & $15$ & $120$ & $0.78$ & nonident. & $132$ & $1.99$ & $2.11$ \\
    200 & $20$ & $160$ & $0.83$ & nonident. & $118$ & $2.64$ & $3.03$ \\
  \addlinespace
    215 & $8$ & $88$ & $0.68$ & nonident. & $164$ & $2.15$ & $1.84$ \\
    215 & $10$ & $110$ & $0.75$ & nonident. & $154$ & $1.90$ & $0.98$ \\
    215 & $12.5$ & $138$ & $0.80$ & nonident. & $145$ & $0.83$ & $1.63$ \\
    215 & $15$ & $165$ & $0.83$ & nonident. & $133$ & $1.88$ & $2.08$ \\
    215 & $20$ & $220$ & $0.87$ & nonident. & $124$ & $2.47$ & $2.45$ \\
  \addlinespace
    240 & $8$ & $112$ & $0.80$ & nonident. & $162$ & $2.62$ & $2.98$ \\
    240 & $10$ & $140$ & $0.84$ & nonident. & $154$ & $1.40$ & $2.41$ \\
    240 & $12.5$ & $175$ & $0.87$ & nonident. & $147$ & $0.40$ & $2.10$ \\
    240 & $15$ & $210$ & $0.90$ & nonident. & $141$ & $1.45$ & $1.63$ \\
    240 & $20$ & $280$ & $0.92$ & nonident. & $132$ & $1.93$ & $1.04$ \\
\bottomrule
\end{tabular}
\end{table} \endgroup

\begingroup \paragraph{Competition}
Table~\ref{tab:exp-real-competition} varies the number of bidders on the
Palm Pilot refit at its three boundaries and on the Xbox refit at
$\lambda=111$, with one censored sample of $N=3000$ per replication
scored at every $n$ ($40$ paired replications). The minimax regret
falls by a factor of $40$ to $165$ between $n=2$ and $n=10$ (before rounding),
and the plug-in's excess over it falls in proportion. The plug-in has the lowest worst-case regret in eight of the twelve Palm Pilot cells, two of them at $n=10$, where it leads the MLEs by at most $0.03$ points, and is within $0.16$ points of the best in the other four; on Xbox the
Weibull MLE stays slightly ahead up to $n=5$ and ties at $n=10$.

\begin{table}[htbp]
\centering\small
\setlength{\tabcolsep}{3pt}
\caption{Minimax regret and mean worst-case and actual regret against
the number of bidders, in percent of $R_G^*$ ($N=3000$, $40$ paired
replications; standard errors at most $0.11$ points for the plug-in and
the MLEs and $0.43$ for the linear-completion reserve).}
\label{tab:exp-real-competition}
\begin{tabular}{lrrc*{6}{c}cc}
  \toprule
  & & & & \multicolumn{6}{c}{Worst-case regret} & \multicolumn{2}{c}{Actual regret} \\
  \cmidrule(lr){5-10}\cmidrule(lr){11-12}
  Item & $\lambda$ & $n$ & $\Delta$ & Plug-in & Weibull & Gamma & Lognormal & Linear compl. & Boundary & Plug-in & Weibull \\
  \midrule
    Palm Pilot & $155$ & $2$ & $0.82$ & $1.15$ & $4.49$ & $5.43$ & $5.70$ & $1.35$ & $1.35$ & $0.09$ & $1.44$ \\
    Palm Pilot & $155$ & $3$ & $0.45$ & $0.62$ & $2.50$ & $2.98$ & $3.11$ & $0.75$ & $0.75$ & $0.06$ & $0.72$ \\
    Palm Pilot & $155$ & $5$ & $0.13$ & $0.17$ & $0.73$ & $0.85$ & $0.88$ & $0.22$ & $0.22$ & $0.02$ & $0.17$ \\
    Palm Pilot & $155$ & $10$ & $0.00$ & $0.01$ & $0.03$ & $0.03$ & $0.03$ & $0.01$ & $0.01$ & $0.00$ & $0.00$ \\
    Palm Pilot & $199$ & $2$ & $4.50$ & $4.95$ & $5.28$ & $5.05$ & $4.79$ & $16.40$ & $42.01$ & $0.20$ & $0.51$ \\
    Palm Pilot & $199$ & $3$ & $2.14$ & $2.34$ & $2.96$ & $2.85$ & $2.71$ & $7.07$ & $25.02$ & $0.01$ & $0.27$ \\
    Palm Pilot & $199$ & $5$ & $0.63$ & $0.70$ & $1.03$ & $1.00$ & $0.96$ & $1.83$ & $10.64$ & $0.02$ & $0.07$ \\
    Palm Pilot & $199$ & $10$ & $0.05$ & $0.05$ & $0.08$ & $0.08$ & $0.08$ & $0.11$ & $1.61$ & $0.01$ & $0.00$ \\
    Palm Pilot & $219$ & $2$ & $7.15$ & $7.60$ & $7.57$ & $8.70$ & $9.76$ & $8.20$ & $76.31$ & $0.10$ & $0.22$ \\
    Palm Pilot & $219$ & $3$ & $3.36$ & $3.56$ & $3.76$ & $3.74$ & $4.24$ & $4.03$ & $51.08$ & $0.02$ & $0.12$ \\
    Palm Pilot & $219$ & $5$ & $1.02$ & $1.09$ & $1.19$ & $1.07$ & $1.17$ & $1.26$ & $28.01$ & $0.00$ & $0.03$ \\
    Palm Pilot & $219$ & $10$ & $0.08$ & $0.09$ & $0.09$ & $0.08$ & $0.09$ & $0.09$ & $8.26$ & $0.00$ & $0.00$ \\
    Xbox & $111$ & $2$ & $0.60$ & $0.95$ & $0.76$ & $1.05$ & $3.13$ & $1.95$ & $6.92$ & $0.74$ & $0.47$ \\
    Xbox & $111$ & $3$ & $0.39$ & $0.60$ & $0.50$ & $0.69$ & $1.93$ & $1.27$ & $4.98$ & $0.48$ & $0.28$ \\
    Xbox & $111$ & $5$ & $0.16$ & $0.24$ & $0.21$ & $0.29$ & $0.72$ & $0.52$ & $2.52$ & $0.20$ & $0.10$ \\
    Xbox & $111$ & $10$ & $0.01$ & $0.02$ & $0.02$ & $0.03$ & $0.06$ & $0.05$ & $0.43$ & $0.02$ & $0.01$ \\
  \bottomrule
\end{tabular}
\end{table} \endgroup

\begingroup \paragraph{The winner's record}
Table~\ref{tab:exp-real-imputation} refits each item with the winner's
record treated in three ways: the Kaplan--Meier draw above the closing
price used throughout, the closing price itself, and dropping the
record. Palm Pilot's regimes and minimax reserves are insensitive to the
closing-price treatment: the
minimax regret at its three boundaries moves from $0.45$, $2.14$ and
$3.36\%$ to $0.47$, $2.27$ and $3.85\%$. Dropping the records lowers its
optimal revenue by $6\%$ and raises the minimax regret to $1.09$, $2.76$
and $4.68\%$. Xbox and Cartier are not: under the closing
price the Xbox cell at $74$ becomes nonidentifiable and the minimax
regret at $111$ rises from $0.39$ to $0.99\%$, and Cartier's optimal
revenue falls by $12\%$.

\begin{table}[htbp]
\centering\small
\setlength{\tabcolsep}{4pt}
\caption{The winner's right-censored record under three treatments:
sample size, optimal revenue and reserve of the refit, and at each of
the item's three boundaries the censored fraction, regime, minimax regret
(percent of $R_G^*$) and minimax reserve (dollars).}
\label{tab:exp-real-imputation}
\begin{tabular}{llrrrrcccc}
  \toprule
  Item & Winner's record & Records & $R_G^*$ & $r_G^*$ & $\lambda$ & $G(\lambda)$ & Regime & $\Delta$ & $r^\dagger$ \\
  \midrule
    Palm Pilot & Kaplan--Meier & $1652$ & $156.7$ & $145.2$ & $155$ & $0.50$ & nonident. & $0.45$ & $149$ \\
     & & & &  & $199$ & $0.70$ & nonident. & $2.14$ & $143$ \\
     & & & &  & $219$ & $0.80$ & nonident. & $3.36$ & $142$ \\
    \addlinespace
    Palm Pilot & closing price & $1652$ & $155.8$ & $145.1$ & $155$ & $0.50$ & nonident. & $0.47$ & $149$ \\
     & & & &  & $199$ & $0.71$ & nonident. & $2.27$ & $143$ \\
     & & & &  & $219$ & $0.82$ & nonident. & $3.85$ & $141$ \\
    \addlinespace
    Palm Pilot & dropped & $1520$ & $146.6$ & $138.0$ & $155$ & $0.54$ & nonident. & $1.09$ & $139$ \\
     & & & &  & $199$ & $0.76$ & nonident. & $2.76$ & $135$ \\
     & & & &  & $219$ & $0.87$ & nonident. & $4.68$ & $132$ \\
    \addlinespace
    Xbox & Kaplan--Meier & $460$ & $85.7$ & $74.2$ & $74$ & $0.50$ & ident. & $0$ & $74$ \\
     & & & &  & $111$ & $0.70$ & nonident. & $0.39$ & $84$ \\
     & & & &  & $141$ & $0.80$ & nonident. & $0.39$ & $84$ \\
    \addlinespace
    Xbox & closing price & $460$ & $80.2$ & $72.4$ & $74$ & $0.52$ & nonident. & $0.04$ & $74$ \\
     & & & &  & $111$ & $0.74$ & nonident. & $0.99$ & $74$ \\
     & & & &  & $141$ & $0.85$ & nonident. & $0.99$ & $74$ \\
    \addlinespace
    Xbox & dropped & $418$ & $75.0$ & $68.7$ & $74$ & $0.55$ & nonident. & $0.29$ & $71$ \\
     & & & &  & $111$ & $0.77$ & nonident. & $0.93$ & $69$ \\
     & & & &  & $141$ & $0.87$ & nonident. & $0.93$ & $69$ \\
    \addlinespace
    Cartier & Kaplan--Meier & $392$ & $717.3$ & $784.9$ & $555$ & $0.50$ & ident. & $0$ & $785$ \\
     & & & &  & $956$ & $0.70$ & nonident. & $<0.01$ & $789$ \\
     & & & &  & $1274$ & $0.80$ & nonident. & $<0.01$ & $790$ \\
    \addlinespace
    Cartier & closing price & $392$ & $634.2$ & $686.8$ & $555$ & $0.55$ & ident. & $0$ & $687$ \\
     & & & &  & $956$ & $0.75$ & nonident. & $<0.01$ & $694$ \\
     & & & &  & $1274$ & $0.84$ & nonident. & $<0.01$ & $693$ \\
    \addlinespace
    Cartier & dropped & $352$ & $585.8$ & $636.1$ & $555$ & $0.58$ & ident. & $0$ & $636$ \\
     & & & &  & $956$ & $0.77$ & nonident. & $<0.01$ & $640$ \\
     & & & &  & $1274$ & $0.86$ & nonident. & $<0.01$ & $640$ \\
  \bottomrule
\end{tabular}
\end{table} \endgroup

\subsection{Behavior near the zero-regret frontiers}\label{app:exp-bracket}

Percentages in this subsection use the generating distribution's
optimal revenue $R_G^*$.

\begingroup \paragraph{Near full identification}\label{app:exp-knife}
As the boundary pair approaches the zero-regret frontier
$A=\lambda h_0$, the linear-completion reserve overtakes the plug-in
(Table~\ref{tab:exp-knife}). The distributions have $\lambda=1.5$,
$h_0=2$ and $A=3(1-\zeta)$, with constant body hazard $A/\lambda$ and
tail hazard $h_0$. Reducing $\zeta$ from $0.5$ to $0.005$ reduces the
minimax regret from $2.376\%$ to $0.001\%$. At the two smallest
distances, $\zeta=0.01$ and $0.005$, the plug-in's mean worst-case regret
is $0.558\%$ and $0.803\%$, the linear-completion reserve's $0.043\%$ and $0.023\%$.
At $N=3000$ the error in the boundary-hazard estimate still dominates,
whereas the linear-completion reserve depends only on the censored count, as
Proposition~\ref{prop:edge} predicts. For $\zeta\le0.1$, the raw estimate
lies beyond the frontier in $40$--$67\%$ of the replications, and the plug-in then posts the linear-completion reserve
at the projected pair, $r_c(\bar A_N)$. The plug-in's grid error at the
true $(A,h_0)$ is below $2\times10^{-4}$ revenue units, so its remaining
regret is estimation error. The auto-selector coincides with the plug-in except
at $\zeta=0.02$, where in one replication the raw $\hat h_0$ fell below
$1/\lambda$ and the tail branch cost about $48$ percentage points.
\begin{table}[htbp]
\centering\small
\setlength{\tabcolsep}{4pt}
\caption{Worst-case regret of the plug-in, the auto-selector and the linear-completion reserve in percent of $R_G^*$, mean (Monte-Carlo standard error), for two-level MHR distributions approaching full identification: $\lambda=1.5$, $h_0=2$, $A=h_0\lambda(1-\zeta)$ ($n=3$, $N=3000$, $60$ paired replications per $\zeta$). ``Frontier'' is the percentage of replications in which the raw estimate lies beyond the zero-regret frontier $A=\lambda h_0$ and the plug-in posts the linear-completion reserve at the projected pair, $r_c(\bar A_N)$. The minimax regret $\Delta$ has body-grid error at most $8.3\times10^{-4}$ percentage points.}
\label{tab:exp-knife}
\begin{tabular}{lccccccc}
\toprule
$\zeta$ & $G(\lambda)$ & $r^\dagger$ & $\Delta$ & Plug-in & Auto-selector & Linear compl. & Frontier \\
\midrule
$0.500$ & $0.78$ & $0.949$ & $2.376\%$ & $2.736$\,{\scriptsize(0.038)} & $2.736$\,{\scriptsize(0.038)} & $3.737$\,{\scriptsize(0.090)} & $0\%$ \\
$0.300$ & $0.88$ & $0.735$ & $1.904\%$ & $2.454$\,{\scriptsize(0.059)} & $2.454$\,{\scriptsize(0.059)} & $2.322$\,{\scriptsize(0.031)} & $2\%$ \\
$0.200$ & $0.91$ & $0.646$ & $1.236\%$ & $1.979$\,{\scriptsize(0.132)} & $1.979$\,{\scriptsize(0.132)} & $1.620$\,{\scriptsize(0.030)} & $23\%$ \\
$0.100$ & $0.93$ & $0.572$ & $0.416\%$ & $1.189$\,{\scriptsize(0.119)} & $1.189$\,{\scriptsize(0.119)} & $0.616$\,{\scriptsize(0.018)} & $40\%$ \\
$0.050$ & $0.94$ & $0.537$ & $0.119\%$ & $0.880$\,{\scriptsize(0.113)} & $0.880$\,{\scriptsize(0.113)} & $0.223$\,{\scriptsize(0.012)} & $47\%$ \\
$0.020$ & $0.95$ & $0.515$ & $0.020\%$ & $0.758$\,{\scriptsize(0.167)} & $1.558$\,{\scriptsize(0.913)} & $0.073$\,{\scriptsize(0.006)} & $47\%$ \\
$0.010$ & $0.95$ & $0.507$ & $0.005\%$ & $0.558$\,{\scriptsize(0.120)} & $0.558$\,{\scriptsize(0.120)} & $0.043$\,{\scriptsize(0.006)} & $67\%$ \\
$0.005$ & $0.95$ & $0.504$ & $0.001\%$ & $0.803$\,{\scriptsize(0.113)} & $0.803$\,{\scriptsize(0.113)} & $0.023$\,{\scriptsize(0.003)} & $57\%$ \\
\bottomrule
\end{tabular}
\end{table}

At the population level, the linear-completion reserve's worst-case regret stays
below the quadratic bound of Proposition~\ref{prop:edge} as $A$
approaches $\lambda h_0$ (Table~\ref{tab:exp-edge-chord}): the ratio of
$\mathsf W(r_c)$ to the bound lies between $0.10$ and $0.93$ in every
row. The ratio $\mathsf W(r_c)/\Delta$ falls from $5$--$6$ at
$A=\lambda h_0/2$ to $1.1$ at $A=0.99\,\lambda h_0$ when $\lambda h_0=2$,
and rises from $1.3$--$1.4$ to $2.2$ when $\lambda h_0=4$.
\begin{table}[htbp]
\centering\small
\caption{Worst-case regret $\mathsf W(r_c)$ of the linear-completion reserve and minimax regret $\Delta$ at the population level, in revenue units, as $A$ approaches $\lambda h_0$ ($\lambda=1$). $w=A(\lambda h_0-A)/(\lambda h_0)$ is the largest gap between the chord and the backward tangent; $\Delta$ uses $1600\times1600$ bodies (body-grid error at most $5.1\times10^{-6}$); ``bound'' is $k_1d^2/(2h_0)$ of Proposition~\ref{prop:edge} with $d=\lambda h_0-A$ and $k_1=((n-1)/n)^{n-1}$.}
\label{tab:exp-edge-chord}
\begin{tabular}{rrrrrrrrr}
\toprule
$n$ & $\lambda h_0$ & $A/(\lambda h_0)$ & $w$ & $\Delta$ & $\mathsf W(r_c)$ & $\mathsf W(r_c)/\Delta$ & bound & $\mathsf W(r_c)$/bound \\
\midrule
$2$ & $2$ & $0.50$ & $0.500$ & $0.02177$ & $0.11574$ & $5.3$ & $0.12500$ & $0.93$ \\
$2$ & $2$ & $0.80$ & $0.320$ & $0.00430$ & $0.00775$ & $1.8$ & $0.02000$ & $0.39$ \\
$2$ & $2$ & $0.90$ & $0.180$ & $0.00111$ & $0.00150$ & $1.3$ & $0.00500$ & $0.30$ \\
$2$ & $2$ & $0.95$ & $0.095$ & $0.00028$ & $0.00033$ & $1.2$ & $0.00125$ & $0.26$ \\
$2$ & $2$ & $0.99$ & $0.020$ & $0.00001$ & $0.00001$ & $1.1$ & $0.00005$ & $0.24$ \\
$2$ & $4$ & $0.50$ & $1.000$ & $0.02612$ & $0.03417$ & $1.3$ & $0.25000$ & $0.14$ \\
$2$ & $4$ & $0.80$ & $0.640$ & $0.00774$ & $0.01263$ & $1.6$ & $0.04000$ & $0.32$ \\
$2$ & $4$ & $0.90$ & $0.360$ & $0.00215$ & $0.00393$ & $1.8$ & $0.01000$ & $0.39$ \\
$2$ & $4$ & $0.95$ & $0.190$ & $0.00056$ & $0.00111$ & $2.0$ & $0.00250$ & $0.45$ \\
$2$ & $4$ & $0.99$ & $0.040$ & $0.00002$ & $0.00005$ & $2.2$ & $0.00010$ & $0.50$ \\
$3$ & $2$ & $0.50$ & $0.500$ & $0.01381$ & $0.08434$ & $6.1$ & $0.11111$ & $0.76$ \\
$3$ & $2$ & $0.80$ & $0.320$ & $0.00356$ & $0.00621$ & $1.7$ & $0.01778$ & $0.35$ \\
$3$ & $2$ & $0.90$ & $0.180$ & $0.00099$ & $0.00131$ & $1.3$ & $0.00444$ & $0.29$ \\
$3$ & $2$ & $0.95$ & $0.095$ & $0.00026$ & $0.00030$ & $1.2$ & $0.00111$ & $0.27$ \\
$3$ & $2$ & $0.99$ & $0.020$ & $0.00001$ & $0.00001$ & $1.1$ & $0.00004$ & $0.25$ \\
$3$ & $4$ & $0.50$ & $1.000$ & $0.01644$ & $0.02316$ & $1.4$ & $0.22222$ & $0.10$ \\
$3$ & $4$ & $0.80$ & $0.640$ & $0.00538$ & $0.00982$ & $1.8$ & $0.03556$ & $0.28$ \\
$3$ & $4$ & $0.90$ & $0.360$ & $0.00178$ & $0.00336$ & $1.9$ & $0.00889$ & $0.38$ \\
$3$ & $4$ & $0.95$ & $0.190$ & $0.00050$ & $0.00100$ & $2.0$ & $0.00222$ & $0.45$ \\
$3$ & $4$ & $0.99$ & $0.040$ & $0.00002$ & $0.00005$ & $2.2$ & $0.00009$ & $0.53$ \\
\bottomrule
\end{tabular}
\end{table}

The auto-selector's excess worst-case regret is small on both sides of both
zero-regret frontiers (Table~\ref{tab:exp-frontier}). The $21$ cells
use Weibull and two-level distributions at $\lambda=1$, with $200$
replications at $N=1000$ and $10{,}000$. The largest excess over the
cells falls from $1.33\%$ to $0.21\%$ as $N$ increases. At
$\lambda h_0=1$ the auto-selector's branch disagrees with the true regime in
$46$--$54\%$ of the replications, yet a wrong branch at a frontier costs
little because the two candidate reserves are close: branch accuracy
and regret are different quantities. At $N=10{,}000$ the tail rule has
the smallest excess of the auto-selector, the plug-in and the tail rule in $9$
of the $11$ tail-optimal cells, but it loses $5.4$--$5.8\%$ at
$\lambda h_0=1.5$; the auto-selector's plug-in branch prevents this
loss. On the identifiable side, projection moves every estimate to the
frontier, so the plug-in's excess falls only from $0.139\%$ to $0.121\%$
at $\lambda h_0=0.85$, while the auto-selector's falls from $0.118\%$ to
$0.027\%$.
\begin{table}[htbp]
\centering\footnotesize
\setlength{\tabcolsep}{3pt}
\caption{Excess worst-case regret $W_G(r)-\Delta$ of four rules in percent of $R_G^*$, mean (Monte-Carlo standard error), in cells on both sides of the zero-regret frontiers at $\lambda=1$ ($n=3$, $200$ paired replications per cell and $N$). ``Side'' is the case of the true boundary pair $(A,h_0)$: tail-optimal or fully identified, both in the identifiable regime, or nonidentifiable; ``Misclass.'' is the percentage of replications in which the auto-selector's branch disagrees with that regime (binomial standard error). Tail is the empirical revenue maximizer on $[\lambda,10\lambda]$. In identifiable cells $\Delta=0$ and $W_G$ includes the tail gain \eqref{eq:tail-gain}. Entries without a standard error are constant across replications.}
\label{tab:exp-frontier}
\begin{tabular}{lrrlrccccc}
\toprule
Family & $\lambda h_0$ & $A/(\lambda h_0)$ & Side & $N$ & Misclass. & Plug-in & Auto-selector & Tail & Linear compl. \\
\midrule
Weibull($k{=}2$) & $0.85$ & $0.5$ & tail-optimal & $1{,}000$ & $22.0$\,{\scriptsize(2.9)} & $0.139$\,{\scriptsize(0.005)} & $0.118$\,{\scriptsize(0.011)} & $0.090$\,{\scriptsize(0.011)} & $0.120$\,{\scriptsize(0.000)} \\
Weibull($k{=}2$) & $0.85$ & $0.5$ & tail-optimal & $10{,}000$ & $3.0$\,{\scriptsize(1.2)} & $0.121$\,{\scriptsize(0.000)} & $0.027$\,{\scriptsize(0.003)} & $0.023$\,{\scriptsize(0.003)} & $0.120$\,{\scriptsize(0.000)} \\
Weibull($k{=}2$) & $0.9$ & $0.5$ & tail-optimal & $1{,}000$ & $34.5$\,{\scriptsize(3.4)} & $0.094$\,{\scriptsize(0.008)} & $0.122$\,{\scriptsize(0.016)} & $0.091$\,{\scriptsize(0.015)} & $0.055$ \\
Weibull($k{=}2$) & $0.9$ & $0.5$ & tail-optimal & $10{,}000$ & $11.5$\,{\scriptsize(2.3)} & $0.056$\,{\scriptsize(0.001)} & $0.020$\,{\scriptsize(0.002)} & $0.016$\,{\scriptsize(0.001)} & $0.055$ \\
Weibull($k{=}2$) & $0.95$ & $0.5$ & tail-optimal & $1{,}000$ & $36.5$\,{\scriptsize(3.4)} & $0.072$\,{\scriptsize(0.009)} & $0.127$\,{\scriptsize(0.013)} & $0.095$\,{\scriptsize(0.012)} & $0.014$\,{\scriptsize(0.000)} \\
Weibull($k{=}2$) & $0.95$ & $0.5$ & tail-optimal & $10{,}000$ & $32.5$\,{\scriptsize(3.3)} & $0.026$\,{\scriptsize(0.003)} & $0.024$\,{\scriptsize(0.003)} & $0.013$\,{\scriptsize(0.001)} & $0.014$\,{\scriptsize(0.000)} \\
Weibull($k{=}2$) & $0.98$ & $0.5$ & tail-optimal & $1{,}000$ & $48.5$\,{\scriptsize(3.5)} & $0.096$\,{\scriptsize(0.012)} & $0.164$\,{\scriptsize(0.014)} & $0.109$\,{\scriptsize(0.013)} & $0.002$\,{\scriptsize(0.000)} \\
Weibull($k{=}2$) & $0.98$ & $0.5$ & tail-optimal & $10{,}000$ & $45.0$\,{\scriptsize(3.5)} & $0.026$\,{\scriptsize(0.004)} & $0.036$\,{\scriptsize(0.004)} & $0.017$\,{\scriptsize(0.002)} & $0.002$\,{\scriptsize(0.000)} \\
Weibull($k{=}2$) & $1$ & $0.5$ & tail-optimal & $1{,}000$ & $49.5$\,{\scriptsize(3.5)} & $0.102$\,{\scriptsize(0.012)} & $0.164$\,{\scriptsize(0.014)} & $0.103$\,{\scriptsize(0.013)} & $0.000$ \\
Weibull($k{=}2$) & $1$ & $0.5$ & tail-optimal & $10{,}000$ & $50.0$\,{\scriptsize(3.5)} & $0.026$\,{\scriptsize(0.004)} & $0.039$\,{\scriptsize(0.004)} & $0.019$\,{\scriptsize(0.002)} & $0.000$ \\
Weibull($k{=}2$) & $1.02$ & $0.5$ & nonident. & $1{,}000$ & $47.0$\,{\scriptsize(3.5)} & $0.113$\,{\scriptsize(0.013)} & $0.179$\,{\scriptsize(0.013)} & $0.097$\,{\scriptsize(0.010)} & $0.000$\,{\scriptsize(0.000)} \\
Weibull($k{=}2$) & $1.02$ & $0.5$ & nonident. & $10{,}000$ & $42.0$\,{\scriptsize(3.5)} & $0.038$\,{\scriptsize(0.005)} & $0.051$\,{\scriptsize(0.005)} & $0.018$\,{\scriptsize(0.002)} & $0.000$\,{\scriptsize(0.000)} \\
Weibull($k{=}2$) & $1.05$ & $0.5$ & nonident. & $1{,}000$ & $46.0$\,{\scriptsize(3.5)} & $0.133$\,{\scriptsize(0.015)} & $0.213$\,{\scriptsize(0.016)} & $0.117$\,{\scriptsize(0.012)} & $0.003$\,{\scriptsize(0.000)} \\
Weibull($k{=}2$) & $1.05$ & $0.5$ & nonident. & $10{,}000$ & $30.5$\,{\scriptsize(3.3)} & $0.058$\,{\scriptsize(0.007)} & $0.072$\,{\scriptsize(0.006)} & $0.022$\,{\scriptsize(0.002)} & $0.003$\,{\scriptsize(0.000)} \\
Weibull($k{=}2$) & $1.1$ & $0.5$ & nonident. & $1{,}000$ & $36.5$\,{\scriptsize(3.4)} & $0.178$\,{\scriptsize(0.016)} & $0.265$\,{\scriptsize(0.016)} & $0.152$\,{\scriptsize(0.014)} & $0.020$\,{\scriptsize(0.000)} \\
Weibull($k{=}2$) & $1.1$ & $0.5$ & nonident. & $10{,}000$ & $18.0$\,{\scriptsize(2.7)} & $0.093$\,{\scriptsize(0.009)} & $0.101$\,{\scriptsize(0.009)} & $0.040$\,{\scriptsize(0.003)} & $0.020$\,{\scriptsize(0.000)} \\
Weibull($k{=}2$) & $1.2$ & $0.5$ & nonident. & $1{,}000$ & $22.5$\,{\scriptsize(3.0)} & $0.274$\,{\scriptsize(0.017)} & $0.341$\,{\scriptsize(0.018)} & $0.259$\,{\scriptsize(0.013)} & $0.137$\,{\scriptsize(0.000)} \\
Weibull($k{=}2$) & $1.2$ & $0.5$ & nonident. & $10{,}000$ & $4.0$\,{\scriptsize(1.4)} & $0.130$\,{\scriptsize(0.008)} & $0.133$\,{\scriptsize(0.008)} & $0.154$\,{\scriptsize(0.002)} & $0.137$\,{\scriptsize(0.000)} \\
\midrule
two-level & $0.9$ & $0.9$ & tail-optimal & $1{,}000$ & $35.0$\,{\scriptsize(3.4)} & $0.307$\,{\scriptsize(0.015)} & $0.400$\,{\scriptsize(0.035)} & $0.360$\,{\scriptsize(0.044)} & $0.212$\,{\scriptsize(0.000)} \\
two-level & $0.9$ & $0.9$ & tail-optimal & $10{,}000$ & $15.0$\,{\scriptsize(2.5)} & $0.218$\,{\scriptsize(0.002)} & $0.089$\,{\scriptsize(0.008)} & $0.059$\,{\scriptsize(0.007)} & $0.212$\,{\scriptsize(0.000)} \\
two-level & $0.9$ & $0.99$ & tail-optimal & $1{,}000$ & $34.5$\,{\scriptsize(3.4)} & $0.323$\,{\scriptsize(0.014)} & $0.384$\,{\scriptsize(0.038)} & $0.325$\,{\scriptsize(0.045)} & $0.234$\,{\scriptsize(0.000)} \\
two-level & $0.9$ & $0.99$ & tail-optimal & $10{,}000$ & $13.5$\,{\scriptsize(2.4)} & $0.242$\,{\scriptsize(0.002)} & $0.100$\,{\scriptsize(0.008)} & $0.074$\,{\scriptsize(0.007)} & $0.234$\,{\scriptsize(0.000)} \\
two-level & $0.9$ & $1$ & tail-optimal & $1{,}000$ & $32.0$\,{\scriptsize(3.3)} & $0.333$\,{\scriptsize(0.016)} & $0.432$\,{\scriptsize(0.042)} & $0.382$\,{\scriptsize(0.044)} & $0.237$\,{\scriptsize(0.000)} \\
two-level & $0.9$ & $1$ & tail-optimal & $10{,}000$ & $16.5$\,{\scriptsize(2.6)} & $0.249$\,{\scriptsize(0.003)} & $0.114$\,{\scriptsize(0.009)} & $0.075$\,{\scriptsize(0.007)} & $0.237$\,{\scriptsize(0.000)} \\
two-level & $1$ & $0.9$ & tail-optimal & $1{,}000$ & $46.0$\,{\scriptsize(3.5)} & $0.183$\,{\scriptsize(0.022)} & $0.379$\,{\scriptsize(0.033)} & $0.316$\,{\scriptsize(0.035)} & $0.000$\,{\scriptsize(0.000)} \\
two-level & $1$ & $0.9$ & tail-optimal & $10{,}000$ & $53.0$\,{\scriptsize(3.5)} & $0.062$\,{\scriptsize(0.008)} & $0.093$\,{\scriptsize(0.009)} & $0.050$\,{\scriptsize(0.007)} & $0.000$ \\
two-level & $1$ & $0.99$ & tail-optimal & $1{,}000$ & $51.5$\,{\scriptsize(3.5)} & $0.163$\,{\scriptsize(0.020)} & $0.326$\,{\scriptsize(0.029)} & $0.327$\,{\scriptsize(0.040)} & $0.017$\,{\scriptsize(0.003)} \\
two-level & $1$ & $0.99$ & tail-optimal & $10{,}000$ & $54.0$\,{\scriptsize(3.5)} & $0.045$\,{\scriptsize(0.006)} & $0.084$\,{\scriptsize(0.008)} & $0.068$\,{\scriptsize(0.008)} & $0.001$\,{\scriptsize(0.000)} \\
two-level & $1$ & $1$ & tail-optimal & $1{,}000$ & $46.5$\,{\scriptsize(3.5)} & $0.112$\,{\scriptsize(0.015)} & $0.300$\,{\scriptsize(0.037)} & $0.383$\,{\scriptsize(0.054)} & $0.015$\,{\scriptsize(0.002)} \\
two-level & $1$ & $1$ & tail-optimal & $10{,}000$ & $54.0$\,{\scriptsize(3.5)} & $0.037$\,{\scriptsize(0.005)} & $0.068$\,{\scriptsize(0.007)} & $0.056$\,{\scriptsize(0.007)} & $0.002$\,{\scriptsize(0.000)} \\
two-level & $1.1$ & $0.9$ & nonident. & $1{,}000$ & $43.0$\,{\scriptsize(3.5)} & $0.265$\,{\scriptsize(0.019)} & $0.435$\,{\scriptsize(0.026)} & $0.487$\,{\scriptsize(0.031)} & $0.121$\,{\scriptsize(0.004)} \\
two-level & $1.1$ & $0.9$ & nonident. & $10{,}000$ & $17.5$\,{\scriptsize(2.7)} & $0.122$\,{\scriptsize(0.008)} & $0.141$\,{\scriptsize(0.009)} & $0.215$\,{\scriptsize(0.008)} & $0.147$\,{\scriptsize(0.002)} \\
two-level & $1.1$ & $0.99$ & nonident. & $1{,}000$ & $41.5$\,{\scriptsize(3.5)} & $0.211$\,{\scriptsize(0.013)} & $0.448$\,{\scriptsize(0.032)} & $0.638$\,{\scriptsize(0.038)} & $0.049$\,{\scriptsize(0.004)} \\
two-level & $1.1$ & $0.99$ & nonident. & $10{,}000$ & $28.5$\,{\scriptsize(3.2)} & $0.085$\,{\scriptsize(0.006)} & $0.148$\,{\scriptsize(0.012)} & $0.312$\,{\scriptsize(0.008)} & $0.009$\,{\scriptsize(0.001)} \\
two-level & $1.1$ & $1$ & fully ident. & $1{,}000$ & $39.0$\,{\scriptsize(3.4)} & $0.216$\,{\scriptsize(0.013)} & $0.465$\,{\scriptsize(0.040)} & $0.648$\,{\scriptsize(0.046)} & $0.044$\,{\scriptsize(0.004)} \\
two-level & $1.1$ & $1$ & fully ident. & $10{,}000$ & $23.5$\,{\scriptsize(3.0)} & $0.077$\,{\scriptsize(0.006)} & $0.130$\,{\scriptsize(0.012)} & $0.306$\,{\scriptsize(0.007)} & $0.004$\,{\scriptsize(0.000)} \\
two-level & $1.5$ & $0.9$ & nonident. & $1{,}000$ & $16.0$\,{\scriptsize(2.6)} & $0.817$\,{\scriptsize(0.087)} & $1.328$\,{\scriptsize(0.162)} & $5.708$\,{\scriptsize(0.048)} & $0.220$\,{\scriptsize(0.015)} \\
two-level & $1.5$ & $0.9$ & nonident. & $10{,}000$ & $0.5$\,{\scriptsize(0.5)} & $0.188$\,{\scriptsize(0.016)} & $0.207$\,{\scriptsize(0.031)} & $5.354$\,{\scriptsize(0.005)} & $0.166$\,{\scriptsize(0.005)} \\
two-level & $1.5$ & $0.99$ & nonident. & $1{,}000$ & $9.5$\,{\scriptsize(2.1)} & $0.365$\,{\scriptsize(0.043)} & $0.824$\,{\scriptsize(0.138)} & $6.146$\,{\scriptsize(0.038)} & $0.052$\,{\scriptsize(0.005)} \\
two-level & $1.5$ & $0.99$ & nonident. & $10{,}000$ & $0.0$ & $0.070$\,{\scriptsize(0.007)} & $0.070$\,{\scriptsize(0.007)} & $5.820$\,{\scriptsize(0.005)} & $0.009$\,{\scriptsize(0.001)} \\
two-level & $1.5$ & $1$ & fully ident. & $1{,}000$ & $14.5$\,{\scriptsize(2.5)} & $0.468$\,{\scriptsize(0.065)} & $1.168$\,{\scriptsize(0.177)} & $6.215$\,{\scriptsize(0.044)} & $0.026$\,{\scriptsize(0.003)} \\
two-level & $1.5$ & $1$ & fully ident. & $10{,}000$ & $0.0$ & $0.073$\,{\scriptsize(0.009)} & $0.073$\,{\scriptsize(0.009)} & $5.844$\,{\scriptsize(0.004)} & $0.004$\,{\scriptsize(0.000)} \\
\bottomrule
\end{tabular}
\end{table} \endgroup

\begingroup \paragraph{The population minimax regret}\label{app:exp-floor-map}
Figure~\ref{fig:delta-heatmap} maps the minimax regret and the minimax
reserve over the boundary pair for $n=3$, on a $36\times36$ grid with
$1<\lambda h_0\le6$ that contains $786$ feasible nonidentifiable cells.
The largest computed $\Delta/\lambda$ on this grid is $2.6\%$, at
$(\lambda h_0,A)=(5.9,2.1)$, and every computed minimax reserve is below
$\lambda$.

\begin{figure}[htbp]
\centering\includegraphics[width=\textwidth]{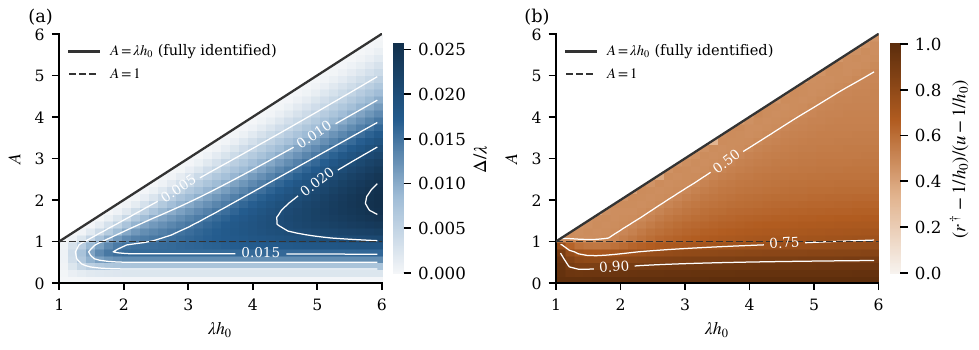}
\caption{(a) Minimax regret $\Delta/\lambda$ and (b) the position $(r^\dagger-1/h_0)/(u-1/h_0)$ of the minimax reserve in the interval \eqref{eq:rdagger-interval}, $u=\min\{\lambda,\lambda-(A-1)/h_0\}$, over the boundary pair $(\lambda h_0,A)$, $n=3$, computed with $200\times200$ bodies, $401$ reserves and three local refinement rounds. No pair lies above the line $A=\lambda h_0$; white lines are contours.}\label{fig:delta-heatmap}
\end{figure} \endgroup

\subsection{When MHR is false}\label{app:exp-misspec}

Lognormal distributions are not MHR: their hazard eventually decreases.
When the observed tail violates MHR, the compatible MHR set is empty, so
$W_G$ under that set is not a useful regret measure. We instead evaluate
actual regret $R_G^*-R_G(r)$ against the distribution's own optimum.
Table~\ref{tab:exp-misspec} uses $400$ shared replications per cell at
$N=3000$. The plug-in continues to post a reserve after projection; it does
not test or certify MHR.

Under light censoring ($\lambda=3$) the projected plug-in rule misses the
observed-tail optimum, and the auto-selector and Drop perform better. At the
larger reserve the ordering changes. The plug-in's mean actual regrets are
$0.00$--$1.69\%$; a small loss under one distribution supplies no worst-case
guarantee when MHR is false.
\begin{table}[htbp]
\centering\small
\caption{Mean actual regret in percent of each lognormal distribution's optimal
revenue ($n=3$, $N=3000$, $400$ paired replications).
Parentheses show Monte-Carlo standard errors.}
\label{tab:exp-misspec}
\begin{tabular}{llccc}
\toprule
Distribution & $\lambda$ & Plug-in & Auto-selector & Drop\\
\midrule
Lognormal$(1.6,0.5)$ & $3$ & $0.34$\,{\scriptsize(0.00)} & $0.03$\,{\scriptsize(0.00)} & $0.03$\,{\scriptsize(0.00)}\\
Lognormal$(1.6,0.5)$ & $5$ & $0.66$\,{\scriptsize(0.02)} & $0.66$\,{\scriptsize(0.02)} & $2.40$\,{\scriptsize(0.01)}\\
Lognormal$(1.6,0.8)$ & $3$ & $1.69$\,{\scriptsize(0.00)} & $0.09$\,{\scriptsize(0.01)} & $0.11$\,{\scriptsize(0.01)}\\
Lognormal$(1.6,0.8)$ & $5$ & $0.00$\,{\scriptsize(0.00)} & $0.06$\,{\scriptsize(0.01)} & $0.29$\,{\scriptsize(0.01)}\\
\bottomrule
\end{tabular}
\end{table}

\subsection{Reserves under different shape constraints}\label{app:exp-regular-more}
Table~\ref{tab:exp-reg-floors} gives the reserves used in
Table~\ref{tab:exp-shape}, with the same observed tail for all three classes.

We compute minimax regret for the regular class using
Theorem~\ref{thm:regular-reduction}. For the
unrestricted class, tail revenue is decreasing above $\lambda$ on all
three instances, so $B_\lambda=0$. Corollary~\ref{cor:unrestricted-computation}
then gives
$r^\dagger_{\mathrm{all}}=\lambda k_\alpha/(\alpha^n+k_\alpha)$ and
$\Delta_{\mathrm{all}}=\alpha^n r^\dagger_{\mathrm{all}}$.

For actual regret in Table~\ref{tab:exp-shape}, we integrate the revenue
derivative of Section~\ref{sec:shared-lemmas} to compute
$100[R_G^*-R_G(r)]/R_G^*$. The true optimum for Weibull shape $k$ and
scale $s$ is $s k^{-1/k}$; Rayleigh$(1)$ uses $k=2$, $s=\sqrt2$.
Direct subtraction of auction revenues agrees within $10^{-11}$ revenue
units. These are population comparisons with exact boundary information
and no sampling error. MHR and regular reserves are numerical;
unrestricted reserves use the exact formula above.
\begin{table}[htbp]
\centering\small
\caption{Population minimax reserves used in Table~\ref{tab:exp-shape}
($n=3$).}
\label{tab:exp-reg-floors}
\begin{tabular}{llccc}
\toprule
Instance & $\lambda$ & $r^\dagger_{\mathrm{MHR}}$ & $r^\dagger_{\mathrm{reg}}$ & $r^\dagger_{\mathrm{all}}$\\
\midrule
Rayleigh(1) & $1.5$ & $1.088$ & $1.132$ & $0.886$ \\
Weibull(1.5,1) & $1$ & $0.827$ & $0.797$ & $0.636$ \\
Weibull(2.5,1.5) & $2$ & $1.128$ & $1.030$ & $0.803$ \\
\bottomrule
\end{tabular}
\end{table}

\subsection{Numerical accuracy}\label{app:exp-oracles}

Table~\ref{tab:exp-oracles-floors} reports numerical MHR minimax regret
from the default $400\times400$ body grid and $801$ reserves with four local refinement
rounds. The last column is the observed change when the number of bodies
per side increases from $150$ to $800$. It is a stabilization check, not a
rigorous error bar. The largest change on the three instances is
below $0.01\%$ of the computed minimax regret.

Corollary~\ref{cor:two-level-oracle} gives the rigorous interval
\[
\Delta^{\mathcal T}-\mathrm{Lip}_r\varepsilon_r
 \le\Delta\le\Delta^{\mathcal T}
 +3\lambda^2\max\{\varepsilon_a,\varepsilon_b\}.
\]
This bound is much more conservative than the observed refinement changes.
The tables use refined numerical estimates, not tightly certified MHR
minimax regret values. The regular scalar solver uses quantile mesh $10^{-7}$; its
certified intervals for minimax regret have width at most $5\times10^{-5}$ percentage
points on these instances.
\begin{table}[htbp]
\centering\small
\caption{Numerical MHR minimax regret in revenue units ($n=3$).
Refinement change compares $150$ with $800$ bodies per side and is
not an error bound.}
\label{tab:exp-oracles-floors}
\begin{tabular}{llcc}
\toprule
Instance & $\lambda$ & Computed minimax regret & Refinement change\\
\midrule
Rayleigh(1) & $1.5$ & $0.02039$ & $0.000001$ \\
Weibull(1.5,1) & $1$ & $0.00699$ & $0.000000$ \\
Weibull(2.5,1.5) & $2$ & $0.04468$ & $0.000002$ \\
\bottomrule
\end{tabular}
\end{table}
\begingroup \paragraph{Conservatism of stability constants}\label{app:exp-constants}
Finite differences of the two-level worst case were compared with the bounds
of Lemma~\ref{lem:stability} and the constants of Appendix~\ref{app:learning-details}, on tested instances with $n\in\{2,3,5\}$ and
$\lambda\in\{0.5,1,2\}$. Table~\ref{tab:exp-constants} gives the largest
measured-to-bound ratios. These diagnostics indicate slack on the tested
instances; they do not establish uniform sharpness or replace a proof.

\begin{table}[htbp]
\centering\small
\caption{Measured finite-difference slopes relative to the proved MHR bounds.}
\label{tab:exp-constants}
\begin{tabular}{lll}
\toprule
Quantity & Bound & Largest measured / bound\\
\midrule
$\mathrm{Lip}_\Theta$ & $3\lambda\max\{1,\lambda\}$ & $0.0486$\\
$\mathrm{Lip}_r$ & $\max\{1,\lambda h_0\}$ & $0.169$\\
\bottomrule
\end{tabular}
\end{table} \endgroup

\begingroup \subsection{Restricted randomized games}\label{app:exp-randomized}

The two-level reduction concerns a single posted reserve and does not
justify restricting nature to two levels against a mixture of reserves.
In an exploratory finite game, the seller mixes over $61$ reserves in
$[0.3\lambda,\lambda]$ plus the computed $r^\dagger$, and nature chooses
among $150\times150$ two-level bodies; a finer seller grid uses $121$
reserves plus $r^\dagger$. At the resulting mixture, a separate seeded
search allows $k$-level hazards for $k\in\{2,3,4,6,8\}$, initialized at
the game's best two-level responses, and every returned response is
re-evaluated with an independent closed form.

On the three synthetic instances of Table~\ref{tab:exp-shape}, the
restricted game's value is $0.9999$--$1.0000$ times the minimax regret
$\Delta$, and the $k$-level best-response search leaves the ratio within
$10^{-4}$ of one. A separate $6\times6$ grid of boundary pairs at
$\lambda=1$ gives restricted values down to $0.911$ of the computed
minimax regret, without the $k$-level check. These finite searches do not determine
$\Delta_{\mathrm{rand}}$: restricting the seller raises the value and
restricting nature lowers it, so the experiments do not show that
randomization has no value. The only guarantee the paper uses is
Proposition~\ref{prop:randomization-bound}. \endgroup

\begingroup \subsection{Data and code availability}\label{app:exp-provenance}

The eBay bid data of Jank and Shmueli~\citep{jank2010modeling} are public, under a
Creative Commons Attribution 4.0 license, at Zenodo record 11259259
(originally distributed at \texttt{modelingonlineauctions.com}); we use
its file \texttt{auction.csv} unchanged. Appendix~\ref{app:exp-real} describes the
records we form from it, the winner imputation, the refit, and the study
that repeats the comparisons of Sections~\ref{sec:exp-shape}
to~\ref{sec:exp-truths} on the refits. Seeds
and sample sharing are described in Appendix~\ref{app:exp-seeds}.
We will archive the experiment code, synthetic-data
generation routines and per-replication outputs on Zenodo. \endgroup

\end{document}